\documentclass[12pt,oneside,openany,a4paper]{report}
\usepackage[english]{babel}
\usepackage[latin1] {inputenc}
\usepackage{amsmath,amsfonts,amssymb,bm,fancyhdr,makeidx}
\usepackage[dvips]{graphicx}
\usepackage{makeidx}
\makeindex

\newcommand{\bgar}{\begin{eqnarray}}
\newcommand{\enar}{\end{eqnarray}}
\newcommand{\eq}[1]{(\ref{eq:#1})}

\newcommand{\be}{\begin{equation}}
\newcommand{\ee}{\end{equation}}
\newcommand{\ket}[1]{ \left. | #1 \right\rangle }
\newcommand{\bra}[1]{\left\langle  #1 | \right.}
\newcommand\ws{{\widetilde\rho}}

\newcommand{\omp}{\omega_{p}}
\newcommand{\omc}{\omega_{c}}
\newcommand{\dtp}{\delta_{p}}
\newcommand{\dtc}{\delta_{c}}

\newcommand{\sca}[2]{\langle#1 \ket{#2}}
\newcommand{\tr}[1]{\ensuremath{{\rm Tr}\left[#1\right]}}
\newcommand{\proj}[1]{\ket{#1}\bra{#1}}

\def\bd{\begin{displaymath}}
\def\ed{\end{displaymath}}

\advance\textwidth by -.5truecm \advance\oddsidemargin by 1.truecm
\advance\baselineskip by \baselineskip

\renewcommand{\chaptermark}[1]{\markboth{Chapter\,\thechapter\ {\bf
#1}}{}}

\fancypagestyle{plain}{\fancyhead{}}

\begin{document}

\begin{titlepage}
\begin{center}
\large{\sc Universit\`a degli Studi di Catania}\\
{\sc Scuola Superiore di Catania}\\
\hbox to \textwidth{\hrulefill}

\vspace{2truecm}

{\sc Filippo Caruso}

\vspace{3truecm}

\large{\sc Storing Quantum Information \\ via Atomic Dark
Resonances}

\vspace{3truecm}

\centerline{\hbox to 2.5truecm{\hrulefill}}

\medskip

{\sc diploma di licenza}\\

\centerline{\hbox to 2.5truecm{\hrulefill}}

\vspace{2truecm}
\begin{minipage}{\textwidth}
\begin{flushright}
\begin{minipage}{0.3\textwidth}
\begin{tabbing}
\= Prof. F. S. Cataliotti \kill
Relatore: \> \\
\> Chiar.mo Prof. F. S. Cataliotti \\

\end{tabbing}
\end{minipage}
\end{flushright}
\end{minipage}

\vspace{2truecm}

\hbox to \textwidth{\hrulefill} {\sc Anno Accademico 2004/2005}

\end{center}
\end{titlepage}

\tableofcontents

\chapter*{Introduction} \pagestyle{fancy}\chaptermark{Introduction}
\addcontentsline{toc}{chapter}{Introduction}

It is now widely accepted that quantum mechanics allows for
fundamentally new forms of communication and
computation\index{quantum computation}. Indeed recently many
interesting new concepts in the field of quantum information such
as quantum computation, quantum cryptography, quantum cloning and
teleportation \cite{q-comp,q-crypt,q-teleport,Furusawa98} have
left the theoretical domain to become commercial prototypes like
quantum key distribution systems (QKD) \cite{idquantique}. In
these protocols information is encoded in delicate quantum states,
like the polarization state of single-photons, and subsequently it
should be manipulated and transported \cite{deVinzenco} without
being destroyed.
\par
In this context on one hand atoms or similar systems like quantum
dots represent reliable and long-lived storage and processing
units. On the other hand photons are ideal carriers of quantum
information \cite{deVinzenco,Zoller-Varenna}: they are fast,
robust but they are difficult to localize and process. Actually
photons play a key role in network quantum computing
\cite{Mabuchi-PRL-1997}, in long-distance, secure quantum
communication and quantum teleportation
\cite{Duan-PRL-2000,Duan-Nature-2001,Julsgaard-Nature-2001,Kuzmich-Nature-2003,van-der-Wal-Science-2003}.
As an example we can cite the application to teleportation which
is of particular interest because of its potentials for quantum
information processing with linear optical elements
\cite{Gottesman-Nature-1999,KLM-Nature-2001}. Nevertheless,
photons normally behave as non-interacting particles. This
property ensures that information encoded in optical signals will
be insensitive to environmental disturbances. For this reason
optics has emerged as the preferred method for communicating
information. In contrast, the processing of information requires
interactions between signal carriers, that is, either between
different photons or photons and electrons. Therefore one of the
main challenges of nonlinear optical science is the ``tailoring''
of material properties to enhance such interactions, while
minimizing the role of destructive processes such as photon
absorption.
\par
Therefore today's challenge is to interface the photons to the
atoms in order to realize a quantum network. One of the essential
ingredients for this idea is a reliable quantum memory capable of
a faithful storage and a prompt release of the quantum states of
the photons. We need to develop a technique for coherent transfer
of quantum information carried by light to atoms and vice versa
and in order to achieve a unidirectional transfer (from field to
atoms or vice versa) an explicit time dependent control mechanism
is required.
\par
Optical storage has been already investigated for classical data.
Particularly interesting are techniques based on Raman photon
echos \cite{LeungOptCom82} as they combine the long lifetime of
ground-state hyperfine or Zeeman coherences for storage with data
transfer by light at optical frequencies \cite{HemmerOL94}.
Nevertheless, while these techniques are very powerful for
high-capacity storage of {\it classical} optical data, they cannot
be used for {\it quantum} memory purposes; indeed they employ
direct or dressed-state optical pumping and thus contain
dissipative elements or have other limitations in the transfer
process between light and matter. As a consequence they do not
operate on the level of individual photons and cannot be applied
to quantum information processes.
\par
The conceptually simplest approach to a {\it quantum} memory for
light is to ``store'' the state of a single photon in an
individual atom. This approach involves a coherent absorption and
emission of single photons by single atoms and it is very
inefficient because the single-atom absorption cross-section is
very small. A very elegant solution to this problem is provided by
cavity QED \cite{cavity-QED}. Indeed placing an atom in a high-$Q$
resonator effectively enhances its cross-section by the number of
photon round-trips during the ring-down time and thus makes an
effective transfer possible \cite{Briegel-LectNotes-1999}. Raman
adiabatic passage techniques \cite{STIRAP} with time-dependent
external control fields can be used to implement a directed but
reversible transfer of the quantum state of a photon to the atom
(i.e. {\it coherent} absorption). However, despite the enormous
experimental progress in this field \cite{cavity-QED-prob}, it is
technically very challenging to achieve the necessary
strong-coupling regime. In addition the single-atom system is by
construction highly susceptible to the loss of atoms and the speed
of operations is limited by the large $Q$-factor. On the other
hand if atomic ensembles are used rather than individual atoms no
such requirements exists and coherent and reversible transfer
techniques for individual photon wavepackets
\cite{Grobe-PRL-1997,Lukin-PRL-2000,Fleischhauer-PRL-2000,Phillips-PRL-2001,Liu-Nature-2001,Fleischhauer-PRA-2002,Lukin-RMP-2003}
and cw light fields
\cite{Kuzmich-PRL-1997,Hald-PRL-1999,Kuzmich-PRL-2000,Schori-PRL-2002}
have been proposed and in part experimentally implemented.
\par
Recently the authors in
\cite{Lukin-PRL-2000,Fleischhauer-PRL-2000,Fl00-OptCom} have
proposed a technique based on an adiabatic transfer of the quantum
state of photons to collective atomic excitations
(\textit{dark--state polaritons}, \textbf{DSP}) \textit{using
electromagnetically induced transparency} (\textbf{EIT}) in
three--level atomic schemes \cite{EIT}.
\par In this thesis we investigate how to obtain a quantum
memory of a coherent state with atomic systems and we point out
that it is possible to compensate the unavoidable losses using the
amplification without inversion in the EIT regime. For this aim we
analyze in detail the propagation of a coherent light pulse
through a medium under the conditions for gain without inversion.
Moreover we introduce a quantum memory for polarized photons with
a four--level system and investigate the scattering of dark--state
polaritons in a tripod configuration.
\newline
\newline
\par
The layout of this thesis is as follows.
\newline
\par
Chapter 1 shows a brief review of some concepts of Quantum Optics
and, in particular, the essence of the electromagnetically induced
transparency. We analyze a three--level system, interacting with
two laser fields, in which destructive quantum interference
appears and no atomic population is promoted to the excited
states, leading to a vanishing light absorption. In these
conditions, the narrow transparency resonance is accompanied by a
very steep variation of the refractive index with frequency and
therefore a strong variation of the group velocity in light
propagation in an EIT medium. Then we summarize some results of
classical electromagnetic theory and from the Scroedinger equation
we derive expressions for density matrix, i.e. optical Bloch
equations, and for the expression for the susceptibility.
\newline
\par
Chapter 2 is devoted entirely to the propagation of a gaussian
pulse along a cigar-shaped cloud of atoms in EIT regime; we derive
expressions for the group velocity and we calculate the
expectation value after transmission of the probe pulse
normal-order Poynting vector. When the central frequency of the
pulse is resonant with an atomic transition, we show that it is
possible to amplify a slow propagating pulse without population
inversion. We also analyze the regime of anomalous light
propagation showing that it is possible to observe superluminal
energy propagation\index{superluminal energy propagation}.
Particularly we show these results for both cold and hot atoms. In
this last case we analyze a realistic system in a 10 cm long cell
containing $^{87}$Rb at a temperature of $35^\circ$C and with a
density equal to $5.296 \cdot 10^7$ $atoms/cm^3$.
\newline
\par
Chapter 3 discusses how to imprint the information carried by the
photons onto the atoms, specifically as a coherent pattern of
atomic spins. The procedure is reversible and the information
stored in the atomic spins can later be transferred back to the
light field, reconstituting the original pulse. Therefore we
analyze the propagation of a quantum field in an EIT medium
sustaining ``dark state polaritons'' in a quasi-particle picture.
Moreover we study the decoherence effects in this quantum memory
for photons, by analyzing the fidelity of the quantum state
transfer.
\newline
\par
Chapter 4 discusses the emergence of parastatistics in the
quasi-particle picture in gain medium. Indeed the dark--state
polaritons obey generalized bosons commutation relations that
describe the mapping from bosons to fermions during the stopping
of light and vice verse in the release. A deformation boson scheme
is connected to this mapping and the Pauli principle is described
by an effective repulsive interaction between the dark--state
polaritons.
\newline
\par
Chapter 5 introduces a polarization quantum memory for photons by
using a tripod atomic configuration in which two ideal EIT windows
appear. Therefore we study the scattering of two dark--state
polaritons (DSP) and we show that they present a solitonic
behavior.
\newline
\par
In Appendix \ref{lwi} we show an application of EIT effect, based
on the possibility to lase without population inversion
(\textit{LWI}). Afterwards we emphasize the concept of lasing
without inversion through an original approach; indeed, reviewing
the Einstein theory about light-matter interaction, there is a
important relation between the possibility of lasing without
inversion and a symmetry breaking between the Einstein B
coefficients. In Appendix \ref{einstein1} we discuss causality in
the regime of anomalous light propagation showing that no
contradiction is present. In Appendix \ref{fidelity} we recall an
important concept of the quantum information theory, the
\textit{fidelity}, and finally in Appendix \ref{noclon} we show
the proof of \textit{No-Cloning Theorem}.

\chapter{Dark resonance in three--level atomic systems}

\section{The story of the ``three--level system''}
Three--level systems \index{three--level system} have been the
object of extensive studies, both theoretically and
experimentally, for the past thirty years. The reason for this
prolonged interest must be searched in the fact that a
three--level system is the test model for quantum interference
effects to appear on a macroscopic scale. The possibility of
completely changing the absorptive and dispersive characteristic
of a medium at a given frequency by applying a coherent field at a
different frequency is both intriguing and surprising.
 \par
As early as 1933 \textbf{Weisskopf} \cite{weiss}\index{Weisskopf}
used a three--level model to predict spectral narrowing and
frequency shift of resonant fluorescence due to narrow-band
optical pumping. No experimental confirmation was possible at that
time given the absence of a narrow-band spectral source. In 1955
\textbf{Autler} and \textbf{Townes} \cite{autler} demonstrated
that in presence of a strong coupling microwave field the resonant
absorption of a probe field coupled to a different transition was
split into a doublet (\textbf{AC Stark splitting} or
\textbf{Autler-Townes effect}\index{Autler-Townes effect}) [see
Sec. \ref{ac}].
\par
In 1976 in Pisa, the group of \textbf{A. Gozzini}
\cite{gozzini}\index{Gozzini} observed a sudden drop of
fluorescence in a sodium vapor where a three--level system with
two ground and an excited level was irradiated by two modes of a
dye laser. In the sodium cell an inhomogeneous magnetic field was
applied almost along the laser propagation axis and the
fluorescence drop appeared as a \textbf{dark line} \index{dark
line} in the fluorescent image. The effect was soon recognized to
be due to optical pumping of atoms in a coherent superposition of
the two ground states which was uncoupled from the laser light
(\textit{\textbf{Coherent Population Trapping (CPT) or Dark
resonance}})\index{coherent population trapping}\index{dark
resonance} \cite{arim,ari}. The same effect was independently
investigated first theoretically in a system where the three
levels were arranged in cascade by Whitley \index{Whitley} and
Stroud \index{Stroud}\cite{whit} then experimentally once again in
sodium in a system with two ground and one excited level by Gray,
Whitley and Stroud \cite{gray}. In the last eighties the
\textit{\textbf{Velocity Selective Coherent Population Trapping}}
(VSCPT)\index{VSCPT} \cite{aspect} method took advantage of dark
resonance to cool and trap atoms below the one-photon-recoil
limit. Renewed interest was brought into the field in the same
years when it was recognized that three--levels systems could
provide amplification and eventually lasing without population
inversion (AWI, LWI). In 1991 Harris \cite{EIT} called
``Electromagnetically Induced Transparency'' (EIT) \cite{harris1}
the interference effect leading to a reduction in absorption in
the center of an Autler-Townes doublet.
 \par
Today this physical system is at the basis of all the recent
experiments on slow light propagation
\cite{hau,Phillips-PRL-2001,Liu-Nature-2001} and speculations
about possible realizations of quantum memories
\cite{Fleischhauer-PRA-2002}, quantum phase gates \cite{ottaviani}
and photon--counters with unprecedented efficiency
\cite{Imamoglu-PRL-2002,Lukin-RMP-2003}.

\section{EIT domain}\label{EIT}

\textbf{Electromagnetically induced transparency} (\textit{EIT})
\index{EIT}\index{electromagnetically induced trans-\\parency} is a
quantum interference effect that permits the propagation of light
through an otherwise opaque atomic medium; a ``coupling'' laser is
used to create the interference necessary to allow the
transmission of resonant probe pulses.
\par
In general let us recall that the strength of the interaction
between light and atoms is a function of the wavelength or
frequency of light. When the light frequency matches the frequency
of a particular atomic transition, a resonance condition occurs
and the optical response of the medium is greatly enhanced. Light
propagation is then accompanied by strong absorption and
dispersion, as the atoms are actively promoted into fluorescing
excited states. \cite{harris3}

\begin{figure} [th!]
\begin{center}
\includegraphics[width=.7 \textwidth]{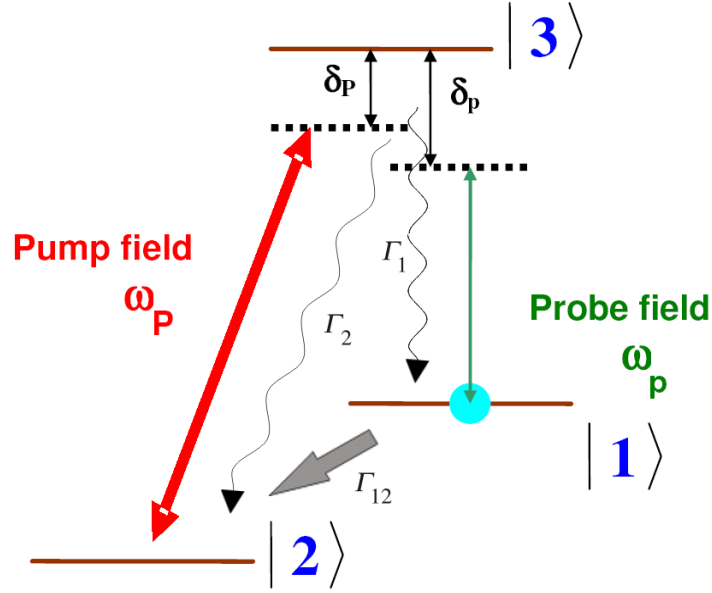}
\caption{The three--level system interacting with two laser fields
($\Lambda$--configuration).} \label{fig1}
\end{center}
\end{figure}

In order to understand in detail EIT effect, let us consider the
situation in which the atoms have a pair of lower energy states
($|1\rangle$ and $|2\rangle$ in Fig. \ref{fig1}) in each of which
the atoms can live for a long time. Such is the case for sublevels
of different angular momentum (spin) within the electronic ground
state of alkali atoms. In order to modify the propagation through
this atomic medium of a light field (\textit{probe field}) that
couples the ground state $|1\rangle$ to an electronically excited
state $|3\rangle$, one can apply a second ``control'' field
(\textit{pump field}) that is quasi resonant with the transition
$|3\rangle \leftrightarrow |2\rangle$. Particularly in the scheme
shown in Fig. \ref{fig1} the states $\ket{3}$ and $\ket {1}$ are
coupled by a weak \textit{probe} beam at frequency $\omega_{p}$
while a stronger \textit{pump} beam at frequency
 $\omega_{P}$ couples the states $\ket {2}$ and $\ket{3}$.
\par
Such a system exhibits a new set of coherent phenomena mainly
correlated with population trapping and quantum interference. For
any combination of intensities of the two fields, there will be
superpositions of the atomic states, $|1\rangle$ and $|2\rangle$,
that is in counterphase with the field, such combination can not
absorb light and is therefore called \textbf{dark state} (there is
also an in-phase component which is named \textbf{bright state}).
In other terms, the two possible pathways in which light can be
absorbed by atoms ($|1\rangle \mapsto |3\rangle $ and $|2\rangle
\mapsto |3\rangle$) can interfere and cancel each other. With such
destructive quantum interference, none of the atoms are promoted
to the excited states, leading to a vanishing light absorption.
\par
From the theoretical point of view it is possible to write the
free-Hamiltonian $\hat{H}_{0}$ and the atom-laser interaction
Hamiltonian $\hat H_I$ as follows:
 \bgar
 \hat{H_{0}}=\sum_{j=1}^{3}\hbar \omega_{j} |j\rangle \langle j|
 + \hbar \omega_{p} \hat{a}_{p}^{\dagger} \hat{a}_{p} + \hbar \omega_{P} \hat{a}_{P}^{\dagger}
 \hat{a}_{P}  \\
 \hat{H}_{I}=\hbar g_{p} \hat{a}_{p} |3 \rangle \langle 1 | +
 \hbar g_{P} \hat{a}_{P} |3 \rangle \langle 2 | +
 h.c. \label{NC}
 \enar
where $\hat a$ and $\hat a^{\dagger}$ are the annihilation and
creation operators for the two fields and $g$ are the relative
coupling coefficients as shown in Sec. \ref{obe}.
\par
Now, let us consider the following two orthogonal linear
combinations of the lower states ($|1\rangle$,\ $|2\rangle$):
 \be
 |C\rangle=\frac{1}{(\Omega_{p}^{2}+\Omega_{P}^{2})^{1/2}}(\Omega_{p} |1\rangle + \Omega_{P} |2
 \rangle)
  \ee
 \be
 |NC\rangle=\frac{1}{(\Omega_{p}^{2}+\Omega_{P}^{2})^{1/2}}(\Omega_{P} |1\rangle - \Omega_{p} |2
 \rangle)
  \ee
where $\Omega_{p}$ and $\Omega_{P}$ are coefficients proportional,
respectively, to $\hbar g_{p} \langle \hat a_p
 \rangle$ and $\hbar g_{P} \langle \hat a_P
 \rangle$.
\par
This defines the uncoupled and coupled states which have the
property that, according to the atom-laser interaction Hamiltonian
of Eq. (\ref{NC}), the transition matrix element between
$|NC\rangle$ and $|3\rangle$ vanishes:
 \be
 \langle 3 |\hat{H}_{I}| NC \rangle = 0
 \ee
 whereas
 \be
  \langle 3 |\hat{H}_{I}| C \rangle \neq 0
  \ee
Consequently, an atom in the uncoupled state $|NC\rangle$ cannot
absorb photons and cannot be excited to $|3 \rangle$. Moreover for
an atom prepared in the $|NC \rangle$ state, the Scroedinger
equation under the Hamiltonian $\hat{H}_{0}+\hat{H}_{I}$ results
in:
 \be
 \frac{d}{dt}|NC \rangle = \frac{1}{i \hbar}
 (\hat{H}_{0}+\hat{H}_{I}) |NC \rangle = 0
  \ee
Thus an atom prepared in $|NC \rangle$ remains in this state and
can leave it neither by the free evolution (effect of the free
Hamiltonian $\hat{H}_{0}$) nor by absorption of a laser photon
(effect of the atom-laser interaction $\hat{H}_{I}$). Besides,
because $|NC \rangle$ is a linear combination of the two ground
states, and is radiatively stable, the atom cannot leave $|NC
\rangle$ either by spontaneous emission. This is the essence of
the \textbf{riga nera} \cite{gozzini} (\textit{dark resonance}) or
\textbf{electromagnetically induced transparency} (\textbf{EIT}).

\begin{figure} [!ht]
\begin{center}
\includegraphics[angle=0, width=.8\textwidth ]{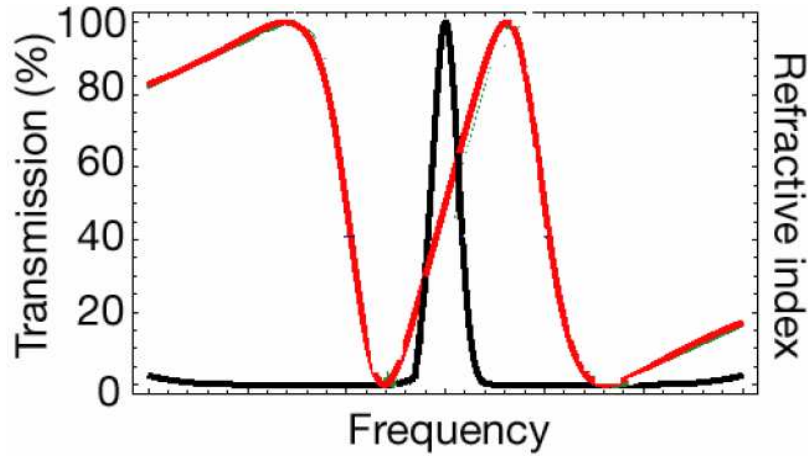}
\caption{Spectrum of transmission and refractive index
corresponding to EIT. Rapid variation of the refractive index (red
curve) causes a reduction of group velocity. \cite{hau}}
\label{fig2}\end{center}
\end{figure}

The reduction in the probe absorption can also be explained
\cite{EIT,ari} as due to a combination of AC-Stark splitting (see
Sec. \ref{ac}) and destructive quantum interference in the
absorption of a probe photon from the two coherent superpositions
of lower states to the excited state $\ket{3}$. This interference
is analogous to that seen if mutually coherent optical fields are
interfered such as in the common Young interferometer. Yet another
way to view this effect is in terms of the creation of a new class
of laser dressed matter in which laser fields and atoms have
become strongly coupled.

\section{AC-Stark Effect} \label{ac}

In general when the laser is resonant with an atomic transition
two effects come into play: the \textbf{\textit{Rabi
oscillations}}\index{Rabi oscillations} and the \textit{\textbf{AC
Stark Effect}} or \textit{\textbf{Autler-Townes Doublet}}\index{AC
Stark effect} \cite{autler}.
\par
First of all, because of coupling laser, the electrons cycle back
and forth between the two levels: these are the Rabi oscillations
and their characteristic frequency is the so--called
\textit{\textbf{Rabi frequency}}.
\par
Due to this rapid oscillation the atom acquires an induced
electric dipole that interacts with the laser electric field
splitting both the upper and the lower level of the transition
into two sub levels - one higher in energy the other lower. This
splitting of the energy levels is caused by the oscillating
electric field of the laser beam \cite{pavone}: it is the
\textit{\textbf{AC Stark effect}}\index{AC Stark effect}.
\par
Both of these effects are dependent on the generalized
\textbf{Rabi frequency} ($\Omega$)\index{Rabi frequency} that
gives us information concerning how effectively the laser can
stimulate transitions in the atom and it is dependent on:

\begin{itemize}

\item[1)]
Laser field strength ($\vec{E}$)

\item[2)] Dipole moment of the transition
($\vec{\mu}$)\index{dipole moment}

\item[3)]
Difference between the laser and the atomic transition
frequencies, i.e. detuning($\delta$).

\end{itemize}

It is defined as:

\be \Omega_R=\sqrt{\left(\frac{\vec{\mu}\cdot
\vec{E}}{\hbar}\right)^2+\delta^2}\label{Rabi} \ee

The new levels are separated by $\Omega_R$ and with a population
dependent on the laser detuning.

\section{Macroscopic theory of absorption}\label{cprobe}

Before considering the density-matrix approach to the three--level
system, it is convenient to summarize the relevant results of
classical electromagnetic theory \cite{loudon}. Let us analyze a
gas of atoms in a cavity as a dielectric medium: the presence of
this dielectric leads to the generation of a polarization
\textbf{P} by an applied electric field \textbf{E}. By definition,
the polarization is equal to: \be
 \textbf{P}=\frac{N}{V} \ \textbf{d}
 \ee
where $\frac{N}{V}$ is the atomic density and $\textbf{d}$ is the
electric dipole moment. For electric fields that are not too
strong, the polarization is proportional to the field, \be
 \textbf{P}=\varepsilon_{0}\chi \textbf{E}
 \ee
where $\chi$ is the linear electric susceptibility and
$\varepsilon_{0}$ is the vacuum electric permittivity. The
susceptibility is a function of the frequency $\omega$ of the
applied field, whose form depends on the energy levels and wave
functions of the atoms that make up the dielectric. Maxwell's
equations still have wavelike solutions, but the relation between
frequency, $\omega$, and wavevector, $k$, has the following
general expression: \be \label{disp}
 \Big(\frac{kc}{\omega} \Big)^{2}=1+\chi
 \ee
which reduces to known dispersion equation $\omega=k c$ in the
free-space limit $\chi=0$; $c$ is the velocity of light in vacuum.
\par
The quantity $1+\chi$ is known as the dielectric constant; of
course, it is constant only in the sense of being independent of
$\textbf{E}$ but its magnitude is a function of the frequency.
Moreover, the susceptibility is generally a complex quantity and
we write \be
 \chi=\chi'+i \chi''
 \ee
where $\chi'$ and $\chi''$ are, respectively, the real and
imaginary parts of $\chi$.
\par
It is conventional to write the square root of Eq. (\ref{disp}) as
\be \label{sub}
 \frac{kc}{\omega}=\eta+i \kappa
 \ee
where $\eta$ and $\kappa$, so defined, are, respectively, the
refractive index and extinction coefficient. Comparison of the
real and imaginary parts of Eq. (\ref{disp}) after substitution of
Eq. (\ref{sub}) yields
 \bgar
 \eta^{2}-\kappa^{2}=1+\chi' \nonumber\\
 2 \eta \kappa= \chi''
 \enar
These equations will be used to determine the frequency dependence
of $\eta$ and $\kappa$ once the frequency-dependent susceptibility
is knows through a off-diagonal term of the density operator.

\section{The Optical Bloch Equations (OBE)}
\label{obe}\index{OBE}\index{optical Bloch equations}

Let us investigate in detail the scheme shown in Fig. \ref{fig1}.
A generic atomic state can be written as a linear superposition of
the atomic eigenstates \be
 |\psi\rangle=C_{1}|1\rangle+C_{2}|2\rangle+C_{3}|3\rangle
 \ee
and the density matrix operator, $\hat \rho$, is given by the
outer product of two wave functions,
\begin{displaymath}
\hat \rho = |\psi\rangle \langle \psi| \Longrightarrow \left(
\begin{array}{ccc}
|C_{1}|^{2} & C_{1}C_{2}^{*} & C_{1}C_{3}^{*} \\
C_{2}C_{1}^{*} & |C_{2}|^{2} & C_{2}C_{3}^{*} \\
C_{3}C_{1}^{*} & C_{3}C_{2}^{*} & |C_{3}|^{2}
\end{array} \right)= \left(
\begin{array}{ccc}
\rho_{11} & \rho_{12} & \rho_{13} \\
\rho_{21} & \rho_{22} & \rho_{23} \\
\rho_{31} & \rho_{32} & \rho_{33}
\end{array} \right)
\end{displaymath}
The diagonal terms give us the probability of finding the atom in
one of the three levels while the transverse terms are
proportional to the complex dipole moments. The off-diagonal
elements are generally complex and they satisfy the following
relations: \be
 \rho_{21}=\rho_{12}^{*} \ \ \ \ \ \ \ \rho_{13}=\rho_{31}^{*}   \ \ \
 \ \ \ \  \rho_{23}=\rho_{32}^{*}
 \ee
The expectation value of any operator ($\hat{A}$) can now be
written in terms
 of $\rho$ as
\be
 \langle \hat{A} \rangle = Tr(\hat A \hat \rho) = \sum_{i,j=1}^{3} \rho_{ij} A_{ij}
 \ee
In our case, the hamiltonian operator of the three--level system
is the following one:
 \bgar \hat{H}=\hat{H_{0}}+\hat{H_{I}} \nonumber
 \enar
where the free- and interaction-hamiltonian are
 \bgar
 \hat{H_{0}}=\sum_{j=1}^{3}\hbar \omega_{j} |j\rangle \langle j|
 + \hbar \omega_{p} \hat{a}_{p}^{\dagger} \hat{a}_{p} + \hbar \omega_{P} \hat{a}_{P}^{\dagger}
 \hat{a}_{P}  \enar
 \bgar
 \hat{H}_{I}=\hbar g_{p} \hat{a}_{p} |3 \rangle \langle 1 | +
 \hbar g_{P} \hat{a}_{P} |3 \rangle \langle 2 | +
 h.c.
 \enar
 where the notation is the same as in Sec. \ref{EIT}.\par
Making a transformation to a rotating frame and performing the
\textit{rotating wave approximation} (\textbf{RWA}), which
consists in neglecting the antiresonant term containing the sum
frequency and therefore very rapidly oscillating, we obtain
 \bgar
 \tilde{H}=\sum_{j=1}^{3} \hbar \omega_{j} |j\rangle \langle j| +
 (\hbar g_{p} e^{- i \omega_{p} t}\hat{a}_{p} |3 \rangle \langle 1 | +
 \hbar g_{P} e^{- i \omega_{P} t} \hat{a}_{P} |3 \rangle \langle 2 | +
 h.c.) \nonumber
 \enar
Note that the approximation is justified because the effect of the
terms that oscillate at frequency $\omega_{p}+\omega_{P}$ is
negligible compared to the effect of the terms that oscillate at
frequency $\omega_{p}-\omega_{P}$ when $\omega_{P}$ is close to
$\omega_{p}$.
\par
For a set of \textit{classical} fields (i.e. \textit{coherent
states}), we have
 \bgar
 \hbar g_{p} \hat{a}_{p} \longrightarrow \hbar g_{p} \langle \hat
a_p
 \rangle \equiv -\frac{\hbar}{2} \Omega_{p} e^{-i \phi_{p}} =
 \vec{p}_{31} \cdot \vec{E} = e \langle3|\vec{x}|1 \rangle \cdot
 \widehat{\epsilon}_{p} \ \epsilon_{p}
 \enar
  \bgar
 \hbar g_{P} \hat{a}_{P} \longrightarrow \hbar g_{P} \langle \hat
a_P
 \rangle \equiv -\frac{\hbar}{2} \Omega_{P} e^{-i \phi_{P}} =
 \vec{p}_{32} \cdot \vec{E} = e \langle3|\vec{x}|2 \rangle \cdot
 \widehat{\epsilon}_{P} \ \epsilon_{P}
  \enar
where $\vec{p}$ is the transition dipole moment, $\Omega_{p}$ and
$\Omega_{P}$ are the Rabi frequencies (real), $\phi_p$ and
$\phi_P$ are the relative phases, $e$ is the electron charge,
$\widehat{\epsilon}_{p}$ and $\widehat{\epsilon}_{P}$ are the unit
polarization vector and $\epsilon_{p}$ and $\epsilon_{P}$ are the
electric field amplitudes. Then the hamiltonian operator is
 \bgar
 \tilde{H}=\sum_{j=1}^{3} \hbar \omega_{j} |j\rangle \langle j| +
  \{-\frac{\hbar}{2} \Omega_{p} e^{- i \phi_{p}} e^{- i \omega_{p} t} |3 \rangle \langle 1
  |- \frac{\hbar}{2} \Omega_{P} e^{- i \phi_{P}} e^{- i \omega_{P} t}  |3 \rangle \langle 2 | +
 h.c.\} \nonumber
 \enar
In order to find the equation of motion for the density matrix
elements, $\rho_{ij}$, starting from the Schr\"{o}dinger equation,
we consider the Liouville equation for $\rho$: \be
\dot{\hat{\rho}}=-\frac{i}{\hbar}[\tilde{H},\hat{\rho}]-\dot
{\hat{\rho}}_{int} - \dot {\hat{\rho}}_{ext}
 \ee
 Generally it is possible to divide relaxation phenomena into two
groups: in the first one the system relaxes towards external
states $\dot {\hat{\rho}}_{ext}$ (\textit{external relaxation})
and in the second one it relaxes towards internal states $\dot
{\hat{\rho}}_{int}$ (\textit{internal relaxation}). These rates
are given by\footnote{\{. , .\} denotes the anti-commutator.}
 \bgar
\dot {\hat{\rho}}^{i \longrightarrow j}_{int}=-\frac{T_{ij}}{2}
\{\ket{i} \bra{i}, \hat \rho \} +T_{ij} \rho_{ii} \ket{j} \bra{j}
 \enar
 \bgar
\dot{\hat{ \rho}}_{ext}= \frac{1}{2} \Big\{\sum_i{\zeta_i
\ket{i}\bra{i}} ,\hat \rho\Big\}
 \enar
where $T_{ij}$ and $\zeta_i$ are, respectively, the internal and
external decay rates; $\dot \rho^{i \longrightarrow j}_{int}$
represents the internal decay from level $\ket{i}$ into level
$\ket{j}$. Note that in Fig. \ref{fig1} we
have\footnote{$\Gamma_1$ and $\Gamma_2$ are, respectively, the
transition linewidths of the levels $\ket{1}$ and $\ket{2}$.}
$\Gamma_1=T_{31}$, $\Gamma_2=T_{32}$ and $\Gamma_{12}=T_{12}$ and
we neglect the external decays. Actually there is no decay between
two lower-states because they are meta-stable but we introduce an
incoherent RF field that simulates a loss from $\ket{1}$ and
 places population back into $\ket{2}$ (as it will
be clear in Sec. \ref{article}).
\par
By treating this three--level system as closed, the equations of
motion for the density matrix elements, $\rho_{ij}$, are:
 \bgar
 \dot{\rho}_{31}&=&-[\frac{1}{2} (\Gamma_{1}+\Gamma_{2}+\Gamma_{12}) + i \omega_{31}]
 \rho_{31} + \frac{i}{2} \Omega_{P} e^{- i \phi_{P}} e^{-
 i \omega_{P} t} \rho_{21}+ \nonumber \\
 &&-\frac{i}{2} \Omega_{p} e^{- i \phi_{p}} e^{-
 i \omega_{p} t} (\rho_{33}-\rho_{11}) \nonumber \\
 \dot{\rho}_{32}&=&-[\frac{1}{2} (\Gamma_{1}+\Gamma_{2}) + i \omega_{32}]
 \rho_{32} + \frac{i}{2} \Omega_{p} e^{- i \phi_{p}} e^{-
 i \omega_{p} t} \rho_{12}+ \nonumber \\
 &&-\frac{i}{2} \Omega_{P} e^{- i \phi_{P}} e^{-
 i \omega_{P} t} (\rho_{33}-\rho_{22}) \ \ \ \ \ \ \ \ \ \label{obeq} \\
 \dot{\rho}_{21}&=&-[\Gamma_{12}+ i \omega_{21}]
 \rho_{21} + \frac{i}{2} \Omega_{P} e^{- i \phi_{P}} e^{-
 i \omega_{P} t} \rho_{31}- \frac{i}{2} \Omega_{p} e^{- i \phi_{p}} e^{-
 i \omega_{p} t} \rho_{23} \nonumber \enar \bgar
 \dot{\rho}_{33}&=&(\frac{i}{2} \Omega_{p} e^{- i \phi_{p}} e^{-
 i \omega_{p} t} \rho_{13}+\frac{i}{2} \Omega_{P} e^{- i \phi_{P}} e^{-
 i \omega_{P} t} \rho_{23}+h.c.)-(\Gamma_{1}+\Gamma_{2}) \rho_{33} \nonumber \\
\dot{\rho}_{22}&=&(-\frac{i}{2} \Omega_{P} e^{- i \phi_{P}} e^{-i
\omega_{P} t} \rho_{23}+h.c.)+\Gamma_{2} \rho_{33}+\Gamma_{12} \rho_{11} \nonumber \\
 \dot{\rho}_{11}&=&-\dot{\rho}_{22}-\dot{\rho}_{33} \nonumber
 \enar
These are known as \textbf{Optical Bloch Equations}
(OBE)\index{OBE}\index{optical Bloch equations}. They are similar
to equations derived by Bloch to describe the motion of a spin in
an oscillatory magnetic field. The quantum mechanics of the
three--level atom considered here is formally identical to that of
a spin $1$ system. Indeed it is possible to draw many analogies
between the influences of oscillatory fields on the two systems.
\par
These equations can be solved without any further approximations.
They represent a set of six simultaneous equations for the six
independent elements of the atomic density
matrix\footnote{Actually they are five simultaneous equations for
five independent elements of the atomic density matrix with the
constraint $\rho_{11}+\rho_{22}+\rho_{33}=1$}. After obtaining the
steady state solution, we will be able to use the off-diagonals
terms to calculate the susceptibility and therefore the frequency
dependence of the refractive index and extinction coefficient;
after these analytical calculations, it is possible to analyze in
detail \textbf{CPT} (\textit{coherent population trapping}) and
\textbf{EIT} effects.

Let us now introduce the slowly varying variables:
($\Delta_{1}^{'}$, $\Delta_{2}^{'}$, $\Delta_{3}^{'}$
$\longrightarrow$ free parameters)
 \bgar
 \rho_{31} \equiv e^{i \Delta_{1}^{'} t} \tilde{\rho}_{31} \ \ \ \ \
 \rho_{32} \equiv e^{i \Delta_{2}^{'} t} \tilde{\rho}_{32} \ \ \ \ \
 \rho_{21} \equiv e^{i \Delta_{3}^{'} t} \tilde{\rho}_{21} \ \ \ \ \
 \enar
and so we can rewrite the optical Bloch equations as: \bgar
\dot{\tilde{\rho}}_{31}&=&-\tilde{\rho}_{31}
[\frac{1}{2}(\Gamma_{1}+\Gamma_{2}+\Gamma_{12})+ i \Delta_{1}^{'}
+ i \omega_{31}] +\frac{i}{2} \Omega_{P} e^{-i \phi_{P} -i
\omega_{P} t -i \Delta_{1}^{'}t+
i\Delta_{3}^{'}t} \tilde{\rho}_{21}+ \nonumber \\
&&-\frac{i}{2} \Omega_{p} e^{-i \phi_{p} -i \omega_{p} t -
i \Delta_{1}^{'}t} (\rho_{33}-\rho_{11})
\nonumber \\
\dot{\tilde{\rho}}_{32}&=&-\tilde{\rho}_{32}
[\frac{1}{2}(\Gamma_{1}+\Gamma_{2})+ i \Delta_{2}^{'} + i
\omega_{32}] +\frac{i}{2} \Omega_{p} e^{-i
\phi_{p} -i \omega_{p} t -i \Delta_{2}^{'}t-i\Delta_{3}^{'}t} \tilde{\rho}_{12}+ \nonumber \\
&&-\frac{i}{2} \Omega_{P} e^{-i \phi_{P} -i \omega_{P} t -i \Delta_{2}^{'}t} (\rho_{33}-\rho_{22})
\\
\dot{\tilde{\rho}}_{21}&=&-\tilde{\rho}_{21} [\Gamma_{12}+i
\Delta_{3}^{'} + i \omega_{21}] +\frac{i}{2} \Omega_{P} e^{i
\phi_{P} +i \omega_{P} t
-i \Delta_{3}^{'}t+i\Delta_{1}^{'}t} \tilde{\rho}_{31}+ \nonumber \\
&&-\frac{i}{2} \Omega_{p} e^{-i \phi_{p} -i \omega_{p} t-i
\Delta_{3}^{'}t- i \Delta_{2}^{'}t} \tilde \rho_{23} \nonumber
\enar \bgar \dot{{\rho}}_{33}&=&(\frac{i}{2} \Omega_{p} e^{-i
\phi_{p} -i \omega_{p} t -i \Delta_{1}^{'}t}
\tilde{\rho}_{13}+\frac{i}{2} \Omega_{P} e^{-i \phi_{P} -i
\omega_{P} t -i \Delta_{2}^{'}t}
\tilde{\rho}_{23}+h.c.)-(\Gamma_{1}+\Gamma_{2}) \rho_{33}
\nonumber \\
\dot{\rho}_{22}&=&(-\frac{i}{2} \Omega_{P} e^{-i \phi_{P} -i
\omega_{P} t - i \Delta_{2}^{'}t} \tilde{\rho}_{23}+h.c.)+\Gamma_2
\rho_{33}+\Gamma_{12} \rho_{11}
\nonumber \\
 \dot{\rho}_{11}&=&-\dot{\rho}_{22}-\dot{\rho}_{33}
 \nonumber
 \enar
The three free frequency parameters $\Delta^{'}$ can be chosen to
eliminate the oscillating exponentials: \bgar
 \Delta_{1}^{'}=-\omega_{p} \ \ \ \ \Delta_{2}^{'}=-\omega_{P}  \
 \ \ \ \ \Delta_{3}^{'}=\omega_{P}+\Delta_{1}^{'} = \omega_{P} -
 \omega_{p}
 \enar
and therefore one obtains:
 \bgar
 \dot{\tilde{\rho}}_{31}&=&-(\gamma_{1}+\gamma_{3}+i \delta_{p})
 \tilde{\rho}_{31}+\frac{i}{2} \Omega_{P} e^{-i \phi_{P}}
 \tilde{\rho}_{21}-\frac{i}{2} \Omega_{p} e^{-i \phi_{p}}
 (\rho_{33}-\rho_{11}) \nonumber \\
 \dot{\tilde{\rho}}_{32}&=&-(\gamma_{3}+i \delta_{P})
 \tilde{\rho}_{32}+\frac{i}{2} \Omega_{p} e^{-i \phi_{p}}
 \tilde{\rho}_{12}-\frac{i}{2} \Omega_{P} e^{-i \phi_{P}}
 (\rho_{33}-\rho_{22}) \nonumber \\
 \dot{\tilde{\rho}}_{21}&=&-(\gamma_1+i \delta)
 \tilde{\rho}_{21}+\frac{i}{2} \Omega_{P} e^{-i \phi_{P}}
 \tilde{\rho}_{31}-\frac{i}{2} \Omega_{p} e^{-i \phi_{p}}
 \tilde{\rho}_{23} \\
 \dot{\tilde{\rho}}_{33}&=&(\frac{i}{2} \Omega_{p} e^{-i \phi_{p}} \tilde{\rho}_{13}+
 \frac{i}{2} \Omega_{P} e^{-i \phi_{P}} \tilde{\rho}_{23}+h.c.)-2 \gamma_{3} \rho_{33}
\nonumber \\
 \dot{\tilde{\rho}}_{22}&=&(-\frac{i}{2} \Omega_{P} e^{-i \phi_{P}} \tilde{\rho}_{23}+h.c.)+
\Gamma_{2} \rho_{33}+2 \gamma_1 \rho_{11}
 \nonumber \\
 \dot{\rho}_{11}&=&-\dot{\rho}_{22}-\dot{\rho}_{33}
 \nonumber \enar
where the detunings are
 \bgar
 \delta_{p}=\omega_{31}-\omega_{p} \ \ \ \ \delta_{P}=\omega_{32}-\omega_{P} \ \ \ \
 \delta=\omega_{21}+\omega_{P}-\omega_{p}=\delta_{p}-\delta_{P}
 \nonumber
 \enar
and $\gamma_{3} \equiv \frac{1}{2}(\Gamma_{1}+\Gamma_{2})$,
$\gamma_1 \equiv \frac{\Gamma_{12}}{2}$.
\par We calculate the steady-state solution for
$\tilde{\rho}_{31}$, after fixing the populations
(\textit{unsaturated-populations solution}, see Sec. \ref{num}).
\newline
 We obtain:
 \bgar
 \dot{\tilde{\rho}}_{32}=0 \ \ \text{yields} \ \
 \tilde{\rho}_{32}&=&\frac{i
 \Omega_{p}e^{-i\phi_{p}}}{2(\gamma_{3}+i\delta_{P})}
 \tilde{\rho}_{12}-\frac{i \Omega_{P}e^{-i\phi_{P}}}{2(\gamma_{3}+i\delta_{P})}
(\rho_{33}-\rho_{22}) \nonumber \\
\ \ \ \ \ \ \ \ \ \ \ \ \tilde{\rho}_{23}&=&\frac{-i
 \Omega_{p}e^{i\phi_{p}}}{2(\gamma_{3}-i\delta_{P})}
 \tilde{\rho}_{21}+\frac{i \Omega_{P}e^{i\phi_{P}}}{2(\gamma_{3}-i\delta_{P})}
(\rho_{33}-\rho_{22}) \nonumber \\
 \dot{\tilde{\rho}}_{21}=0 \ \ \text{yields} \ \
 \tilde{\rho}_{21}&=&\frac{i
 \Omega_{P}e^{i\phi_{P}}}{2(\gamma_{1}+i\delta)}
 \tilde{\rho}_{31}-\frac{i \Omega_{p}e^{-i\phi_{p}}}{2(\gamma_{1}+i\delta)}
 \tilde{\rho}_{23} \ \ \ \ \ \ \ \ \ \ \ \ \ \ \nonumber \\
 \dot{\tilde{\rho}}_{31}=0 \ \ \text{yields} \ \
 \tilde{\rho}_{31}&=&\frac{i
 \Omega_{P}e^{-i\phi_{P}}}{2(\gamma_{1}+\gamma_{3}+i\delta_{p})}
 \tilde{\rho}_{21}-\frac{i
 \Omega_{p}e^{-i\phi_{p}}}{2(\gamma_{1}+\gamma_{3}+i\delta_{p})}
 (\rho_{33}-\rho_{11}) \ \ \nonumber
 \enar
\newline
If we put $\tilde{\rho}_{21}$ back in $\tilde{\rho}_{31}$, we get:

 \bgar
 \tilde{\rho}_{31}=-\frac{\frac{i \Omega_{p}e^{-i \phi_{p}}}{2(\gamma_{1}+\gamma_{3}+i \delta_{p}))}}
 {1+\frac{\Omega_{P}^2(\gamma_{3}-i \delta_{P})}{(\gamma_{1}+\gamma_{3}+i
 \delta_{p})\Big(4 (\gamma_{3}-i \delta_{P}) (\gamma_{1}+i \delta)+ \Omega_{p}^{2} \Big)}}
 \Bigg\{(\rho_{33}-\rho_{11})- \\
 -\frac{\Omega_{P}^{2}}{4 (\gamma_{3}-i \delta_{P}) (\gamma_{1}+i \delta)+
  \Omega_{p}^{2}} (\rho_{33}-\rho_{22})\Bigg\}
 \enar
Finally we are able to calculate the susceptibility for the
transition $3\rightarrow 1$.
\newline
We have learnt in Sec. \ref{cprobe} that in a gas of atoms in a
cavity, regarded as dielectric medium, there is a polarization
\textbf{P} by an applied electric field \textbf{E} equal to:
 \be
 \textbf{P}=\frac{N}{V} \ \langle d \rangle= \frac{N}{V} \Big[
 \tilde{\rho}_{31} X_{31} e^{-i \omega t} + \tilde{\rho}_{13} X_{13} e^{i \omega
 t} \Big]
 \ee
where $\frac{N}{V}$ is the atomic density and $\langle d \rangle$
is the electric dipole moment calculated between the states
$|1\rangle$ and $|3\rangle$ depending on a generic frequency
$\omega$. For electric fields that are not too strong, the
polarization is proportional to the field,
 \be
\textbf{P}= \chi \varepsilon_{0} \textbf{E}(t)=\frac{1}{2}
\varepsilon_{0} \textbf{E}_{0} \Big[\chi(\omega) e^{-i \omega t} +
\chi(-\omega) e^{i \omega t}  \Big]
 \ee
where $\chi$ is the linear electric susceptibility and
$\textbf{E}(t)=\frac{1}{2} \textbf{E}_{0} \{ e^{-i \omega t} +
e^{i \omega t}\}$.
\newline
Comparing the two expressions for $\textbf{P}$, using the previous
relation for $\rho_{13}$ and the approximation $\Omega_{P} \gg
\Omega_{p}$, we obtain the following important expression for the
susceptibility for the transition $3\rightarrow 1$:

 \bgar
 \label{eq:chi1}
 \chi_{p}& = &  3 \pi  {\cal{N}}_{p} \Gamma_{1} \Big[\frac{ (\delta_{p} -i \gamma_{1})} {(\delta_{p} - i
\gamma_{3})(\delta_{p} -i
\gamma_{1})-(\Omega_{P}/{2})^{2}}(\rho_{11}-\rho_{33})+
\nonumber\\
& & {}-\frac{i}{\gamma_{3}} (\Omega_{P}/{2})^{2}
\frac{(\rho_{33}-\rho_{22})} {(\delta_{p} - i
\gamma_{3})(\delta_{p} -i \gamma_{1})-(\Omega_{P}/{2})^{2}} \Big]
 \enar
\newline
\newline
where ${\cal{N}}_{p}$ is the scaled sample average density
$\lambda_{p}^{3} N/V$, $\lambda_{p}$ is the probe resonant
wavelength and $n_i$ is the normalized population of level
$\ket{i}$ ($\rho_{11}=n_{1}, \ \rho_{22}=n_{2}, \
\rho_{33}=n_{3}$). To be complete we had to include a factor $1/3$
to account for the integrals over space ($2/3$ for a dipole) and
polarization ($1/2$ for linear polarization).

\section{Analytical and numerical results}
\label{num}

As showed above it is possible to find a closed form (without
external and collisional relaxations) for the steady state
solution and for susceptibility, after making some approximations.
\par There are two possible ways to achieve this result:

\begin{itemize}

\item[1)]
\textbf{Unsaturated-populations}\index{unsaturated-populations}:
we fix the populations and we look for the coherence terms (the
off-diagonal elements) in the steady state case.

\item[2)]
\textbf{Saturated-populations}\index{saturated-populations}: we
resolve the complete set of six simultaneous equations for the six
independent elements of the atomic density matrix.

\end{itemize}

\begin{figure} [!ht]
\begin{center}
\includegraphics[width=.5\textwidth ]{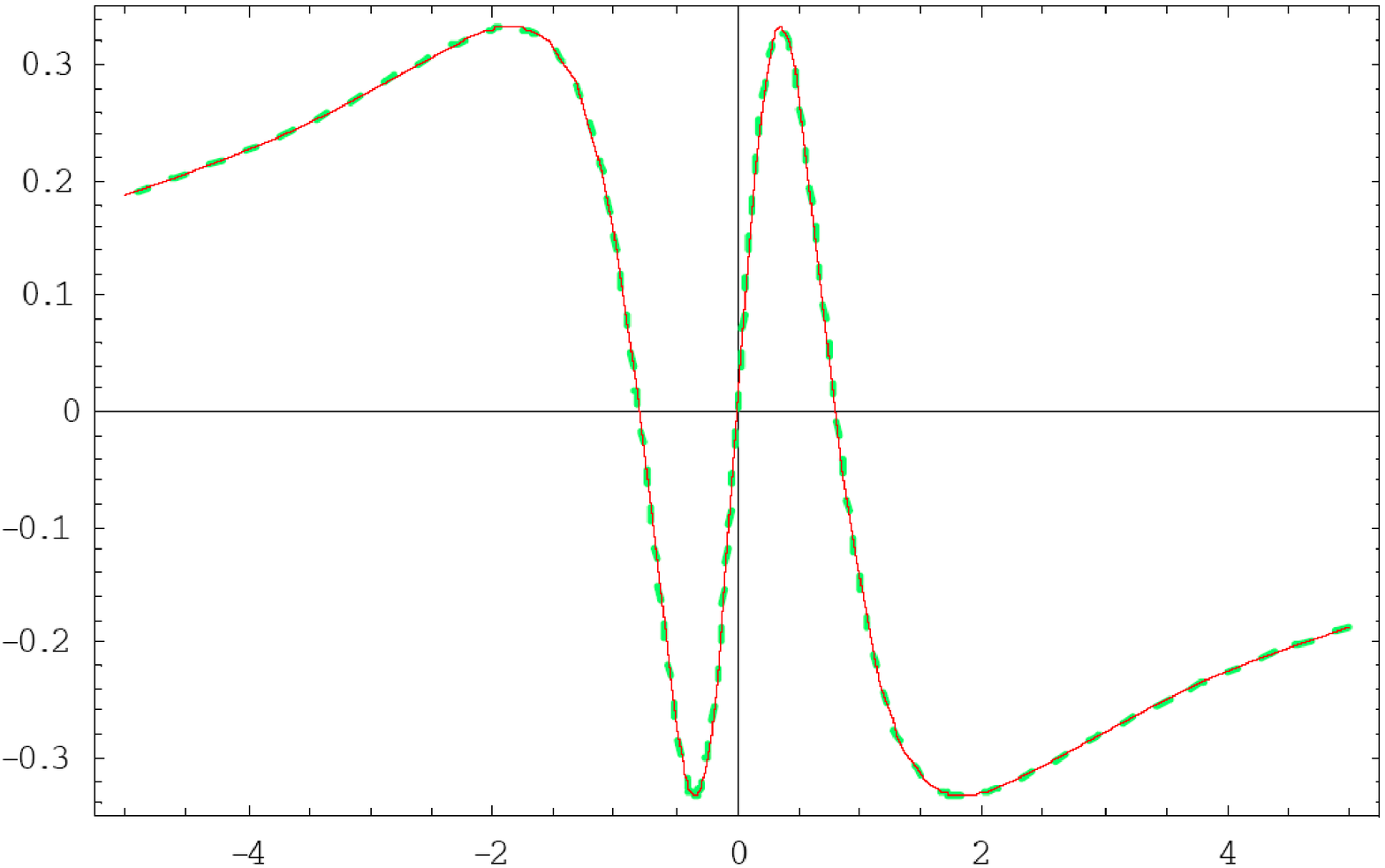}
\includegraphics[width=.49\textwidth ]{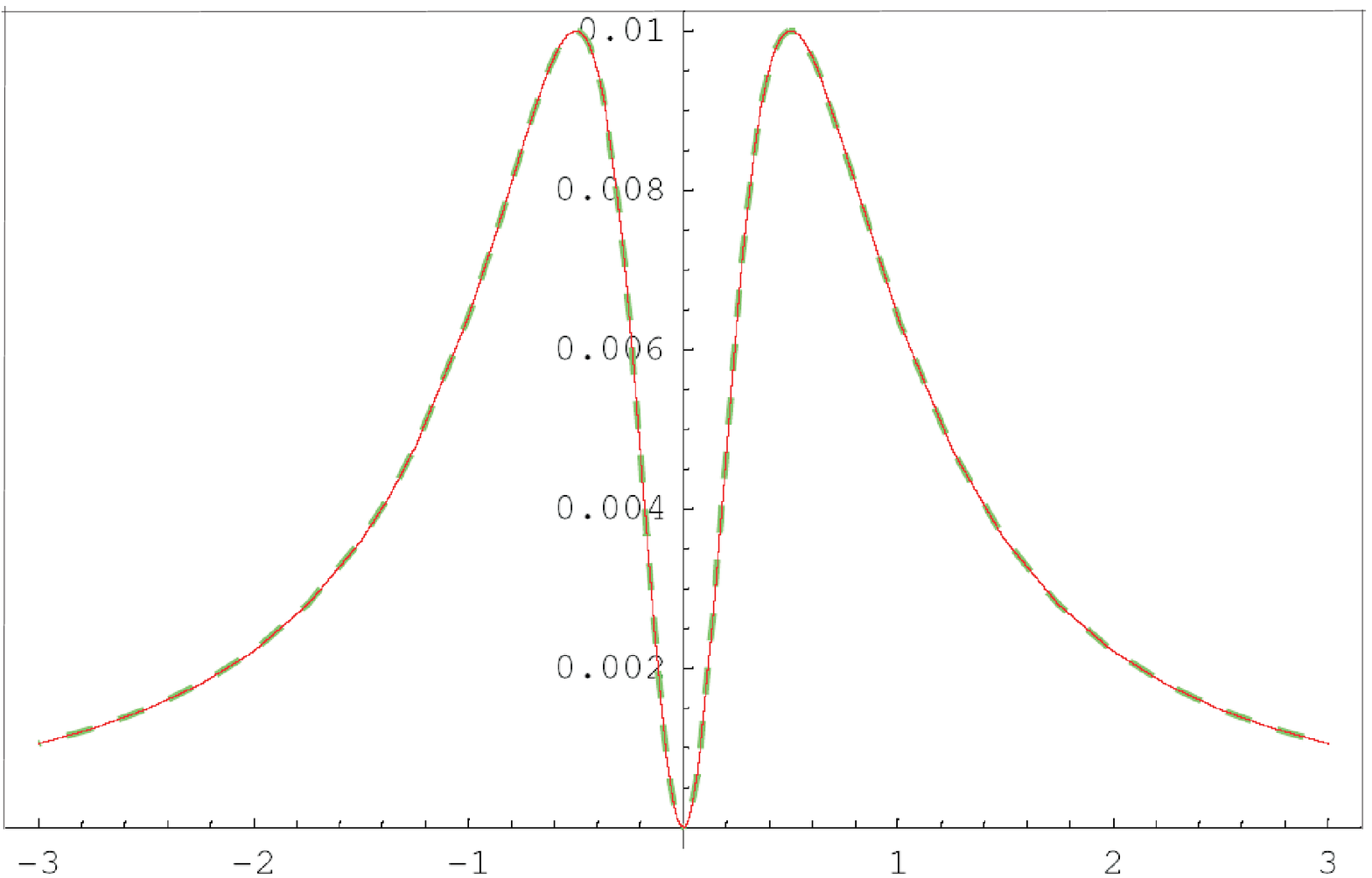}
\caption{Real part (on the left) and imaginary part (on the right)
of the atomic susceptibility from the full density matrix
treatment as a function of the probe detuning $\delta_p$
(continuous line). Atomic susceptibility from equation
(\ref{eq:chi1}) in unsaturated-population approximation when all
population is placed in level $|{1}\rangle$ as a function of probe
detuning $\delta_p$ (dashed line). In both cases the pump Rabi
frequency was $0.8 \gamma_3$.} \label{fig3-4}
\end{center}
\end{figure}

We consider a density matrix which includes all relaxations
channels (internal and external) and we solve it analytically in
the saturated-populations form. However this analytic form is too
tedious to be given here. Therefore we compare the relative
solutions for the refractive index and for the extinction
coefficient with the solutions of our approximated closed form. We
note a perfect agreement in the EIT region, with probe detuning
$\delta_{p}$ close to zero; so in this region we can analyze the
refractive index, the absorption coefficient, the possibility of
having amplification without inversion and the propagation of a
gaussian pulse using the approximated solution to Eqs.
(\ref{obeq}), resolved in the unsaturated-populations form,
considering negligible populations effects and all external and
collisional decay rates.
\par
This comparison is shown in Fig. \ref{fig3-4} for the real and the
imaginary part of the atomic susceptibility on the probe detuning
$\delta_p$ between the exact \textit{saturated-populations}
solution and the \textit{unsaturated-populations} approximation:
the agreement is very good.

\chapter{Propagation in EIT media}\label{cold}\index{cold atoms}

Since the early work of Sommerfeld and Brillouin
\cite{somm}\index{Sommerfeld}\index{Brillouin} on light pulse
propagation, a great deal of attention has been devoted to the
subject of anomalous propagation \index{anomalous propagation} in
absorbing media. In the anomalous dispersion region the group
velocity $v_{g}$ may exceed the speed of light in vacuum $c$ or
become even negative. Negative $v_{g}$'s require a group advance
as first predicted by Garrett and McCumber~\cite{garrett} and then
verified experimentally by Chu and Wong \cite{chu} for optical
pulses propagating through layers of GaP:N. Group advances due to
negative group velocities for other frequency domains have also
been anticipated by Chiao and coworkers
\cite{chiao1,chiao2,kurizki} to occur in the nearly transparent
spectral region of an amplifying medium. In this chapter we will
show how to obtain anomalous propagation in EIT domain but mainly
we will focus on the possibility to slow light pulses, useful to
realize quantum memory as will be shown in chapter \ref{qm}.
\par
Many of the important properties of \textbf{EIT} result from the
fragile nature of quantum interference in a material that is
initially opaque. Indeed the ideal transparency is attained only
if the frequency difference between the two laser fields precisely
matches the frequency separation between the two lower states. If
matching is not perfect, the interference is not ideal and the
medium becomes absorbing. Hence the transparency spike that
appears in the absorption spectrum is typically very narrow. The
tolerance to frequency mismatch can be increased by using stronger
coupling fields, because then interference becomes more robust.
\par
In an ideal EIT atoms are decoupled from the light fields, so the
refractive index \index{refractive index} at resonance is nearly
equal to unity. This means that the propagation velocity of a
phase front (that is, the phase velocity\index{phase velocity})
[see Sec. \ref{group}] is equal to that in vacuum. However, the
narrow transparency resonance is accompanied by a very steep
variation of the refractive index with frequency. As a result the
envelope of a wavepacket propagating in the medium moves with a
group velocity $v_{g}$ \index{group velocity} \cite{harris3} that
is much smaller than the speed of light in vacuum, $c$: so it is
possible to slow light pulses. Actually $v_{g}$ depends on the
control field intensity and the atomic density: decreasing the
control power or increasing the atom density makes $v_{g}$ slower.

\begin{figure} [th!]
\begin{center}
\includegraphics[width=.8\textwidth ]{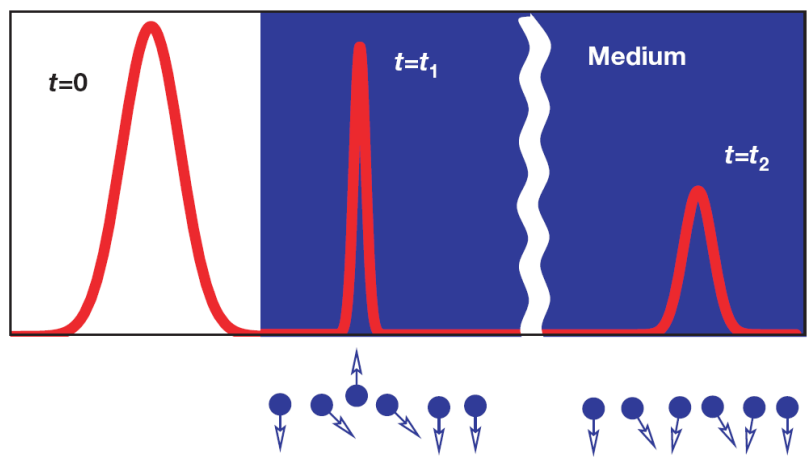}
\caption{Schematic of spatial compression exhibited when a light
pulse (red curve) enters the slow medium (blu). Photons are
converted into flipped spins (blu arrowed circles) and the slow
photonic and spin waves then propagate together. For long distance
($t_2 \gg t_1$), the lossless propagation is limited by the
spreading of the pulses owing to the narrow bandwidth of the
transparency window. \cite{hau}} \label{fig5}\end{center}
\end{figure}

Figure \ref{fig5} illustrates pictorially the dynamics of light
retarded propagation in an EIT medium. Initially the pulse is
outside the medium in which all atoms are in their ground states
($|1\rangle$). The front edge of the pulse then enters the medium
and is rapidly decelerated. Because it is still outside the
medium, the back edge propagates with vacuum speed c. Thus, upon
entrance into the cell the spatial extent of the pulse is
compressed by the ratio $\frac{c}{v_{g}}$, whereas its peak
amplitude remains unchanged. Clearly the energy of the light pulse
is much smaller when it is inside the medium. The rest of the
photons are being expended to establish the coherence between the
states $|1\rangle$ and $|2\rangle$, or in other words to flip
atomic spins, with any excess energy carried away by the control
field. The wave of flipped spins now propagates together with the
light pulse. As the pulse exits the medium its spatial extent
increases again and the atoms return to their original ground
state; the pulse however, is delayed as a whole by $\Delta
t=(1/v_{g} - 1/c)L$ where $L$ is the length of the medium. For
example, in the experiments by \textbf{Hau} \cite{hau} and
collaborators, a pulse that is 760 m long in free space is
compressed by a factor of $10^7$ to a length of 43 $\mu$m. \par
One might expect that the EIT \index{EIT} technique could take the
group velocity all the way to zero but it is not really true.
Indeed to reduce the velocity more one first must make the
coupling laser weaker, but reducing the coupling laser intensity
also reduces the bandwidth, or frequency spread, of the incoming
signal light that can be affected by\index{EIT}. If the probe beam
has a wider bandwidth, much of it will be absorbed by the atomic
medium and not slowed. As the coupling intensity approaches zero -
the condition for zero group velocity \index{group velocity} - the
allowed bandwidth for the incoming signal beam is zero, and no
light can propagate. Nevertheless we will see in the next chapter
that it is possible to stop the probe field with an adiabatic
control of the pump laser because in this case there is a
simultaneous narrowing of transmission and pulse spectrum.

\section{Definitions of wave velocity}
\label{group}

In their classic treatment of the propagation of light in
dispersive media, Sommerfeld and Brillouin \cite{somm} introduced
five different kinds of wave velocities:

\begin{itemize}

\item[1)] The \textit{\textbf{phase velocity}},\index{phase
velocity} which is the speed at which the \ zero crossings \ of \
the carrier wave move.

\item[2)] The \textit{\textbf{group velocity}},\index{group
velocity} at which the peak of a wave packet moves.

\item[3)] The \textbf{\textit{energy velocity}},\index{energy
velocity} at which the energy is transported by the wave.

\item[4)] The \textbf{\textit{signal velocity}},\index{signal
velocity} at which the half-maximum wave amplitude moves.

\item[5)] The \textbf{\textit{front velocity}},\index{front
velocity} at which the first appearance of a discontinuity moves.

\end{itemize}

All five velocities can differ from each other. In linear response
dispersive media, the group, energy, signal, and front velocities
coincide and are usually less than the phase velocity. However,
recent experimental demonstrations that the group velocity of
light can be reduced by 10-100 million compared with its phase
velocity have fuelled many studies and exciting discussions. As
noted before, these remarkable results are based on usage of very
steep frequency dispersion in the vicinity of narrow resonance of
electromagnetically induced transparency\index{EIT}.
\par
It is well known that any reactive medium leads to a delay of
electromagnetic pulses and a system as simple as an infinite chain
of \textit{RLC} circuits can significantly reduce the speed of an
electromagnetic pulse resonant within the circuit. Nevertheless
from the \textit{RLC} analogy it might seem that to have slow
light one does not need any optical nonlinearity.
\par
An interesting aspect of ultraslow light is that this essentially
``linear'' phenomenon appears in coherently driven atomic media
with extremely nonlinear optical behavior where the usual optical
laws concerning dispersion and absorption are no longer valid.
Surprisingly, media prepared for the conditions giving raise to
slow light provide new regimes of nonlinear interaction with
highly increased efficiency even for very weak light fields. These
media hold promise for high-precision spectroscopy and
magnetometry. Besides the ability to manipulate single photon
states is extremely important for the future telecommunications,
as shown in this thesis.

\section{A semi-classical approach}

\begin{figure} [!ht]
\begin{center}
\includegraphics[width=.52 \textwidth]{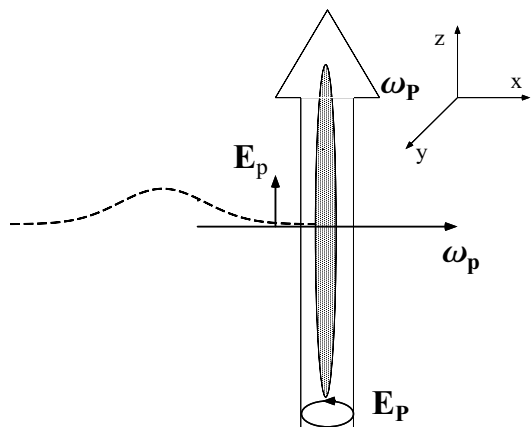}
\caption{A gaussian and linearly polarized weak
\textit{probe}--pulse propagating along the radial $x$-axis of a
cigar--shaped cloud of trapped atoms. A stronger
\textit{pump}--pulse right circularly polarized propagates along
the longitudinal $z$-axis.} \label{fig6}
\end{center}
\end{figure}

Let us calculate the effects of propagation of a gaussian pulse
through a thin trapped ``cloud'' of atoms referring to a
particular experiment on $^{87}Rb$ near Bose-Einstein condensation
proposed in \cite{artoni}.

\begin{figure} [!ht]
\begin{center}
\includegraphics[width=.5 \textwidth]{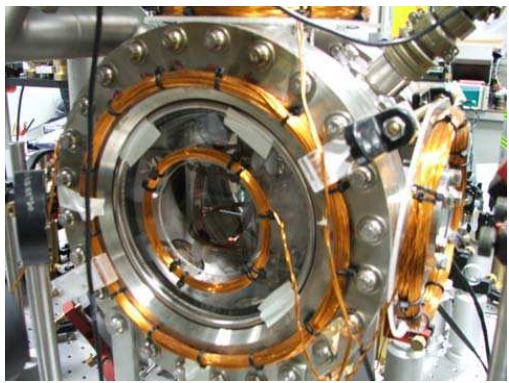}
\caption{Experimental apparatus to confine $^{87}Rb$ atoms in a
temporal dark \textbf{SPOT}\index{SPOT} (\textit{SPontaneous-force
Optical Trap})\index{spontaneous-force optical trap} in order to
realize light propagation in cold atoms. This is a
\textit{magneto-optical trap}\index{magneto-optical trap}
(\textbf{MOT})\index{MOT} where the repumping beam has been
temporarily shut off. (Quantum Information Laboratory, Scuola
Superiore di Catania)} \label{fig7}
\end{center}
\end{figure}

We take the pulse to be gaussian in form and propagating across
the radial width ($d = 10 \ \mu m$) of the cloud (see Fig.
\ref{fig6}) whose optical thickness is much smaller than the
incident pulse length $\cal {L}$. The medium is modelled by the
square density profile of a ``slab'' having uniform density. In
the EIT region reflection is almost vanishing \cite{EIT} and the
sharp boundaries of the slab of thickness $d$ don't introduce
substantial errors.
\par
To analyze the pulse shape modifications, we assume a complex
refractive index $n(\omega)$ that varies slowly over the frequency
band-width $c/\cal {L}$ of the pulse so as to expand the optical
wave vector around the pulse carrier frequency $\omega_c$,
 \bgar k(\omega)&=&\frac{\omega \eta(\omega)}{c}+i \frac{\omega
\kappa(\omega)}{c} \simeq k_{c}+(k'_{cr}+i k'_{ci})
(\omega-\omega_{c})+
\nonumber\\
&&+\frac{1}{2}(k''_{cr}+i k''_{ci}) (\omega-\omega_{c})^{2}
 \label{kesp} \enar
where $k_c=\omega_c n (\omega_{c})/c \equiv \omega_c  (\eta_{c} +
i\kappa_{c})/c $. Here $\eta(\omega)$ and $\kappa(\omega)$ denote
respectively the real refractive index and the extinction
coefficient while $\eta_c$ and $\kappa_c$ are their values at
$\omega_c$. In the linear term the prime denotes the frequency
derivative, which is divided into its real ($r$) and imaginary
($i$) parts, i.e. the dispersion of the real refractive index,
which is related to the group velocity according to
$k'_{cr}=1/v_g$, and the dispersion $k'_{ci}$ of the extinction
coefficient. Moreover in the quadratic term the real part,
$k''_{cr}$, characterizes the lowest order contribution to the
$v_{g}$'s dispersion.
\par
Defining $\Delta x$ the shift in the peak position relative to the
position it would have in free-space, we note that for positive
$v_{g}$'s which are smaller than c the peak position is retarded
($\Delta x<0$) with respect to the free--space value whereas peak
position advances occur for positive $v_{g}$'s which are larger
than c ($0<\Delta x<d$) or negative $v_{g}$'s ($\Delta x>d$)
leading to apparent superluminal propagation.
\par
\begin{figure}[ht!]
\begin{center}
\includegraphics[width=.4\textwidth ]{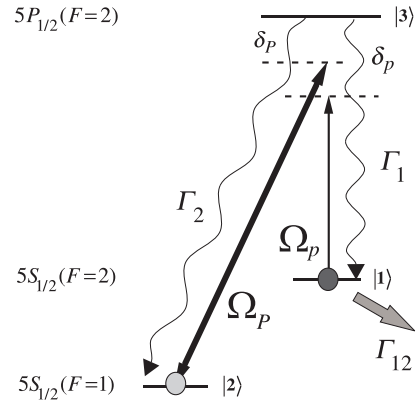} \caption{Level
scheme for the $^{87}Rb$ $D_{1}$ line. The three levels are
coupled by a pump laser with a Rabi frequency $\Omega_{P}$ and a
probe field with a Rabi frequency $\Omega_{p}$. Only the hyperfine
ground sub-levels $|{1} \rangle$ and $|{2} \rangle$ can be
initially populated. The relevant linewidths are $\Gamma_{1}/2\pi
\simeq \Gamma_{2}/2\pi \simeq \gamma_{3}/2 \pi= 5.75 $ MHz while
$\Gamma_{12}/2\pi \simeq 1$ KHz.} \label{fig8}
\end{center}
\end{figure}

We will derive the analytical expressions for the group velocity
$v_{g}$, its dispersion $d_{g}$ and the probe transmission
intensity $G_{T}$ specific to very cold alkali atomic vapors, as
in the paper \cite{artoni}. In Fig. \ref{fig8} we show the level
scheme for the $^{87}Rb$ $D_1$ line as concrete example of Fig.
\ref{fig1}. In particular the excited state $|{3}\rangle$ can
decay to the ground states $|{1}\rangle$ and $|{2}\rangle$ with
rates $\Gamma_1$ and $\Gamma_2$ respectively, and the level
$|{1}\rangle$ decays to levels outside those considered here with
rate $\Gamma_{12}$.

In Eq. (\ref{kesp}) the group velocity $v_g$, its dispersion $d_g$
and the probe transmission intensity $G_T$ are represented by:
 \begin{equation}
\label{eq:vg} v_g=\frac{c}{\eta(\omega)+\omega
\frac{\partial\eta(\omega)}{\partial\omega}}
 \end{equation}
 \begin{equation}
\label{eq:dg} d_{g}(\omega)= -\frac{v_{g}^{2}(\omega)}{c}
\left(\omega \ \frac{\partial^2\eta(\omega)}{\partial\omega^2}+2
\frac{\partial\eta(\omega)}{\partial\omega}\right)=-
v_{g}^{2}(\omega)\frac{ \cal{D}(\omega)}{\Gamma_1}
\end{equation}
 \begin{equation}
 G_{T}(\omega)=|T(\omega)|^{2}=\Bigg|\frac{4
n(\omega) e^{i \left[n(\omega)-1\right]\omega
d/c}}{\left[n(\omega)+1\right]^{2}-\left[n(\omega)-1\right]^{2}
e^{ 2i n(\omega)\omega d/c}}\Bigg|^{ \ 2} \label{eq:T}
 \end{equation}
where $\eta(\omega)$ is the real part of the refractive index
$n(\omega)$, $d$ is the length of the atomic medium and $c$ the
speed of light in vacuum. We have defined a group velocity
dispersion function $\cal{D}(\omega)$ that has the dimension of a
reciprocal velocity.
\par
Now we use a density matrix treatment as in Sec. \ref{obe} and
expand in powers of the probe Rabi frequency. For weak probe
intensities the complex steady-state atomic susceptibility
exhibited to the probe can be fully accounted for by the first
order expansion,
\begin{eqnarray}
\label{eq:chi} \chi_{p}& = &  3 \pi  {\cal{N}}_{p} \Gamma_{1}
\Big[\frac{ (\delta_{p} -i \gamma_{1})} {(\delta_{p} - i
\gamma_{3})(\delta_{p} -i
\gamma_{1})-(\Omega_{P}/{2})^{2}}(n_{1}-n_{3})+
\nonumber\\
& & {}-\frac{i}{\gamma_{3}} (\Omega_{P}/{2})^{2}
\frac{(n_{3}-n_{2})} {(\delta_{p} - i \gamma_{3})(\delta_{p} -i
\gamma_{1})-(\Omega_{P}/{2})^{2}} \Big]
 \end{eqnarray}
where ${\cal{N}}_{p}$ is the scaled sample average density
$(\lambda_{p}/2 \pi)^{3} N /V$, $\lambda_{p}$ is the probe
resonant wavelength and $n_i$ is the normalized population of
level $i$. The overall dephasings
$\gamma_{3}=(\Gamma_{1}+\Gamma_{2})/2$ and
$\gamma_{1}=\Gamma_{12}/2$ of levels $|{3}\rangle$ and
$|{1}\rangle$ are expressed in terms of the respective levels
linewidths (see Fig. \ref{fig8}). We denote by $\delta_{p}
=\omega_{31}-\omega_{p}$ the probe detuning while the pump beam is
taken to be exactly at resonance ($\omega_{P}=\omega_{32}$). In
equation (\ref{eq:chi}) we also neglect contributions due to the
atomic velocity distribution, since these can be eliminated by
choosing a co--propagating pump and probe laser configuration.
\par
In obtaining equation (\ref{eq:chi}) we have assumed that the
populations of the levels do not vary much with time. This
assumption is justified since the Rabi frequency of the pump laser
$\Omega_{P}$ is smaller than the excited state linewidth
$\gamma_3$ ($\Omega_{P}< \gamma_{3}$) and the probe laser is taken
to be much weaker than the pump. Furthermore, because  the excited
state $|{3}\rangle$ may decay to other atomic levels outside the
two considered here, in all the following we will assume
$n_{3}=0$. In order to check the validity of these assumption we
have solved the full system of optical Bloch equations for an open
three--level system including the possibility to pump population
in any of the three levels and the possibility for the population
to decay to levels outside those considered here from any of the
three levels.
\par
The dispersive behavior of the real refractive index $\eta_{p}$
fully determines the group velocity $v_{g}$ which acquires a
positive minimum at resonance where the slope of $\eta_{p}$ is
largest. We consider the case $n_2=n_3=0$ and $n_1=1$
analytically. With the help of Eq. \eq{chi} and a power expansion
of $\eta_{p}$ around probe resonance we obtain for this minimum
 \bgar \frac{c}{v_{g,  min}^{(+)}}=1- \frac{3 \pi {\cal{N}}_{p}}
{2} \frac{\omega_{31} \Gamma_{1}} {\gamma_{3}^{2} }
 \frac{\gamma_{1}^{2}-\frac{\Omega_{P}^{2}}{4}}
 {\left(\gamma_{1}  + \frac{\Omega_{P}^{2} }{4 \gamma_{3} } \right)^{2}}
\label{eq:vmax}
 \enar

The group velocity exhibits inversion at points where the slope of
$\eta_{p}$ vanishes. For small extinctions $\kappa_{p}$ this
occurs approximately at the extrema of the real part of $\chi_{p}$
where, to the lowest order in $\gamma_{1}$, the probe detuning
takes the value $\delta_{\infty}\simeq \pm (\gamma_{3}
-\sqrt{\gamma_{3}^{2}+\Omega_{P}^{2}})/2$.
\par
Under those experimental conditions in \cite{artoni}, minima of
the negative group velocity occur at about twice $\delta_{\infty}$
and (in the lowest order in $\gamma_{1}$)

 \bgar \frac{c}{v_{g, min}^{(-)}}&=&-\frac{3 \pi {\cal{N}}_{p}}{2}
\frac{ \omega_{31} \Gamma_{1} } {\gamma_{3}^{2} }
\left(1+\frac{\Omega_{P}^{2}}{4 \ r^{2} \gamma_{3}^{2}}\right)
 \frac{
1-r^{2} (1-\frac{\Omega_{P}^{2}}{4 \  r^{2} \gamma_{3}^{2}})^{2}}
{\left[ 1+r^{2} (1-\frac{\Omega_{P}^{2}}{4 \ r^{2}
\gamma_{3}^{2}})^{2} \right]^{2}} \label{eq:vmin} \enar where
$r=1-\sqrt{1+(\Omega_{P}/\gamma_{3})^{2}}$ and the last factor on
the right hand side of Eq. \eq{vmin} is of the order of unity. In
correspondence to these minima the group velocity dispersion in
Eq. (\ref{eq:dg}) vanishes and small group velocity deviations
about $v_{g, min}^{(-)}$ are essentially determined by the
dimensionless dispersion function $\cal{D}$.
\par
Within the transparency region, for which $\eta_{p} \simeq 1$ and
$\kappa_{p} \ll 1$, under these particular approximations, in
terms of the detuning, Eq. (\ref{eq:T}) for the probe transmission
intensity $G_{T}$ reduces to \be G_{T}(\dtp)= e^{-2 \kappa_{p}
\omega_{31} d /c} \simeq e^{-\delta_{p}^{2}/2 \Gamma_{G}^{2}}
\label{eq:GT} \ee The approximate expression on the right hand
side holds only for appropriately small detunings ($\delta_{p}
\lesssim \gamma_{3}$) since in this case $\kappa_{p} \simeq 24 \pi
{\cal{N}}_{p} ( \gamma_{3} \Gamma_{1}
\delta_{p}^{2}/\Omega_{P}^{4})$ to the lowest order in
$\gamma_{1}$. The transmission bandwidth
 \be \Gamma_{G}\simeq 0.06
\times \frac{\Omega_{P}^{2}}{\sqrt{\gamma_{3} \Gamma_{1}} } \sqrt{
\frac{c }{\omega_{31} d \ {\cal{N}}_{p}} } \label{eq:bandw} \ee
increases with the pump intensity and with decreasing values of
the atomic density and radial width $d$.
\par
Let us make some remarks about the probe propagation in classical
electromagnetic theory. In the presence of absorption the
steady-state macroscopic atomic polarization, induced by a
time-independent electric field amplitude, would simply radiate a
field that cancels part of the incident field with a subsequent
decrease in transmission. The situation changes for a field
envelope that varies in time: during the leading half of the pulse
the field amplitude rises which results into polarizations that
are smaller than the steady-state value while the reverse takes
place during the trailing half of the pulse. The macroscopic
polarization is responsible for the absorption of energy from the
probe so that the larger polarizations induced during the trailing
half of the probe imply that more energy will be absorbed from the
tail than from the front of the pulse. This asymmetric absorption
of energy will reshape the incident gaussian pulse into a smaller,
but practically undistorted, wavepacket whose peak appears to have
moved faster than $c$, i.e. advanced with respect to the one that
has travelled in vacuum.
\par
Moreover, we will show that it is also possible that the peak of
the pulse appears to have moved slower than $c$, i.e. retarded
with respect to the one that has travelled in vacuum, when we
consider zero detuning, as obtained experimentally by Hau and
coworkers.
\par
For our numerical analysis of pulse propagation across the atomic
sample, we also calculate the expectation value after transmission
of the probe pulse normal-order Poynting vector
\cite{rl1,jackson}, which is

 \bgar S(x,t)&=& \left| \int _{0}^{\infty} d\omp \left(\frac{\hbar
\omp}{2 \pi S}\right)^{1/2} f(\omp) T(\omega_{p})  e^{i
\omp(\frac{x}{c}-t) } \right|^{2}
 \nonumber\\
 &\simeq &
\frac{S_{o}}{4 \pi \sigma_{p}^{2}}
 \left|
\int _{-\infty}^{\infty} d\dtp T(\dtp) e^{-i \dtp (\frac{x}{c}-t)
} e^{- \frac{(\dtp-\dtc)^{2}}{4 \sigma_{p}^{2}}} \right|^{2}
\label{eq:poynting} \enar
 where $S$ is a reference area in the $yz$-plane, $\dtc=\omega_{31}-\omc$ and $S_{o}$ is the peak--power density
 of the incident pulse.
\par
Here

 \be
 f(\omp)=\left(\frac{1}{2 \pi \sigma_{p}^{2}} \right)^{1/4}
 exp\left\{{- \frac{(\omp-\omc)^{2}}{4 \sigma_{p}^{2}}}\right\}
 \label{eq:gauss}
\ee denotes the frequency distribution of the incident gaussian
wavepacket with carrier frequency $\omega_{c}$ and mean--square
spatial length ${\cal{L}}_{p}^{2}=c^{2}/\sigma_{p}^{2}$ while
$T(\omp)$ is the exact transmission coefficient defined in \eq{T}.
The transmitted average power density \eq{poynting} is evaluated
with the help of the transmission amplitude \eq{T} and results are
illustrated in the following sections, in which we integrate
numerically this integral and we observe a reshaping and a shift
of the propagated pulse, under different conditions.

\section{Normal dispersion region}

In the following figures, using the dependence of the real and
imaginary parts of the optical coherence $\rho_{13}$ on the probe
detuning $\delta_{p}$, we show the lineshapes of the index of
refraction and of the absorption coefficient of the medium
depending on $\delta_{p}$. Particularly, the gray line indicates
the situation in which all population is in the level 1, the red
lines show the cases in which there are the following populations:

\begin{itemize}

\item[1)] $n_{1}=0.9 \ \ n_{2}=0.1 \ \ n_{3}=0$

\item[2)] $n_{1}=0.7 \ \ n_{2}=0.3 \ \ n_{3}=0$

\item[3)] $n_{1}=0 \ \ n_{2}=1 \ \ n_{3}=0$ (complete inversion)

\end{itemize}

Note that for $n_{2}>n_{1}$ we have population inversion; of
course, we impose the normalization of the populations, i.e.
$n_1+n_2+n_3=1$.
\par
Firstly, we consider the ``normal'' situation in which there is no
pump field, no dark state and therefore there isn't EIT.

\begin{figure} [!ht]
\begin{center}
\includegraphics[width=.45\textwidth ]{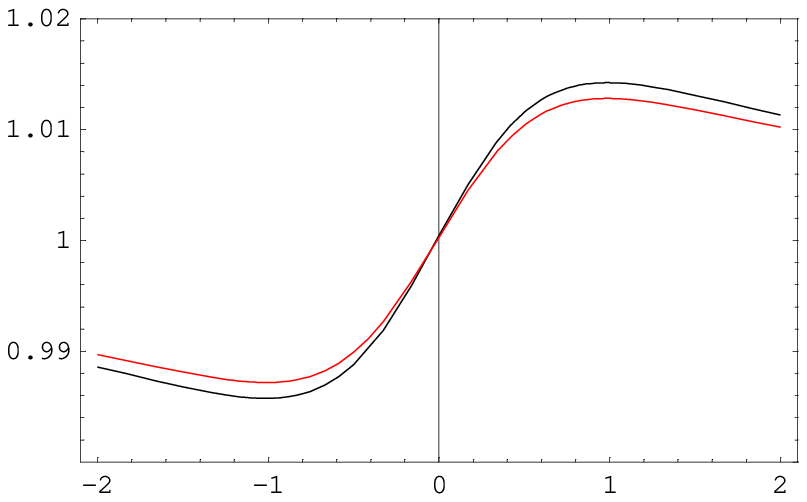}
\includegraphics[width=.45\textwidth ]{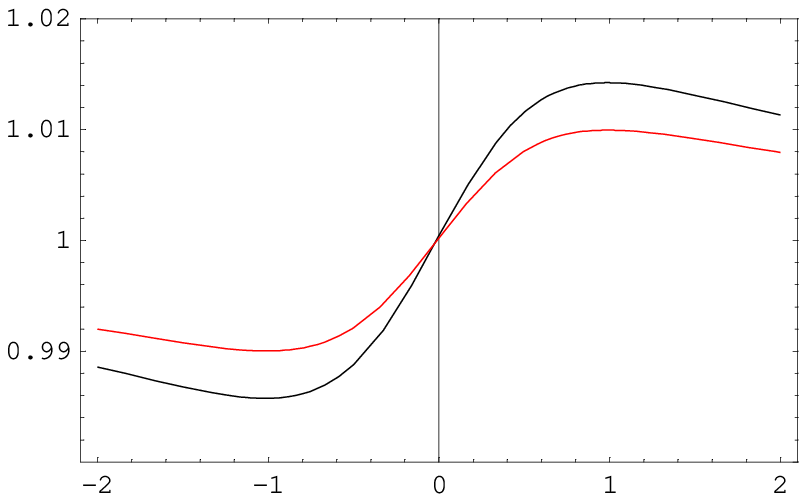}
\includegraphics[width=.45\textwidth ]{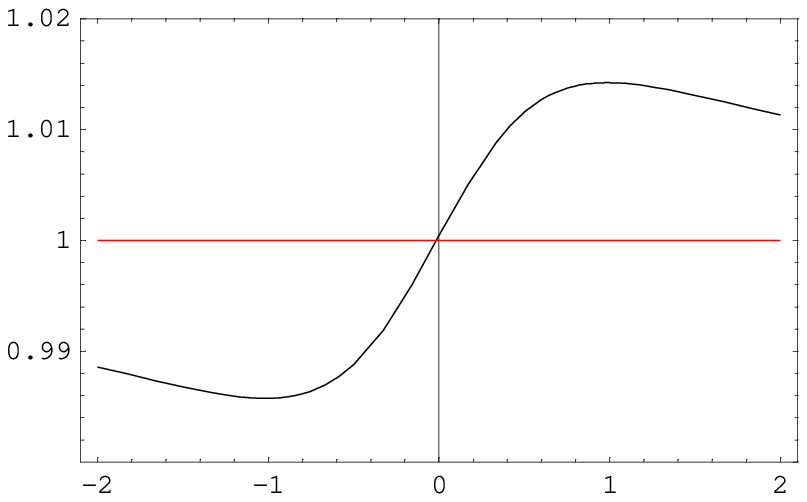}
\caption{Refractive index vs detuning $\delta_{p}$ (in units of
$\gamma_{3}$).}\label{fig9-11}
\end{center}
\end{figure}

\begin{figure} [p]
\begin{center}
\includegraphics[width=.43\textwidth ]{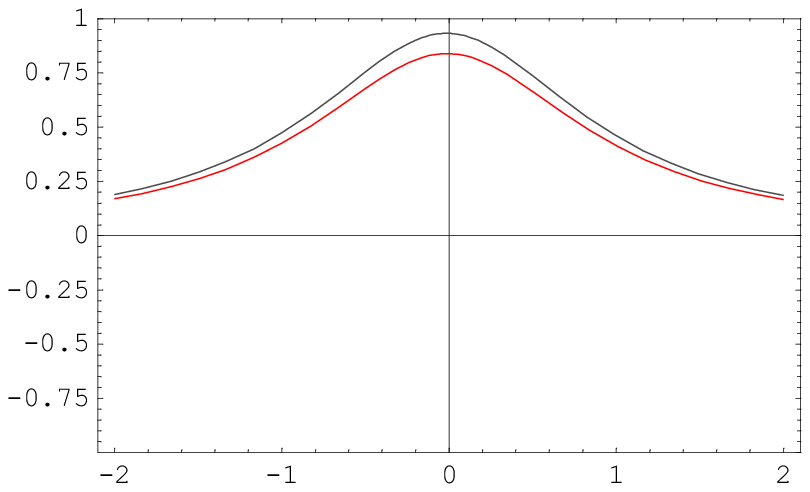}
\includegraphics[width=.43\textwidth ]{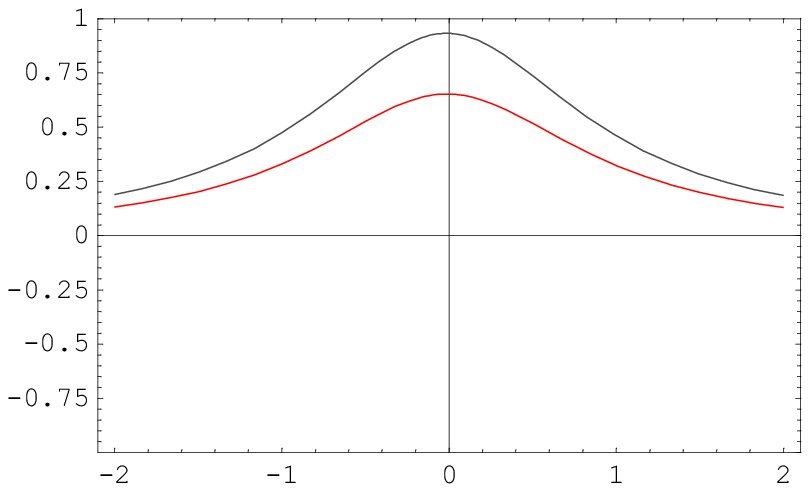}
\includegraphics[width=.43\textwidth ]{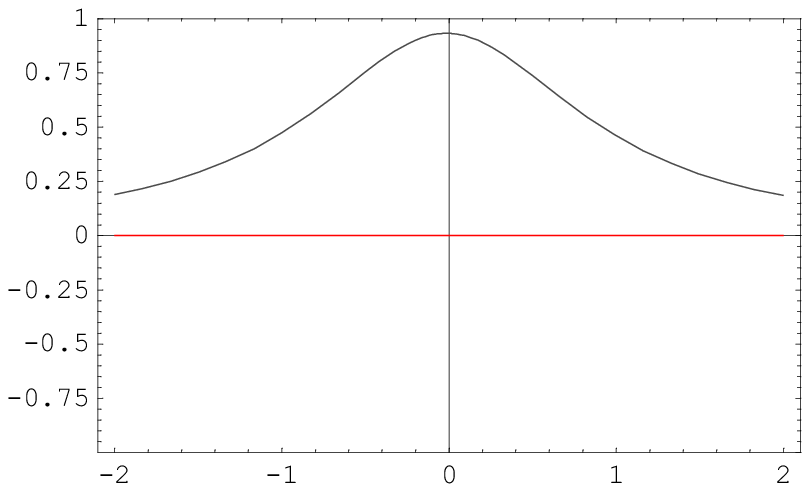}
\caption{Absorption coefficient vs detuning $\delta_{p}$ (in units
of $\gamma_{3}$). There is absorption, depending on $n_{1}$
(ground state), and if we take a gaussian pulse propagating across
this medium with $n_{1}\neq 0$ it's absorbed and
disappears.}\label{fig12-14}
\end{center}
\end{figure}

\begin{figure} [p]
\begin{center}
\includegraphics[width=.43\textwidth ]{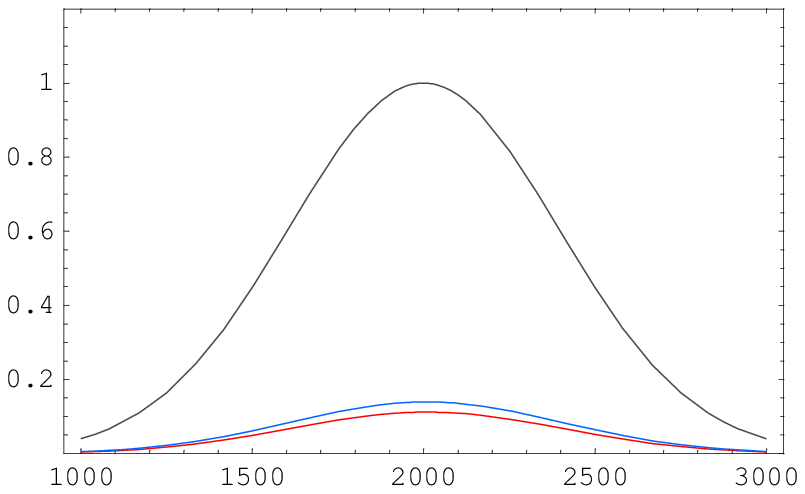}
\includegraphics[width=.43\textwidth ]{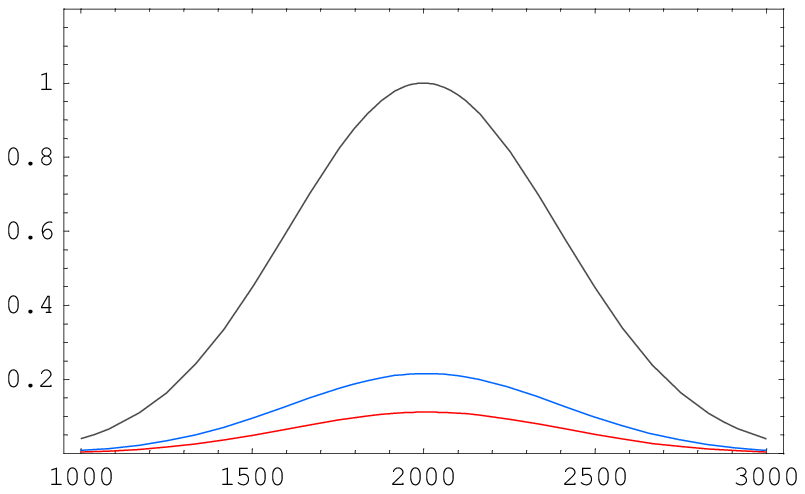}
\includegraphics[width=.43\textwidth ]{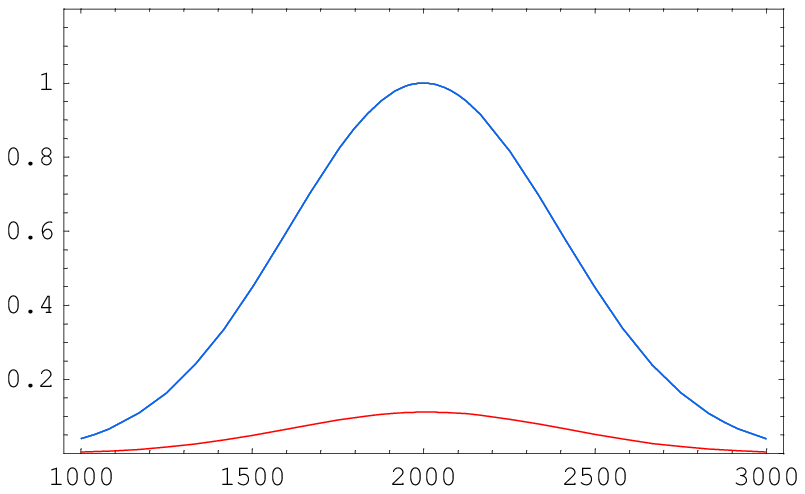}
\caption{Power density $S(x,t)$ for a gaussian probe pulse in
units of the incident peak power density $S_{0} (=1)$ with
$x_{0}=2000$. The curves represent $S(x,t)$ as a function of x(m)
after propagation in free-space (gray line) and across a
cigar-shaped atomic cloud of radial width $d=10 \ \mu m$ (blu
lines $\rightarrow$ three different populations; red line
$\rightarrow$ $n_{1}=1$). There is no coupling and the probe
frequency detuning is $\delta_{p}=0$.} \label{fig15-17}
\end{center}
\end{figure}

\newpage
\section{Anomalous dispersion region (\textbf{EIT})}

\begin{figure} [th!]
\begin{center}
\includegraphics[width=.49\textwidth ]{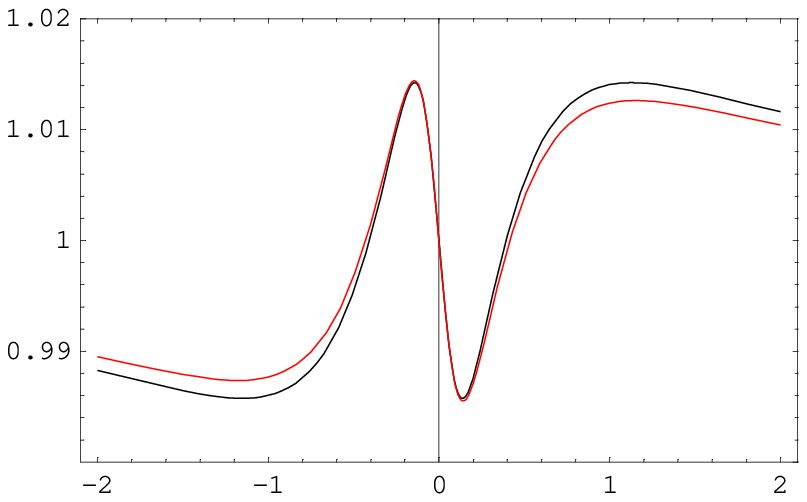}
\includegraphics[width=.49\textwidth ]{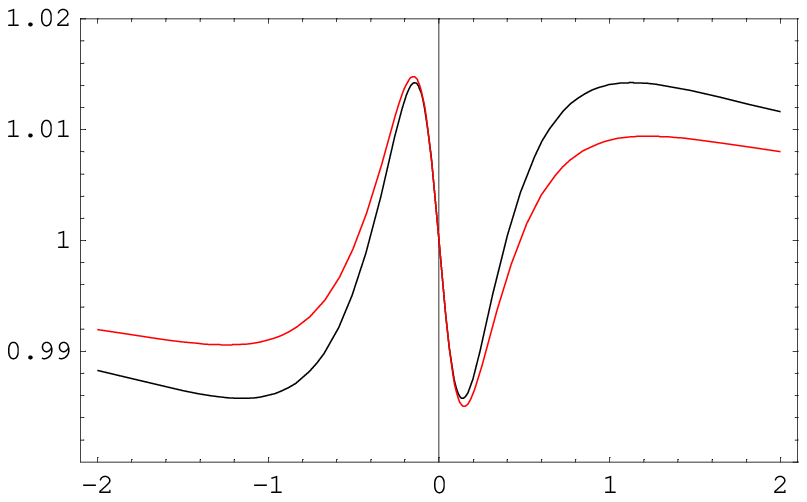}
\includegraphics[width=.49\textwidth ]{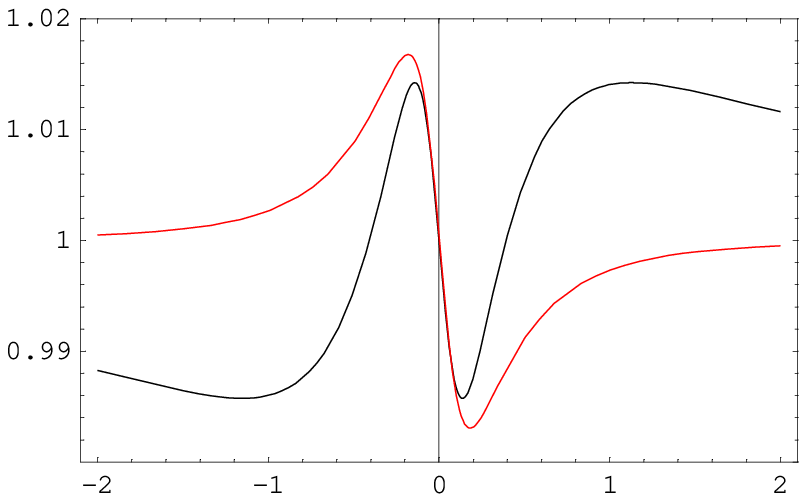}
\caption{Refractive index\index{refractive index} vs detuning
$\delta_{p}$ (in units of $\gamma_{3}$).} \label{fig18-20}
\end{center}
\end{figure}

As noted before, the dependence of the real and imaginary parts of
the optical coherence $\rho_{13}$ on the pump frequency
$\delta_{p}$ determines the lineshapes of the index of refraction
and of the absorption coefficient of the medium under coherent
population-trapping resonance. We repeat the previous procedure
but in the presence of coupling; in the following figures we will
use the same color notation as in the previous section. \par
Figure \ref{fig18-20} shows the lineshape of the refractive index
for three fixed initial populations. We note that the central
region of the detuning $\delta_{p}$ doesn't depend on the
populations, while there is this dependence for the side regions;
this involves an analogous behavior for the group velocity.
\par
\begin{figure} [th!]
\begin{center}
\includegraphics[width=.49\textwidth ]{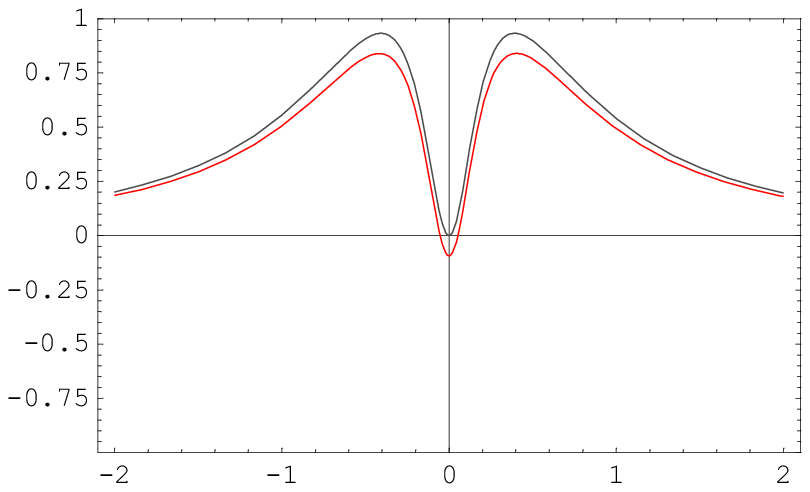}
\includegraphics[width=.49\textwidth ]{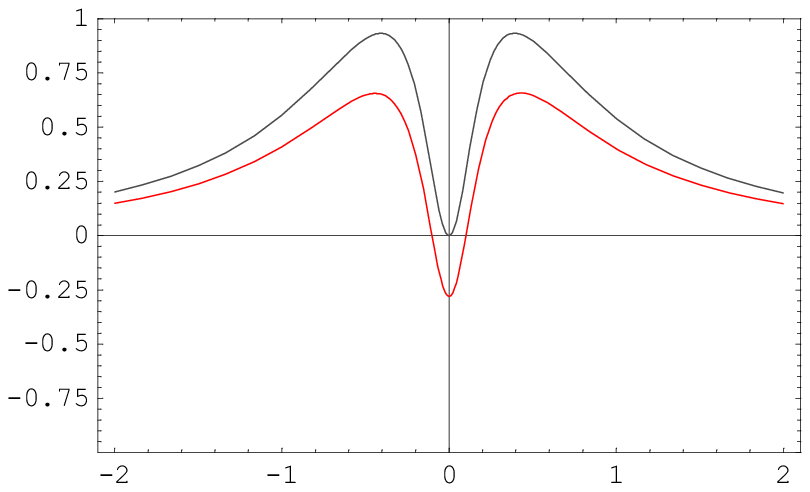}
\includegraphics[width=.49\textwidth ]{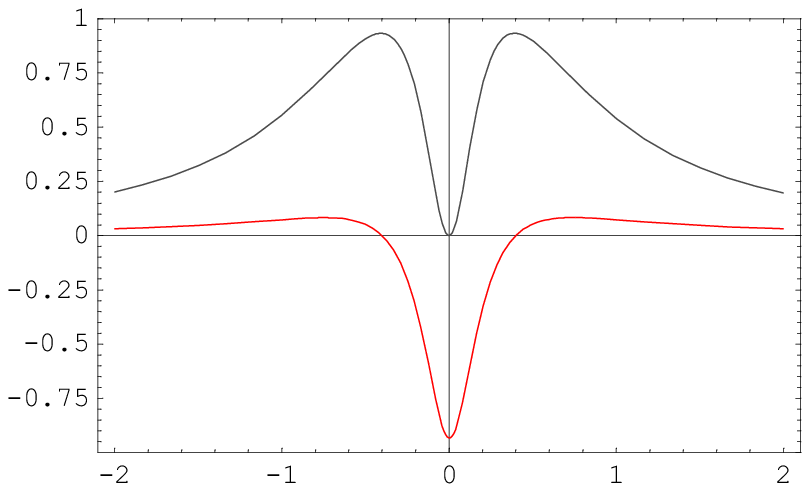}
\caption{Absorption coefficient vs detuning $\delta_{p}$ (in units
of $\gamma_{3}$).} \label{fig21-23}
\end{center}
\end{figure}

In Fig. \ref{fig21-23} we analyze the probe-frequency dependence
of the absorption coefficient and we finally see the phenomenon of
EIT with small detunings $\delta_{p}$ close to zero. Moreover we
observe that with a few population ($30 \ \%$) in the level 2 an
amplification without inversion appears for the transition $1
\leftrightarrow 3$. Indeed there is a change of sign of the
absorption coefficient and with $n_{2}=0.3$ and $n_{1}=0.7$ we
have a gain equal to $28 \% $.

\newpage \newpage We remember that $k=\omega n (\omega)/c \equiv \omega (\eta +
i\kappa)/c $, where $\eta(\omega)$ and $\kappa(\omega)$ denote
respectively the real refractive index and the extinction
coefficient; the dispersion of the real refractive index is
related to the group velocity according to $k'_{r}=1/v_g$ while
$k''_{cr}$ characterizes the lowest order contribution to the
group velocity dispersion. Therefore we can obtain the
probe-frequency dependence of the group velocity and of the
relative dispersion.

\begin{figure} [!ht]
\begin{center}
\includegraphics[width=.49\textwidth ]{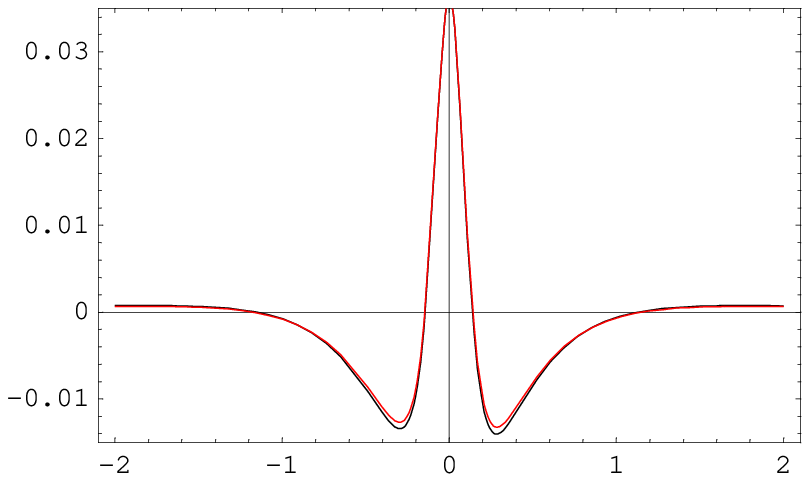}
\includegraphics[width=.49\textwidth ]{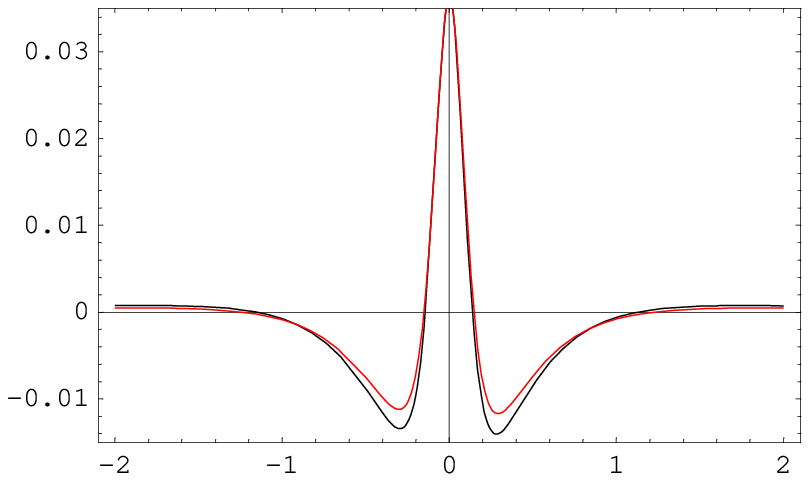}
\includegraphics[width=.49\textwidth ]{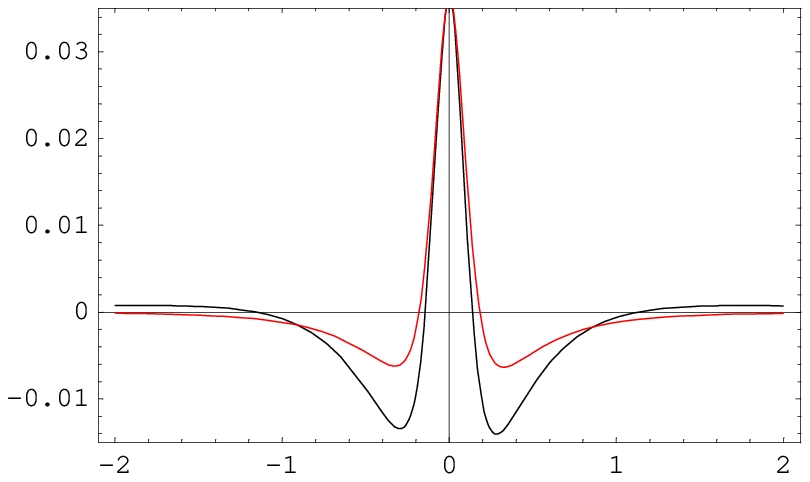}
\caption{Reciprocal group velocity $(m/s)^{-1}$ vs detuning
$\delta_{p}$ (in units of $\gamma_{3}$).} \label{fig24-26}
\end{center}
\end{figure}

As noted above, the group velocity is independent of the
difference of populations for detunings close to zero; in this
region the group velocity is less than $c$ and then the pulse is
retarded, i.e. this is the \textbf{subluminal
propagation}\index{subluminal propagation}. Instead, at the side
regions there is a change of sign and the group velocity is
negative, i.e. there is \textbf{superluminal
propagation}\index{superluminal propagation}, as we will note
explicitly in the propagation of our gaussian pulse.

\begin{figure} [!ht]
\begin{center}
\includegraphics[width=.49\textwidth ]{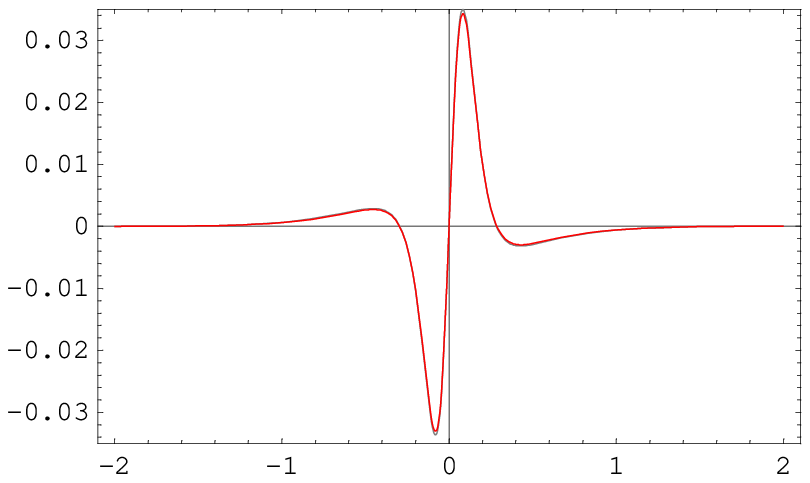}
\includegraphics[width=.49\textwidth ]{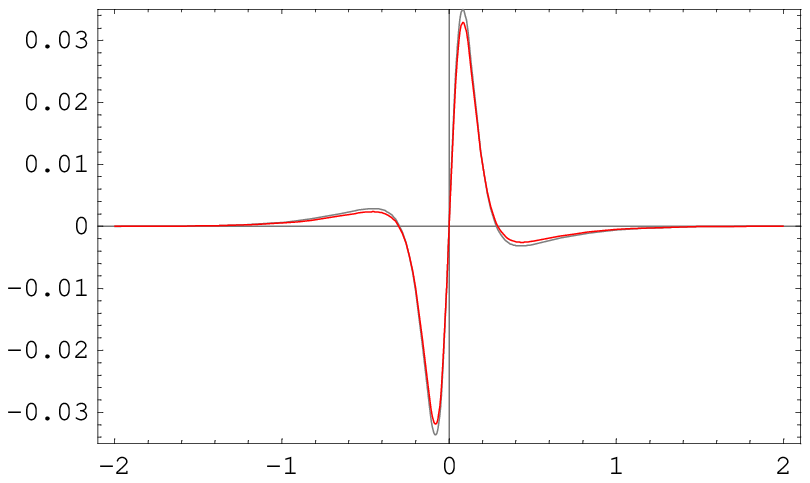}
\includegraphics[width=.49\textwidth ]{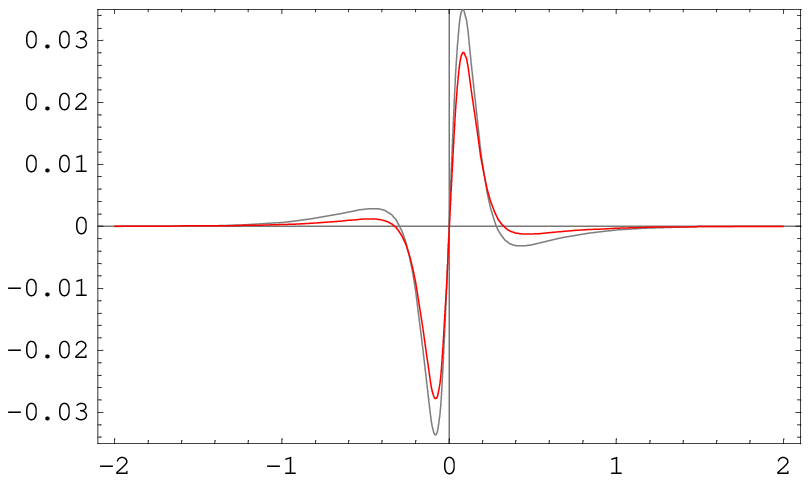}
\caption{Group velocity dispersion, ${\cal D}$ $(m/s)^{-1}$, vs
detuning $\delta_{p}$ (in units of $\gamma_{3}$).}
\label{fig27-29}
\end{center}
\end{figure}

Figure \ref{fig27-29} shows the dispersion of group velocity. It
has an important value because with zero detuning and in
correspondence of minimum negative group velocity, there is no
dispersion. Therefore, in these cases we analyze the propagation
of a gaussian pulse in this dielectric medium.

\section{Gain-assisted and retarded pulse}

We choose the probe frequency detuning equal to zero and we
observe the gain-assisted and retarded propagation of a gaussian
pulse, with the usual previous values of populations. \par In the
following figures, the gray line represents the pulse propagating
in vacuum, the red line refers to the case $n_{1}=1$, while the
blue lines represent the three cases previously indicated. We note
that, rising the population of the level 2, we have much more gain
(we are realizing population inversion) and the group velocity is
always less than $c$, i.e. retarded pulse. This behavior is in
agreement with the frequency dependence of the absorption
coefficient and the group velocity.

\begin{figure} [!ht]
\begin{center}
\includegraphics[width=.49\textwidth ]{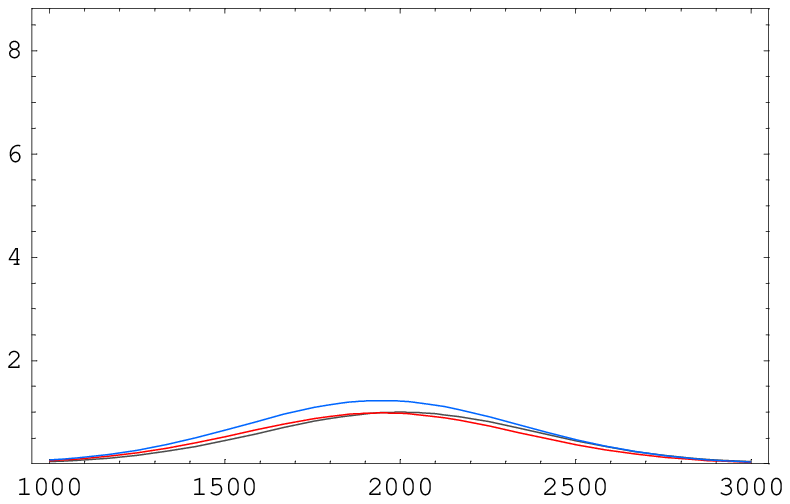}
\includegraphics[width=.49\textwidth ]{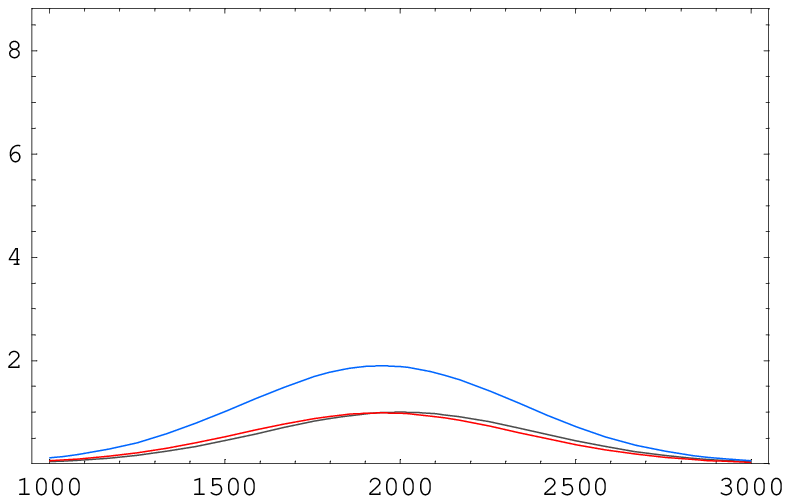}
\includegraphics[width=.49\textwidth ]{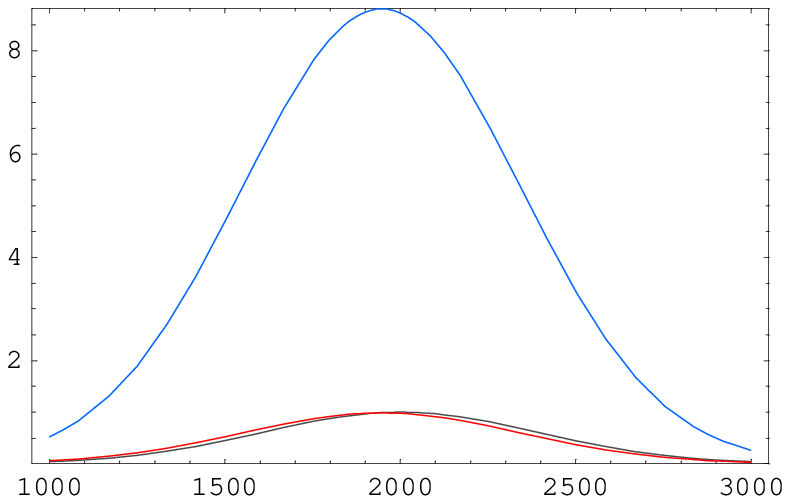}
\caption{Power density $S(x,t)$ as in Fig. \ref{fig15-17}, but in
EIT regime. The pump-beam is resonant while probe pulse has an
initial frequency spread $\sigma_{p}=0.01 \Gamma_{1}$ and a
carrier frequency $\omega_{c}=\omega_{31}+\delta_{p}$, where
$\delta_{p}=0$. The peak of the transmitted pulse has retarded by
an amount $\Delta x \simeq 55 m$. We note that, in second case
($n_{2}=0.3 \ \ \text{and} \ \ n_{1}=0.7$), there is a peak
amplification of $90 \%$, while gain is $28 \%$, i.e. a very slow
propagation involves a very large amplification.} \label{fig30-32}
\end{center}
\end{figure}

\begin{figure} [!ht]
\begin{center}
\includegraphics[width=.49\textwidth ]{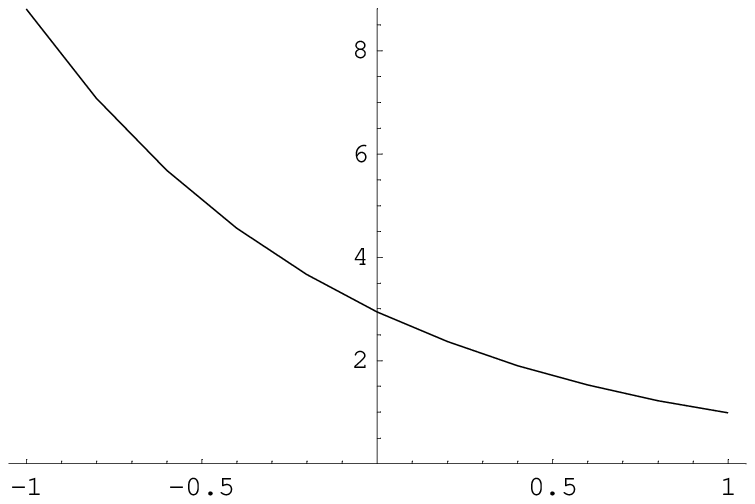}
\includegraphics[width=.49\textwidth ]{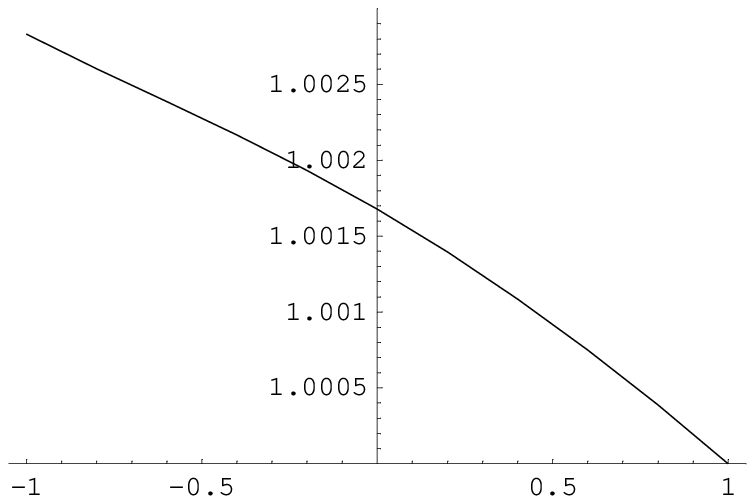}
\caption{Plot of the dependence of the amplitude of the peak and
of the center's shift on the difference of the populations
($n_{1}-n_{2}$) with zero detuning. The shift's values are
normalized to that for $n_{1}=1$.} \label{fig33-34}
\end{center}
\end{figure}

\section{Anomalous propagation}

Let us choose the probe frequency detuning corresponding to the
minimum (negative) group velocity and we observe anomalous
propagation. Note that the gray line represents the pulse
propagating in vacuum, the red line refers to the case $n_{1}=1$,
while the blue lines represent the following ones:
\begin{itemize}

\item[1)] $n_{1}=0.9 \ \ n_{2}=0.1 \ \ n_{3}=0$

\item[2)] $n_{1}=0.7 \ \ n_{2}=0.3 \ \ n_{3}=0$

\item[3)] $n_{1}=0.1 \ \ n_{2}=0.9 \ \ n_{3}=0$

\end{itemize}

\begin{figure} [!ht]
\begin{center}
\includegraphics[width=.49\textwidth ]{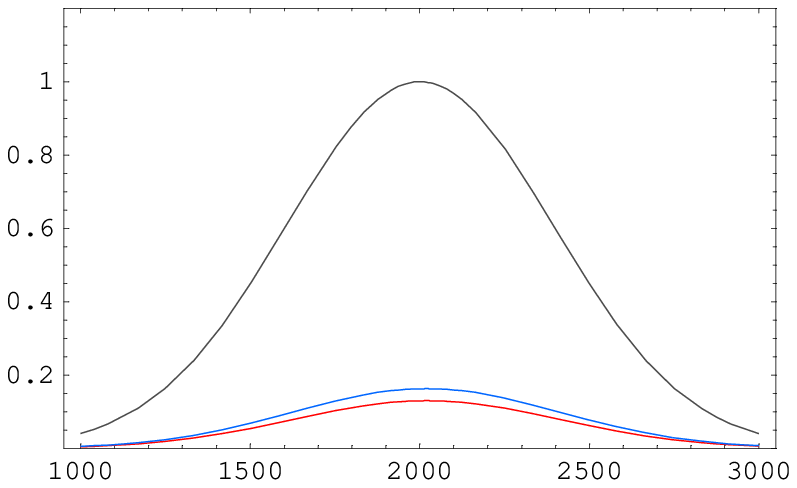}
\includegraphics[width=.49\textwidth ]{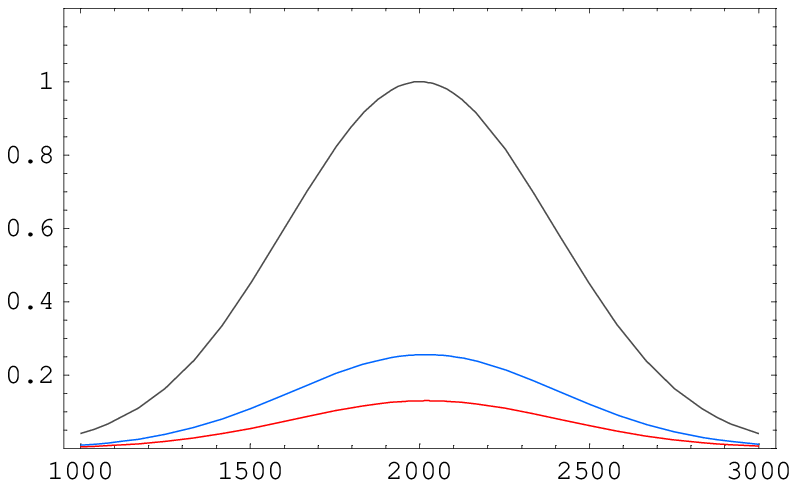}
\includegraphics[width=.49\textwidth ]{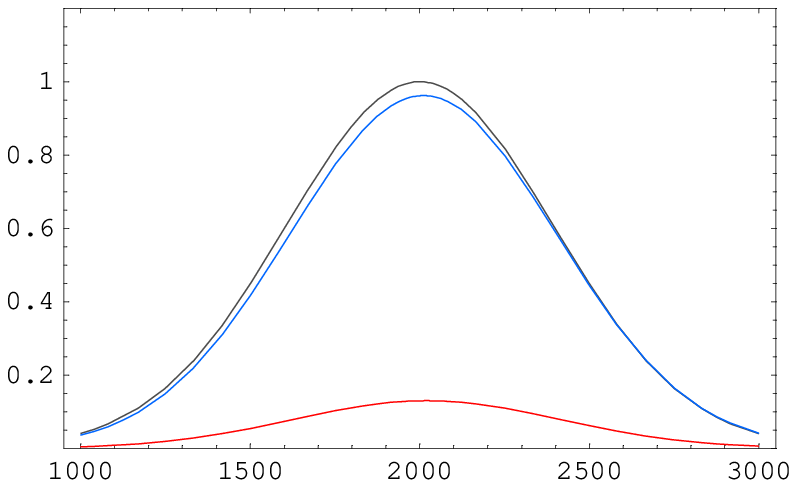}
\caption{Power density $S(x,t)$ as in Fig. \ref{fig30-32}. The
probe pulse has a carrier frequency
$\omega_{c}=\omega_{31}+\delta_{min}$, where $\delta_{min}\simeq
0.28 \gamma_{3}$ is the probe detuning at which $v_{g}$ reaches
the smallest negative value. The peak of the transmitted pulse has
advanced by an amount, respectively, $\Delta x=19.8, 17.5$ and
$10.6 \ m$.\index{negative group velocity}} \label{fig35-37}
\end{center}
\end{figure}

Now we have negative group velocity but there is absorption unless
in the case of complete population inversion. In other words the
propagated pulse advances one propagating in vacuum, but it is
attenuated. However in the case $n_{1}=0 \ \ n_{3}=0 \ \ n_{2}=1$,
the point of minimum negative group velocity falls inside the
gain-region and it seems that we violate the principle of
causality. We discuss this problem in Appendix \ref{einstein1} and
we note that the physics is safe.
\par
Figure \ref{fig40} shows that a gaussian wave packet that enters a
gain medium at the entrance face at $z=0$ generates a transmitted
wave packet at exit face at $z=d$ ($10 \ \mu m$), whose peak
leaves the exit face of the cell \textit{before} the peak of the
incident wave packet arrives at the entrance face.

\begin{figure} [th!]
\begin{center}
\includegraphics[width=.49\textwidth ]{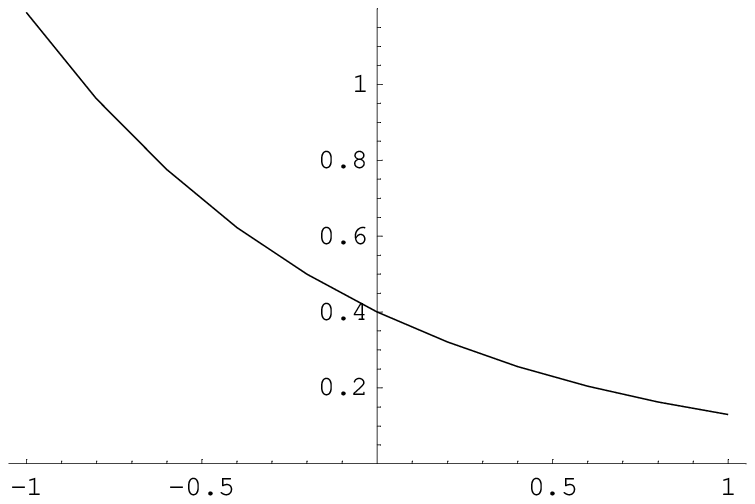}
\includegraphics[width=.49\textwidth ]{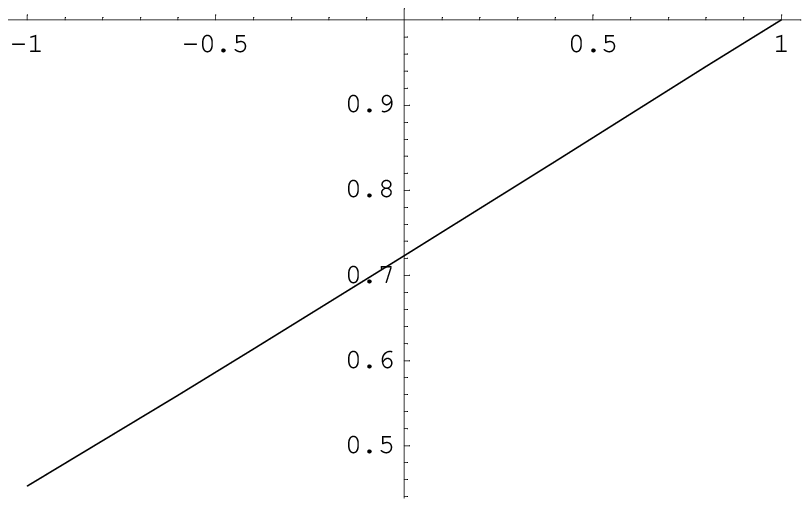}
\caption{Plot of the dependence of the amplitude of the peak and
of the center's shift on the difference of the populations
($n_{1}-n_{2}$) with minimum negative group velocity. The shift's
values are normalized to the case $n_{1}=1$.} \label{fig38-39}
\end{center}
\end{figure}
\
\newline
\newline

\begin{figure} [th!]
\begin{center}
\includegraphics[width=.77\textwidth ]{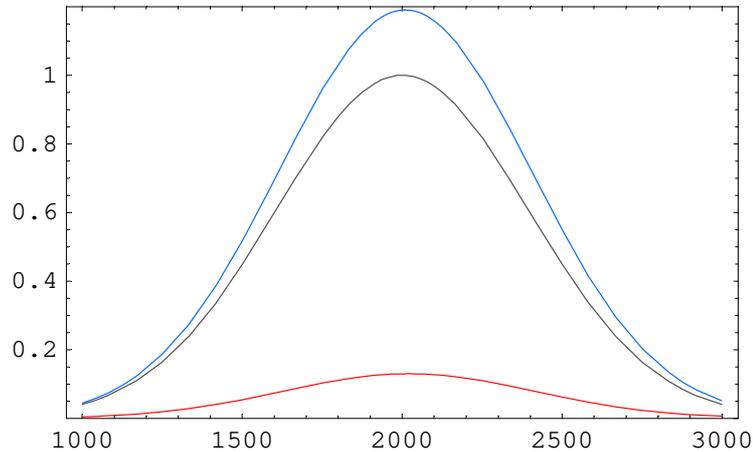}
\caption{Power density $S(x,t)$ as in Fig. \ref{fig35-37}, with
$n_{2}=1$. The peak of the transmitted pulse (blu line) has
advanced by an amount $\Delta x=9.4 \ m$ with a peak amplification
equal to $19 \%$.} \label{fig40}
\end{center}
\end{figure}
\
\newline

\section{A simpler system: hot atoms}\label{article}\index{hot atoms}

Now we perform all the calculations above for a sample of a
$^{87}$Rb vapor at $35^\circ$C in a 10 cm long cell \cite{caruso1}
and examine a realistic model to create the atomic population
ratios needed for amplification without inversion introducing an
incoherent loss rate from one of the ground levels.

\begin{figure} [!ht]
\begin{center}
\includegraphics[width=.5 \textwidth]{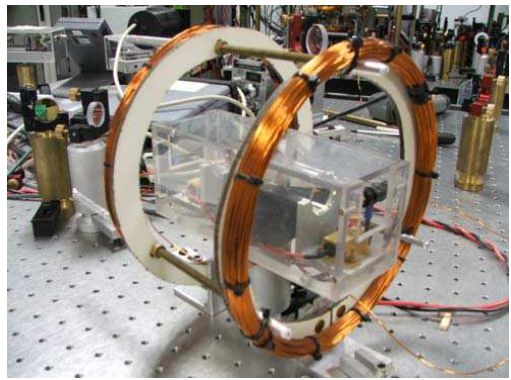}
\caption{A 10 cm long cell containing $^{87}$Rb at a temperature
of $35^\circ$C and with density equal to $5.296 \cdot 10^7
atoms/cm^3$ \cite{caruso1}. (Quantum Information Laboratory,
Scuola Superiore di Catania)} \label{fig41}
\end{center}
\end{figure}

\begin{figure} [p]
\begin{center}
\includegraphics[width=.9\textwidth ]{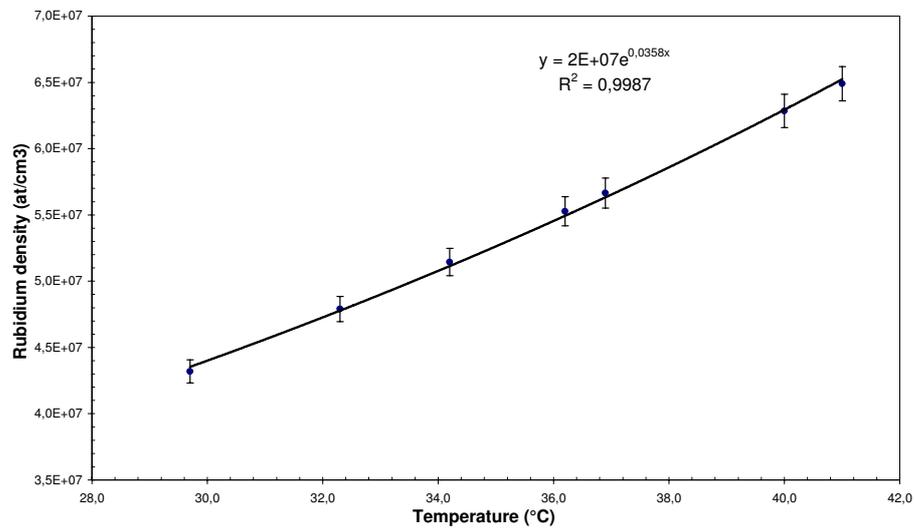}
\caption{Rubidium density as function of the temperature in a 10
cm long cell containing $^{87}$Rb.} \label{fig42}
\end{center}
\end{figure}

\begin{figure} [p]
\begin{center}
\includegraphics[width=.9\textwidth ]{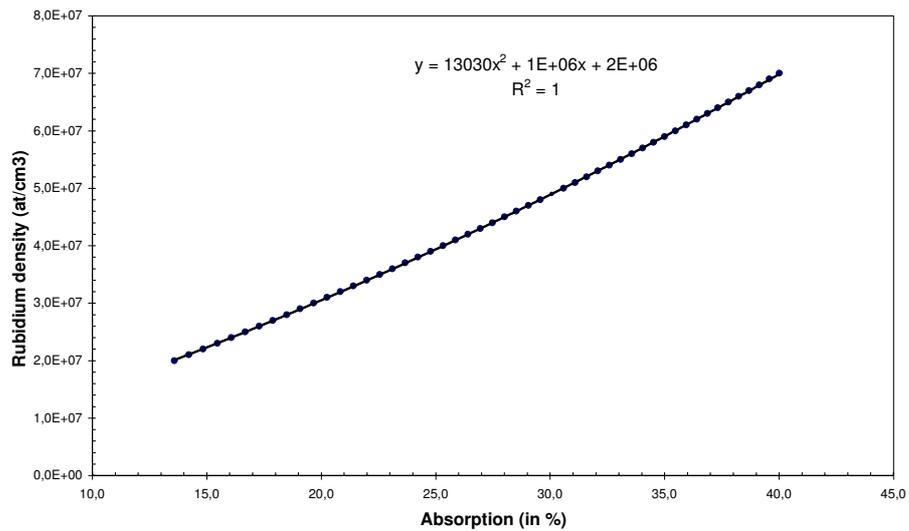}
\caption{Rubidium density as function of absorption in a 10 cm
long cell containing $^{87}$Rb.} \label{fig43}
\end{center}
\end{figure}

In Fig. \ref{fig42} we report the experimental value of the
Rubidium density, contained in a 10 cm long cell, at different
temperatures. These results have been obtained in the Quantum
Information Laboratory (Scuola Superiore di Catania) by measuring
the absorption in the cell at different values of temperature.
Indeed the Rubidium density is strictly connected to the
absorption, as shown in Fig. \ref{fig43}.
\par
In the same level scheme for the $^{87}$Rb $D_1$ line as in Fig.
\ref{fig8} we consider the more realistic situation of a room
temperature cell in which atomic populations are thermally
distributed among all levels. \par
\begin{figure}[ht!]
\begin{center}
\includegraphics[width=.75\textwidth ]{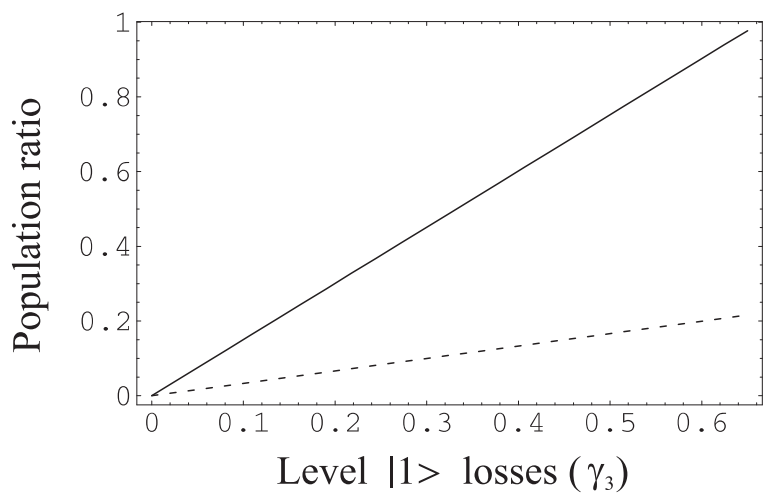}
\caption{Population ratio between levels $|{2}\rangle$ and
$|{1}\rangle$ (continuous line) and $|{3}\rangle$ and
$|{1}\rangle$ (dashed line) as a function of the losses from level
$|{1}\rangle$, $\gamma_1$ (in units of $\gamma_3$)
\cite{caruso1}.} \label{fig44}
\end{center}
\end{figure}

 Therefore the initial
configuration would be a nearly 50/50 distribution between the two
ground states; in addition we introduce a loss mechanism from the
level $|{1}\rangle$ while for level $|{3}\rangle$ we consider the
correct branching ratios for $^{87}$Rb. In Fig. \ref{fig44} we
show how the steady state population ratio between the two ground
levels, even in presence of the laser beams, can indeed be varied
this way while, at the same time, the amount of population placed
in the excited level remains negligible. We note that a similar
loss from the ground state can be easily implemented by an
incoherent RF field stimulating a transition to any other ground
sub--level. However this configuration has the disadvantage that
while the population in level $|{2} \rangle$ increases so does the
dephasing rate $\gamma_1$ of the ground states superposition. This
strongly affects both the gain and the propagation of the pulse as
discussed below.

\begin{figure}[ht!]
\includegraphics[width=.47\textwidth ]{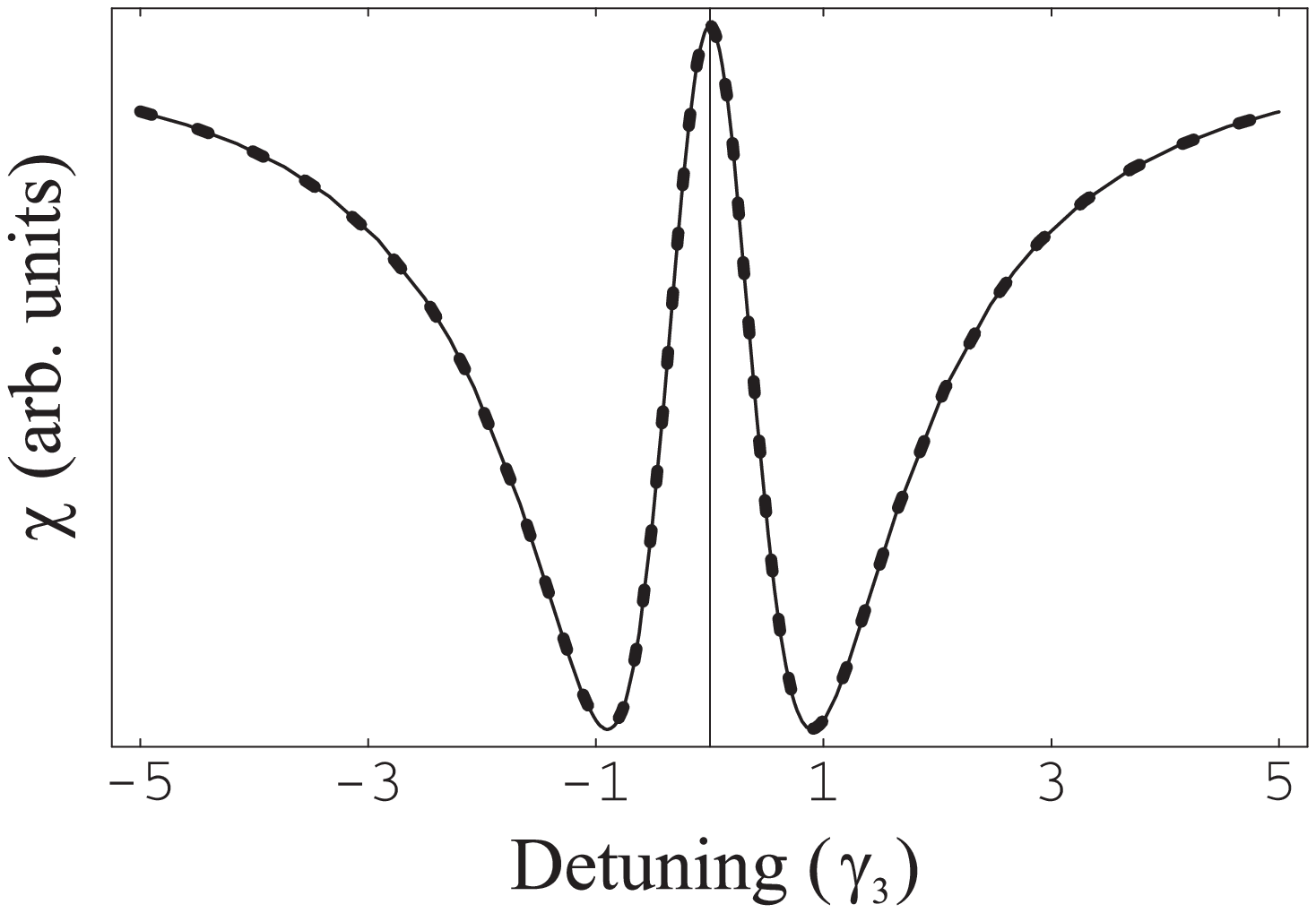}
\includegraphics[width=.52\textwidth ]{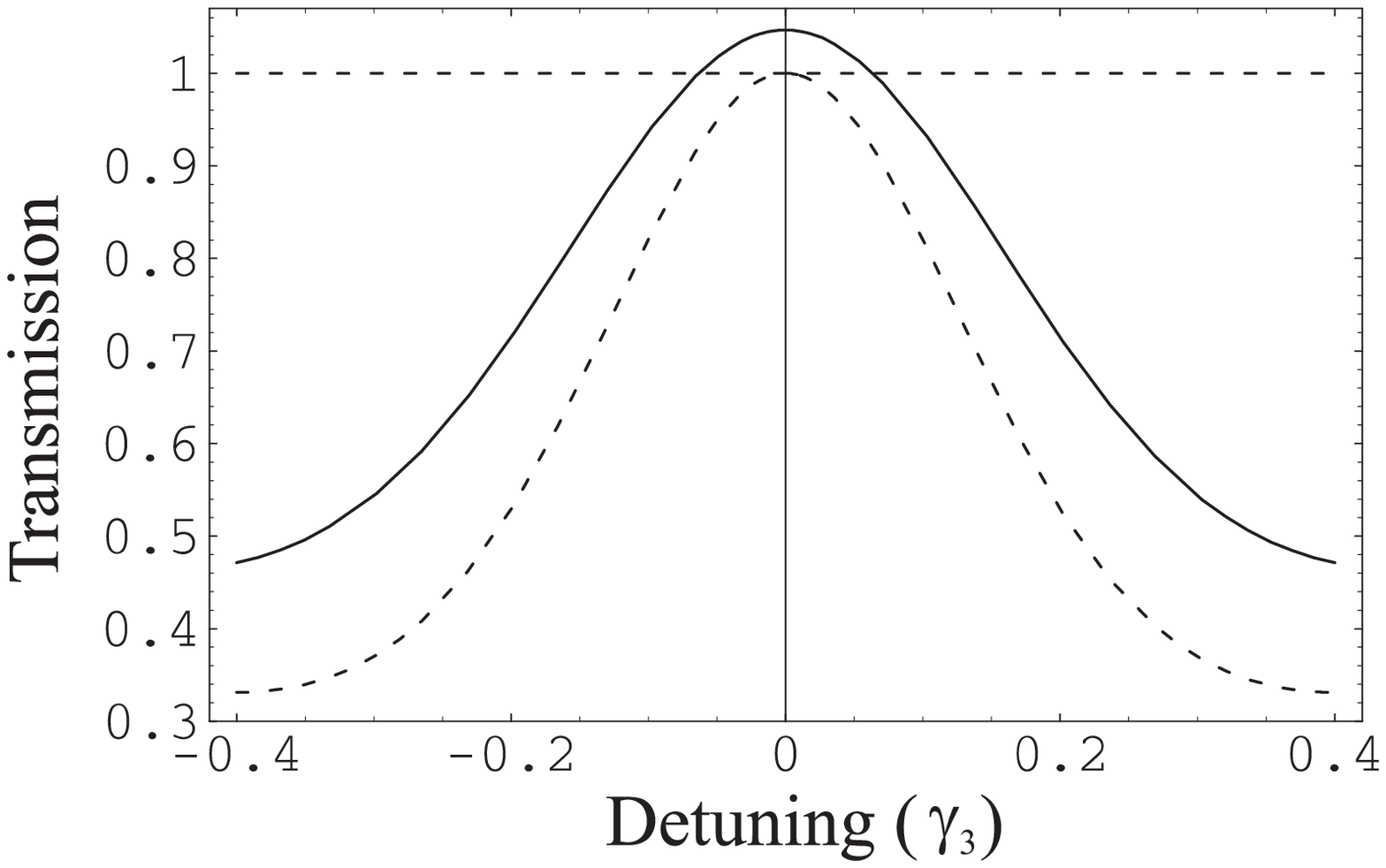}
\caption{Left: Imaginary part of the atomic susceptibility from
the full density matrix treatment for a loss rate from level
$|{1}\rangle$ of $0.2 \gamma_3$ as a function of the probe
detuning $\delta_p$ (continuous line). Atomic susceptibility from
equation (\ref{eq:chi}) when $30\%$ of the population is placed in
level $|{2}\rangle$ as a function of probe detuning $\delta_p$
(dashed line). In both cases the pump Rabi frequency was $0.8
\gamma_3$. Right: Probe transmission as a function of probe
detuning for a 10 cm long cell containing rubidium at $35^\circ$C,
when all population is in level $|{1} \rangle$ (dashed line) and
when $30\%$ of the population is in the level $|{2} \rangle$
(continuous line). In both cases the Rabi frequency of the pump
laser is 0.8 $\gamma_{3}$ \cite{caruso1}. } \label{fig45-46}
\end{figure}

Again we can check that the steady state atomic susceptibility
computed from equation (\ref{eq:chi}) is the same as the one
computed from the full solution of the density matrix for the
parameters considered here. This is shown in Fig. \ref{fig45-46}
left for a pump Rabi frequency of $0.8 \gamma_3$ and $30\%$ of the
atomic population in level $|{2}\rangle$. This corresponds to the
situation where $\gamma_1=0.2 \gamma_3$.

\begin{figure}[ht!]
\begin{center}
\includegraphics[width=.7\textwidth ]{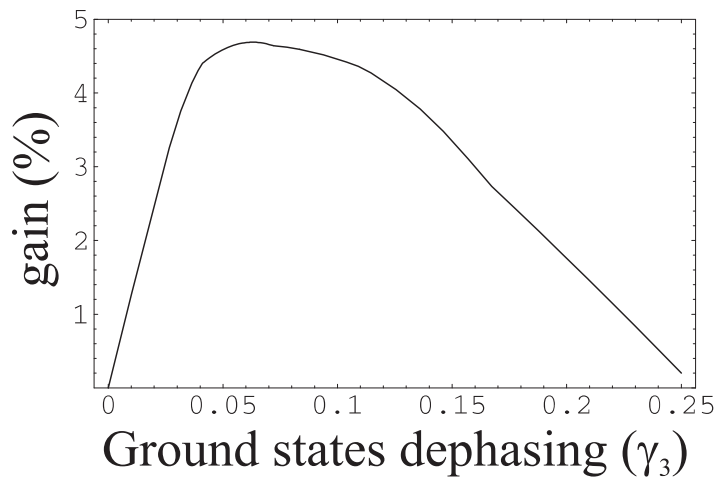} \caption{Percentage gain
as function of the ground state dephasing $\gamma_1$ (in units of
$\gamma_3$) for the same experimental conditions as in Fig.
\ref{fig45-46} \cite{caruso1}.} \label{fig47}
\end{center}
\end{figure}

From equations (\ref{eq:chi}) and (\ref{eq:T}) we obtain the
transmission spectrum around the atomic resonance of the probe
laser through a 10 cm long cell containing $^{87}$Rb at a
temperature of $35^\circ$C. This is reported in Fig.
\ref{fig45-46} right for a pump Rabi frequency of $0.8 \gamma_3$.
The transmission shows a peak in correspondence to the probe
resonance, i.e. when the pump and probe lasers close a Raman
transition between levels $|{1} \rangle$ and $|{2} \rangle$. This
is precisely the electromagnetically induced transparency (EIT) in
which the probe absorption at resonance is cancelled by
destructive quantum interference between the two possible
absorption paths for the probe laser, namely the two--step
transition from $|{1} \rangle$ to level $|{2} \rangle$ through the
excited level $|{3} \rangle$ and the Raman two--photon transition
between levels $|{1} \rangle$ and $|{2} \rangle$
\cite{pavone,ari}. The peak in the spectrum goes all the way to
full transmission when all the atomic population is placed in
level $|{1} \rangle$ (dashed line). On the contrary, when some
population ($30\%$ in Fig. \ref{fig45-46} right) is present in
level $|{2} \rangle$ the transmission goes above unity indicating
the presence of gain (continuous line). We should remark that, as
shown in Fig. \ref{fig44}, no population inversion is present in
the system therefore we are fulfilling the condition for gain
without inversion \cite{ari}. In the situation considered here
increasing the population of level $|{2} \rangle$ does not
necessarily lead to extracting more gain. Indeed we are changing
the population ratio by incoherently removing population from
level $|{1} \rangle$, which in turn increases the dephasing rate
$\gamma_1$ between the ground sublevels. When the dephasing
increases the EIT effect is reduced and no gain is observed. In
Fig. \ref{fig47} we report the centerline gain as a function of
the dephasing rate $\gamma_1$ with all the other experimental
parameters fixed as in Fig. \ref{fig45-46}. As expected the gain
increases to a maximum and then drops down to zero.

\begin{figure}[th!]
\begin{center}
\includegraphics[width=.6\textwidth ]{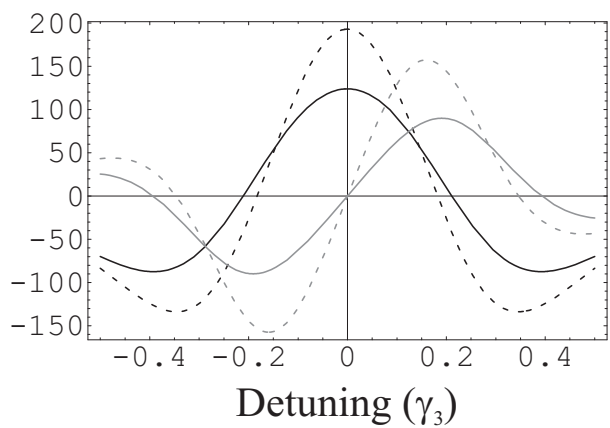}
\caption{Reciprocal group velocity (black line) and group velocity
dispersion function $\cal{D}$ (gray line) [(m/s)$^{-1}$] as a
function of probe detuning $\delta_p$ (in unit of $\gamma_3$) for
the same experimental conditions as in Fig. \ref{fig45-46}. Dashed
lines correspond to the case when all population is in level $|{1}
\rangle$, continuous lines correspond to the case when 30\% of the
population is in level $|{2} \rangle$. Gray curves are reduced by
a factor of 10 \cite{caruso1}.} \label{fig48}
\end{center}
\end{figure}

Combining equations (\ref{eq:vg}) e (\ref{eq:dg}) with equation
(\ref{eq:chi}), we obtain the frequency dependence of the group
velocity and its dispersion around the probe resonance for the
same experimental parameters as before, as reported in Fig.
\ref{fig48}. We note that there are three probe frequencies where
the group velocity dispersion vanishes. These points correspond to
frequency values where a suitable probe pulse can propagate
through the medium without distortion. One of these points
corresponds to the line center where we have retarded propagation.
The other two, which are symmetrical with respect to the first
point, correspond to anomalous propagation, i.e. a negative group
velocity. We note that, when some population is placed in level
$|{2} \rangle$, the positive minimum of the group velocity is
increased. This is not an effect of population but of the increase
in dephasing of the ground sub--levels. At the same time the
negative minimum group velocity is also increased. This means
that, under the conditions of gain without inversion, a
propagating pulse will be slowed down while, at the same time,
undergoing amplification whereas in the anomalous propagation
region the pulse is advanced but not amplified, as shown for cold
atoms in the previous sections. However both the pulse delay and
the pulse advance are reduced with respect to normal EIT.

\begin{figure}[th!]
\begin{center}
\includegraphics[width=.51\textwidth ]{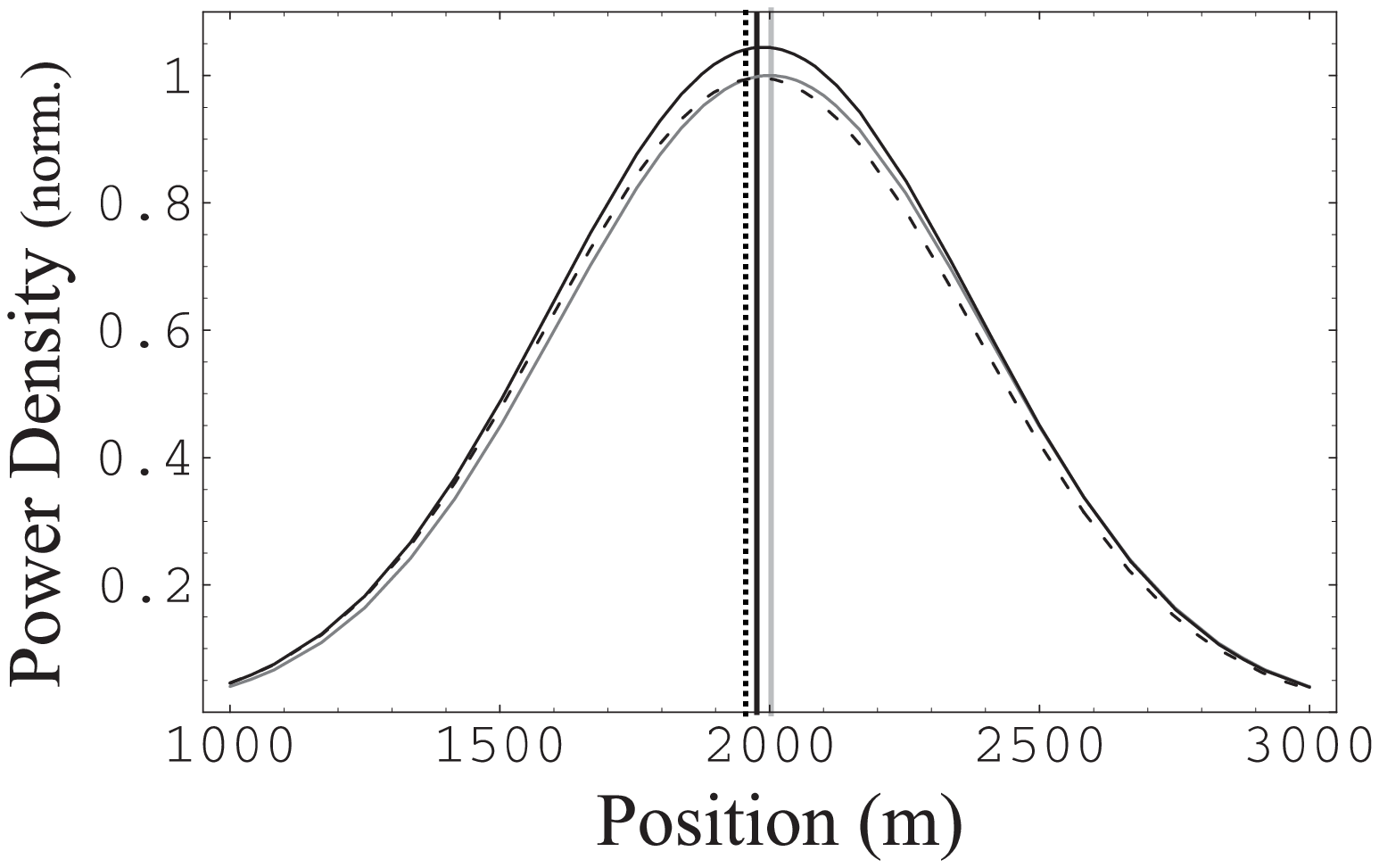}
\includegraphics[width=.48\textwidth ]{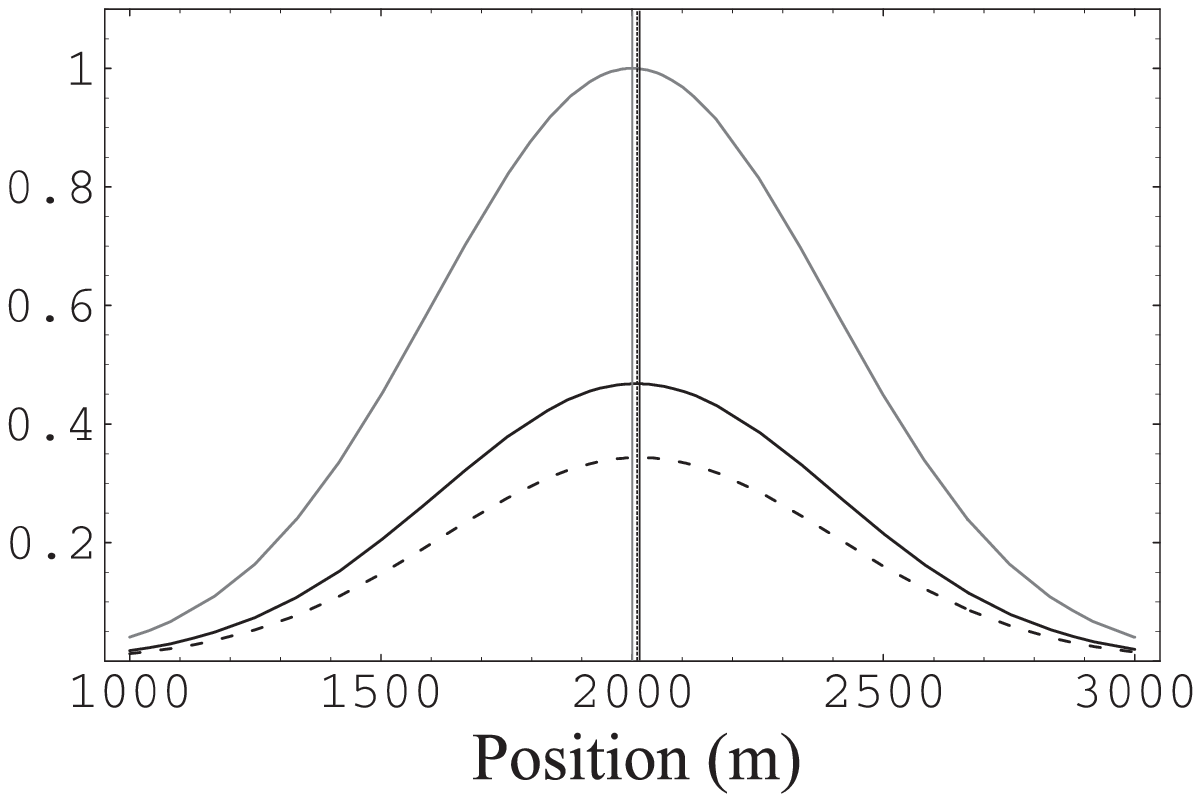}
\caption{Power density for a gaussian probe pulse normalized to
the pulse propagating in vacuum. On the left, the peak central
frequency is on resonance resulting in retarded pulse propagation.
The vertical lines in the plot indicate the center of masses of
the pulses. On the right, the probe detuning corresponds to the
negative minima of reciprocal group velocity showing advanced
propagation. The gray line refers to the pulse propagating in
vacuum, the dashed line to the case where all population is in the
level $|{1} \rangle$ and the black continuous line to the case
where $30\%$ of the population is in the level $|{2} \rangle$.
Experimental parameters are the same as the Fig. \ref{fig45-46}
\cite{caruso1}.} \label{fig49-50}
\end{center}
\end{figure}

\begin{figure}[!]
\begin{center}
\includegraphics[width=.51\textwidth ]{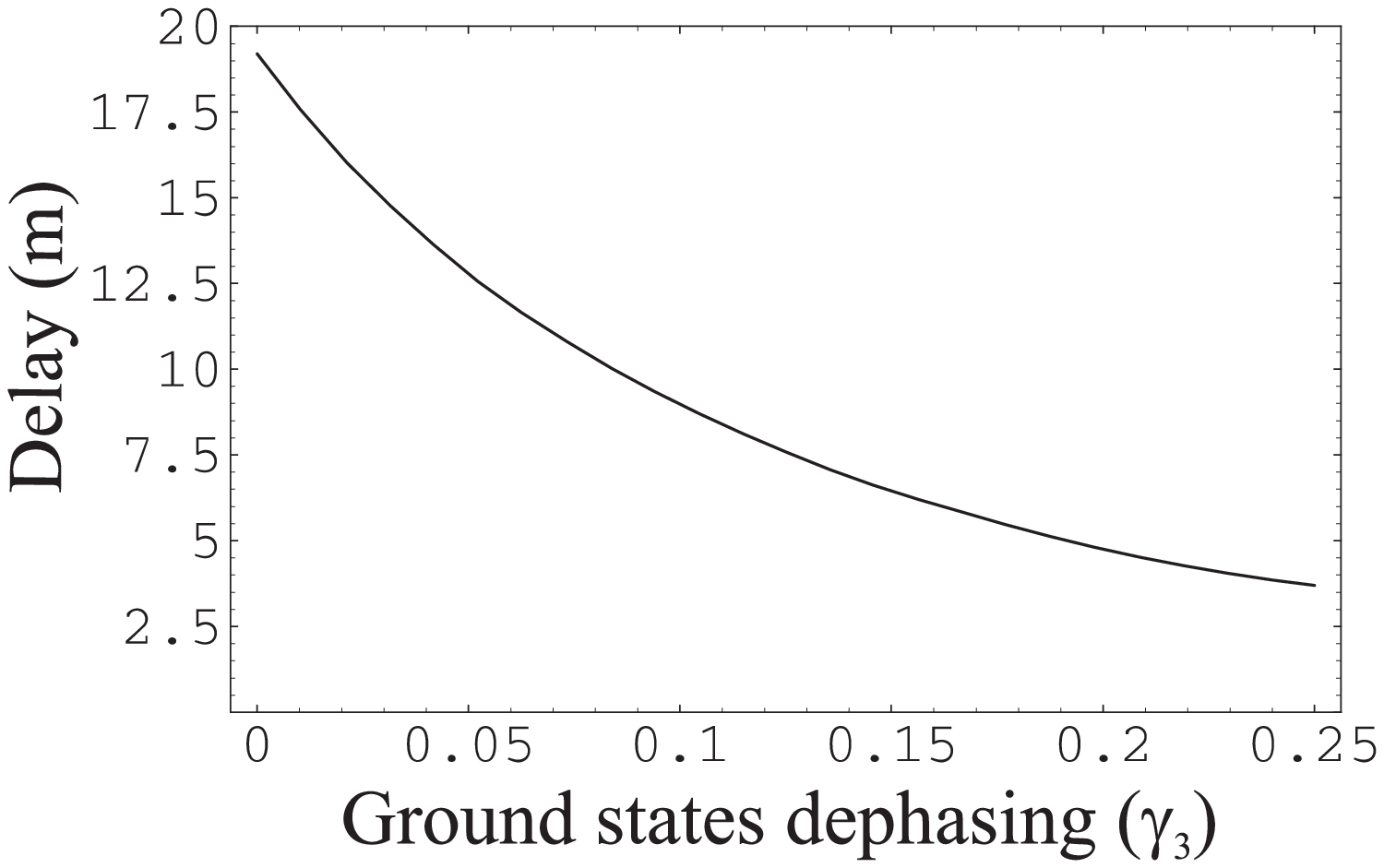}
\includegraphics[width=.48\textwidth ]{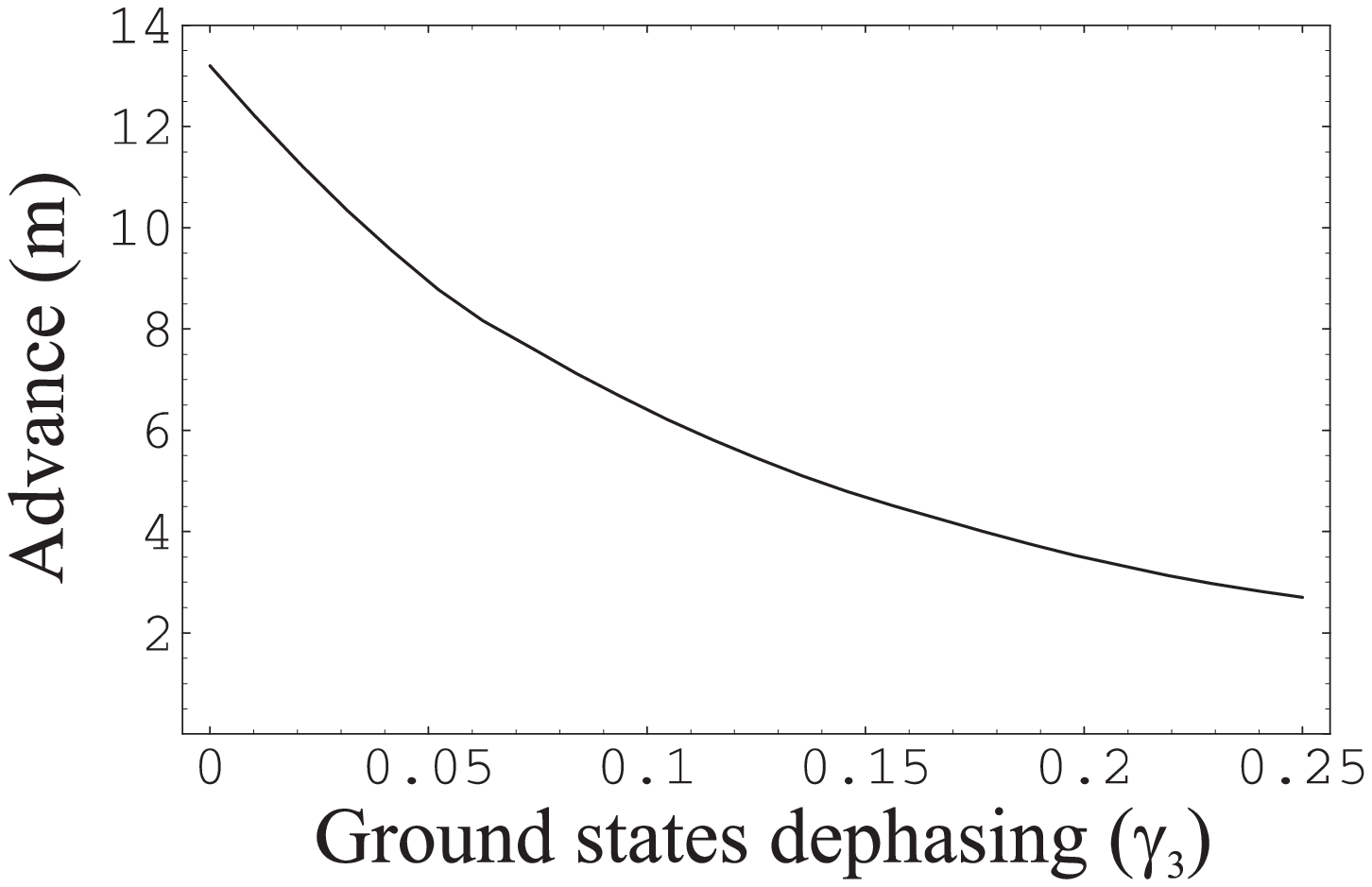}
\caption{On the left pulse delay, on the right pulse advance as a
function of the dephasing rate $\gamma_1$ (in units of $\gamma_3$)
for the same experimental parameters as in Fig. \ref{fig49-50}
\cite{caruso1}.} \label{fig51-52}
\end{center}
\end{figure}

In Fig. \ref{fig49-50} we report the power density spectrum for a
gaussian probe pulse propagating through a 10 cm long cell again
for a pump Rabi frequency of $0.8 \ \gamma_3$. On the left we show
the resonant case where the pulse propagation is retarded. We have
chosen the parameters to be at the maximum amplification (black
line), in such a case the delay with respect to a pulse
propagating in vacuum (grey line) is 12.2 m which amounts to a
velocity of $\frac{c}{120}$ in the cell. In the normal EIT
situation (dotted line) with all the population in level $|{1}
\rangle$ and virtually no dephasing between the ground sublevels
this delay is 18.9 m. As already mentioned, when we increase the
decay rate from level $|{1} \rangle$, there is a reduction of the
delay as a consequence of the larger dephasing between the ground
levels. As reported in Fig. \ref{fig51-52} (left) for our
experimental parameters the delay is reduced below 5 m when the
dephasing is equal to 0.25 $\gamma_3$. As reported in Fig.
\ref{fig47} at the same level of dephasing no amplification is
observable. \par Conversely the detuned case, shown in Fig.
\ref{fig49-50} right, exhibits a pulse advanced of 13.3 m with
respect to the one propagating in vacuum, but the advanced
propagation is accompanied by absorption. The absorption is
reduced when some population is placed in level $|2 \rangle$ but,
at the same time, also the advance is reduced to 8.6 m. We note
that the observed delay and advance can be greatly enhanced by
reducing the pump Rabi frequency $\Omega_P$ but the effect of the
levels population balance on the pulse advance is strongly
suppressed. In the anomalous propagation regime, the pulse advance
is decreased with respect to the normal EIT in a such way that the
pulse edge never appears ahead of the edge of a pulse propagating
in vacuum for the same distance. This effect is enhanced by the
larger dephasing rate as shown in Fig. \ref{fig51-52} (right) and
the advance is below 3 m when the dephasing rate reaches 0.25
$\gamma_3$. We note that in the regime of population inversion the
amplified pulse leading edge can indeed precede that of the vacuum
propagating pulse. This is not surprising since, by inverting the
atomic levels population, we are effectively storing energy in the
medium \cite{artoniPRA}.

Finally this model is easily extendable to the case of weak
coherent fields to study decoherence in quantum memories as well
as to discuss amplification without inversion in connection with
photon cloning. It will be object of the next chapters.

\chapter{Quantum memory for photons}\label{qm}\index{quantum memory}

\section{Definition of Quantum Memory}\label{int.mem1}

A classical memory keeps a bit for a long time by using an huge
redundancy for the two possible bit values, 0 and 1. When we
consider a quantum bit or qubit this technique becomes more
complicated, but still we can provide an appropriate definition of
quantum memory:
\begin{itemize}
\item A quantum memory is a device where a quantum state can be
kept for a long time and be fetched when desired with excellent
\textit{fidelity}.
\end{itemize}
Clearly, both ``a long time'' and ``excellent \textit{fidelity}''
are determined by the desired task for which the state is kept.
Indeed a quantum state develops in time according to some unitary
operation and is inevitably exposed to interactions with the
environment. If redundancy is added in a simple way as in the
classical case, we might lose all the advantages of using quantum
states. Thus, it is not easy to keep a quantum state unchanged for
a long time and the same problem appears when we want to transmit
a state over a long distance. However, with existing technology,
one can talk about transmissions of quantum states (e.g. sending
photons' polarization states to a distance of 100 kilometers),
while it does make much sense to discuss memories where a state
can be kept for a few milliseconds. Let us consider a quantum bit
(a two--level quantum system) in a state
\begin{equation} \psi =  {\alpha \choose \beta}  \end{equation}
If it changes according to some unitary transformation to a state
\begin{equation} \tilde \psi =  {\tilde \alpha \choose \tilde \beta} \end{equation}
we may still be able to use it if we know the transformation, but
if it decoheres due to interactions with other systems, in a way
which we cannot reverse in time, the state is lost. As in the case
of transmission over a long distance --- one can choose some
acceptable error rate, $P_e$, and agree to work with the
experimental system as long as the estimated error rate $p_e$ does
not exceed $P_e$. To estimate the error rate, one first does all
effort to re-obtain the desired state (i.e. take the unitary
transformation into consideration) and then one compares the
expected state $\tilde \psi$ with the obtained state $\rho$ and
defines the error rate as the percentage of failure. In theory, if
we consider some irreversible change to the state (due to
environment, or an eavesdropper, or any other reason) and we can
calculate the obtained state, the error rate is
\begin{equation}
p_e = 1 - F(\tilde \psi,\rho)=1-\langle \tilde \psi | \rho | \tilde \psi \rangle
\end{equation}
where $\rho$ is the final state after the interactions and
$F(\tilde \psi,\rho)$ is the fidelity between the two states,
$\tilde \psi$ and $\rho$ (see Appendix \ref{fidelity}).
\par
The definition of quantum memory contains more that just the
ability to preserve a quantum state for a long time. Other
necessary conditions are the input/output abilities: one must be
able to produce a known quantum state, to input an unknown state
into the memory, to measure the state in some well defined basis.
Moreover one needs to take it out of the memory (without measuring
it) in order to perform any unitary transformation on it or alone
or together with other quantum systems.
\par All these reasons are sufficient to understand that realizing
a quantum memory is a very challenging task.

\newpage
\section{EIT in quantum information science}

The propagation of the light in an EIT medium is associated with
the existence of quasi-particles, which Fleischhauer
\cite{Fleischhauer-PRA-2002} call \textbf{dark--state polaritons}
(\textbf{DSP}). A dark--state polariton is a mixture of
electromagnetic and collective atomic excitations of spin
transitions (spin-wave).\index{dark--state polariton}
\par
Recently the authors in
\cite{Lukin-PRL-2000,Fleischhauer-PRL-2000,Fl00-OptCom} have
showed that it is possible to transfer adiabatically the quantum
state of photons to {\it collective atomic excitations} in an EIT
medium and recent experiments
\cite{Liu-Nature-2001,Phillips-PRL-2001} have already demonstrated
the dynamic group velocity reduction and adiabatic following in
the dark--state polaritons.
\par
When a polariton \index{polaritons} propagates in an EIT medium
\cite{Fleischhauer-PRL-2000}, its properties can be modified
simply by changing the intensity of the control beam and the
polariton group velocity is proportional to the magnitude of its
photonic component. As the control intensity is decreased the
group velocity is slowed, which also implies that the contribution
of photons in the polariton becomes purely atomic, and its group
velocity is reduced to zero\footnote{Perhaps it is a stretch to
talk of stopping light because individual photons are not really
halted. Rather, the excitation carried by the signal pulse,
involving such properties as its angular momentum and pulse shape,
is transferred into a collective atomic spin excitation with the
help of the second, coupling beam (\textit{pump field}) having in
general a different polarization and frequency from the signal
pulse (\textit{probe field}).}. At this point, quantum information
originally carried by photons is mapped onto long-lived spin
states of atoms. As long as the trapping process is sufficiently
smooth (i.e. adiabatic), the entire procedure has no loss and is
completely coherent. The stored quantum state can easily be
retrieved by simply re-accelerating the stopped polariton. \par In
other terms, since the reduction of the group velocity happens in
a linear way, the quantum state of a slowed light pulse can be
preserved. Therefore a non-absorbing medium with a slow group
velocity is a temporary ``storage'' device. However, in principle
such a system has only limited ``storage'' capabilities; in
particular the achievable ratio of storage time to pulse length
can practically attain only values on the order of 10 to 100,
because it depends on the square root of the medium opacity
\cite{Hau99b}. In other words this limitation originates from the
fact that a small group velocity is associated with a narrow
spectral acceptance window of EIT \cite{EIT-spectr} and hence
larger delay times require larger initial pulse length.
\par
\begin{figure} [th!]
\begin{center}
\includegraphics[height=.9\textwidth,width=.95\textwidth ]{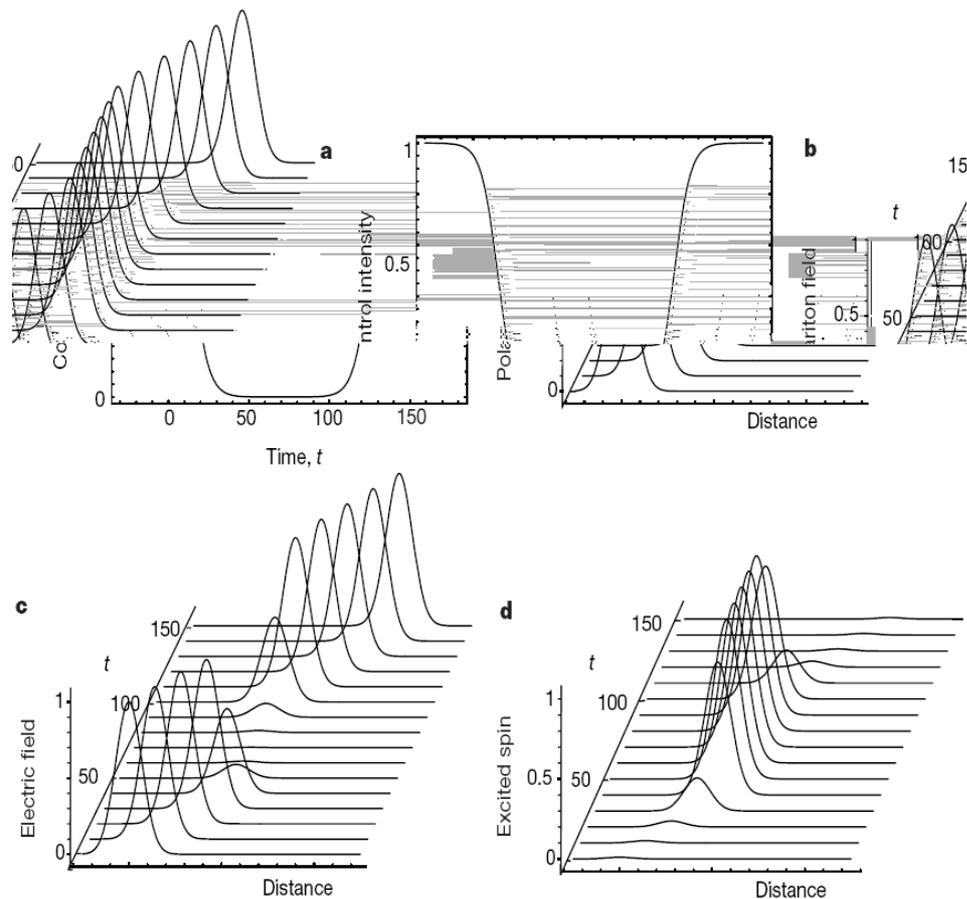}
\caption{Dark--state polaritons. A dark--state polariton can be
stopped and re-accelerating by ramping the control field intensity
as shown in \textbf{a}. The coherent amplitudes of the polariton
$\Psi$, the electric field $E$ and the spin components S are
plotted in \textbf{b} to \textbf{d}. \cite{Fleischhauer-PRA-2002}}
\label{fig53}
\end{center}
\end{figure}
\begin{figure} [!ht]
\begin{center}
\includegraphics[height=.84\textwidth,width=.8\textwidth ]{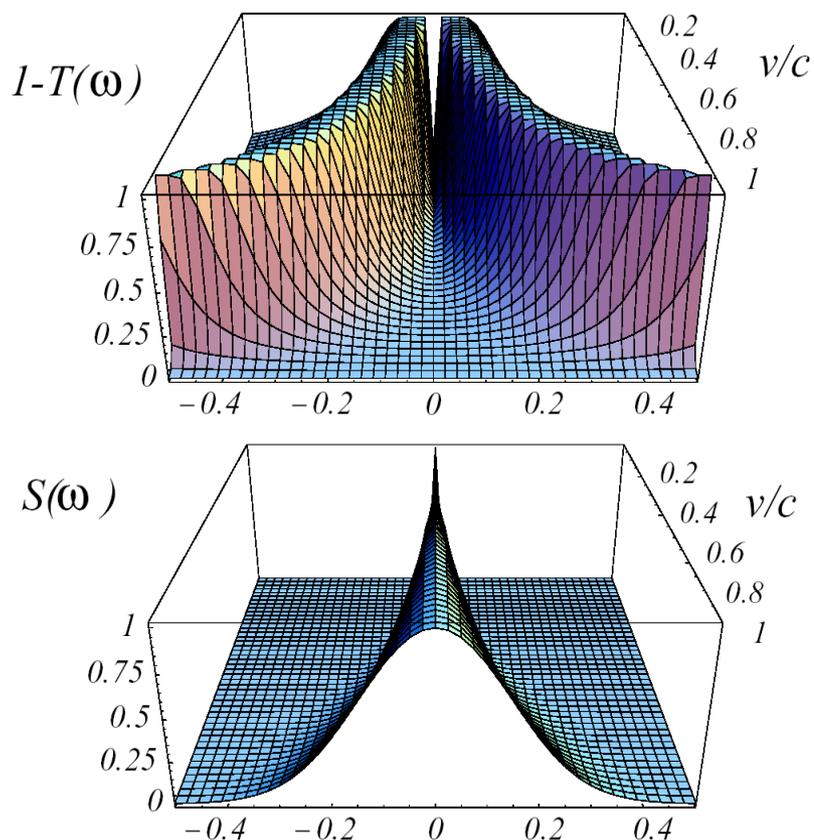}
\caption{Simultaneous narrowing of transmission spectrum (top) and
pulse spectrum (bottom) for time dependent variation of group
velocity $v/c$ in units of the probe detuning $\delta_p$.
\cite{Fleischhauer-PRA-2002}} \label{fig54}
\end{center}
\end{figure}

Figure \ref{fig53} shows the evolution of the ``signal'' light
pulse, spin coherence and polariton \index{polaritons} when the
control beam is turned off and on. The amplitude of the signal
pulse decreases as it is being decelerated whereas the spin
coherence grows; the procedure is reversed when the control beam
is turned back on. Besides during the adiabatic slowing the
spectrum of the pulse becomes narrower in proportion to the group
velocity (Fig. \ref{fig54}); so the limitations on initial
spectral width or pulse length essentially disappear and very
large ratios of storage time to initial pulse length can be
achieved.

We note here that the essential point of this technique is not to
store the energy or momentum carried by photons but to store their
quantum states (\textbf{quantum memory}). Indeed, in practice,
almost no energy or momentum is actually stored in the EIT medium.
Instead, both are being transferred into (or borrowed from) the
control beam in such a way that an entire optical pulse is
coherently converted into a low energy spin wave\index{energy spin
wave}. After some storage time, other coupling photons are sent
through the cell, the information stored in the spin excitations
is transferred back to the radiation field and the original signal
pulse is reconstituted. The information is transferred from purely
photonic to purely atomic excitations under the control of the
coupling laser. This is the key feature that distinguishes the EIT
approach from earlier studies in optics or nuclear physics; it
also makes possible applications in quantum information science. A
different technique, to ``freeze'' light pulses, was suggested in
\cite{Olga-slow}.
\par
Hau \cite{hau} compares the writing process to the formation of a
holographic phase grating on the atomic medium; to read it out,
they turn on the coupling laser and the original light pulse comes
out. Marlan Scully of Texas A\& M University suggests the analogy
of quantum teleportation,\index{quantum teleportation} in which an
atom having a state vector at one point in space is reproduced at
another point in space; in this case it would be a photon state
reproduced at a later time.
\par
Even though EIT has already made a major impact in nonlinear
optical science, commercial applications have not yet emerged. One
potential area is all-optical switching and signal processing in
optical communication. The most serious roadblocks on this front
are materials and speed issues. Good optical control requires long
coherence times and for this reason the majority of experiments
made use of atomic vapors that have relatively slow response. For
practical communication systems, solid state devices are desirable
because of their low cost and the possibility of integration with
existing technologies. Photon-photon interactions enabled by EIT
can fulfil the stringent requirements on precision and efficiency
imposed by quantum information processing. In particular, optical
materials with large nonlinearities and low loss could be
indispensable for the controlled generation of entangled states
and for quantum logic operations. Several avenues for using EIT in
this area have already been explored.

\begin{figure} [th!]
\begin{center}
\includegraphics[width=.6\textwidth ]{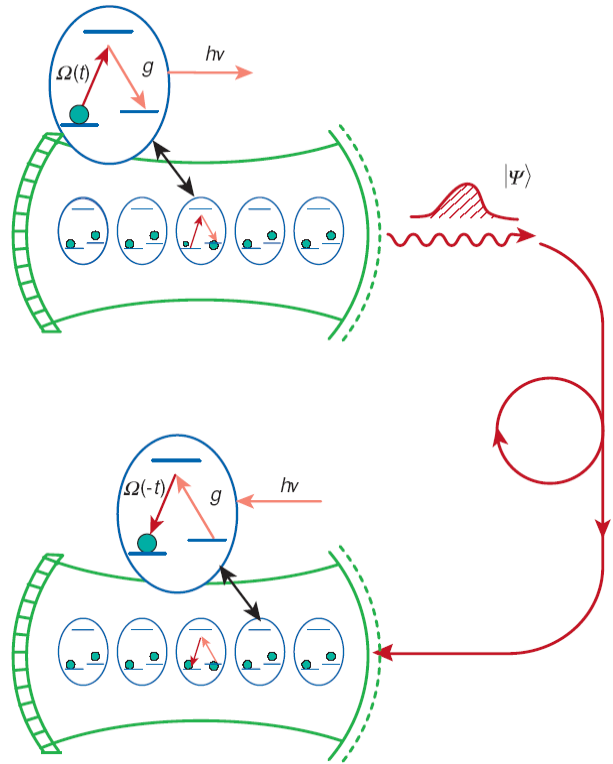}
\caption{Atom-photon quantum network. A selected atom inside the
top optical cavity coherently transfers its internal qubit onto a
single-photon qubit in the cavity through the application of a
classical laser pulse represented by the coupling $\Omega(t)$. The
coupling $g$ is between the single-photon field in the cavity and
the atom. The single photon leaks out of the top cavity, only to
be caught in the lower cavity by a time-reversed and synchronized
classical laser pulse $\Omega(-t). $(H. J. Kimble,
CalTech.)}\label{fig55}\end{center}
\end{figure}

Earlier proposals involved the use of a coherent medium to enhance
photon-photon interactions in optical cavities. The key idea is
that a single photon can shift the resonant frequency such that
the following photon is out of resonance and is therefore
reflected. The resulting ``photon blockade'' effect can form the
basis of a quantum switch \cite{harris}. However, the requirement
of a high-quality cavity is a disadvantage from a practical point
of view. Subsequent work has predicted the efficient generation of
entangled photons on the basis of resonant mixing of four waves.
Using EIT-based phase modulation for two slowly propagating
pulses, the possibility of generating macroscopic quantum states
(so-called ``Schroedinger's cat'' states) of light is predicted
\cite{lukin} but the application of this idea to quantum logic
operations is complicated by the evolution of pulse envelopes in
nonlinear process.
\par
Nevertheless, a scheme for complete quantum teleportation
\index{quantum teleportation} using this technique has recently
been proposed \cite{vitali}. It is also possible to note that once
a dark--state polariton is converted into a purely atomic
excitation in a small-sized sample, logic operations can be
accomplished by promoting atoms into excited states with strong
atom-atom interactions. Here the ability to exchange quantum
information between photons and atoms is essential for performing
operations involving distant units and for the scalability of such
systems. In other words it seems convenient to store and process
quantum information in matter, that forms the nodes of a quantum
network, and to communicate between these nodes using photons.
\newline
\par
In the following sections we will show how to realize a quantum
memory for photons and also how to obtain amplification without
inversion, a typical phenomena of EIT effect. In this way we could
provide a device in which is possible to register efficiently a
quantum state by compensating the unavoidable losses of the
transfer with the photon propagation in gain medium.

\newpage
\section{Quantum memory for a single-mode field}

In order to understand the quantum state mapping transfer, first
of all, as in \cite{Fleischhauer-PRA-2002}, let us consider a
single mode of the radiation field, i.e. a single mode optical
cavity, as a quantum probe. Recall that in Sec. \ref{obe} we used
a classical probe between a meta-stable state and the excited one,
while here we quantize that field, by using a complete quantum
approach.
\par
\begin{figure} [!ht]
\begin{center}
\includegraphics[width=.69\textwidth ]{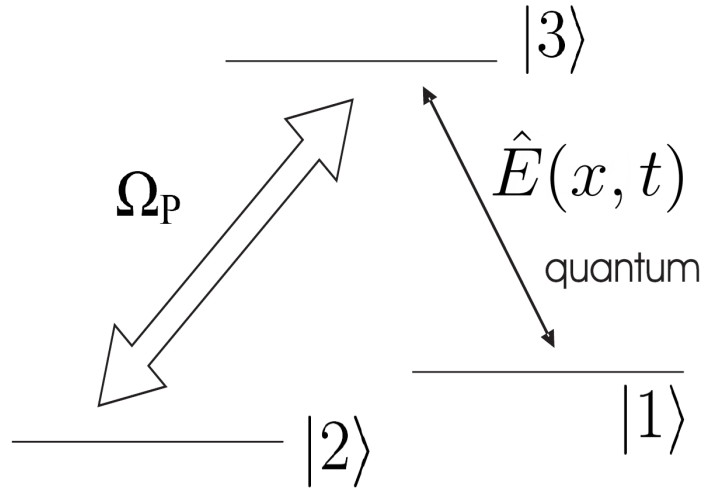}
\caption{Three--level $\Lambda$-type medium resonantly coupled to
a classical field with Rabi-frequency $\Omega_P$ and a quantum
field $\hat E(x,t)$.} \label{fig56}
\end{center}
\end{figure}

Consider a collection of $N$ three--level atoms with two
meta-stable lower states as shown in Fig. \ref{fig56} interacting
with two single-mode optical fields. The transition $|3\rangle \to
|1\rangle$ of each of these atoms is coupled to a quantized
radiation mode, while the transitions from $|3\rangle \to
|2\rangle$ are resonantly driven by a classical control field of
Rabi-frequency $\Omega_P$. Analogously to Sec. \ref{obe}, the
dynamics of this system is described by the interaction
Hamiltonian:

\begin{equation}
\hat H = \hbar g \sum_{i = 1}^N  \hat a \hat{\rho}_{31}^i +
\hbar\Omega_P(t) {\rm e}^{-i\omega_P t} \sum_{i = 1}^N
\hat{\rho}_{32}^i + {\rm h.c.} \label{ham1}
\end{equation}
where $\hat{\rho}_{\mu\nu}^i = |\mu_i\rangle\langle \nu_i|$ is the
flip operator of the $i$th atom between states $|\mu\rangle$ and
$|\nu\rangle$, $\hat a$ is the annihilation operator for the
quantum field, $\omega_P$ is the pump frequency and $g$ is the
coupling constant between the atoms and the quantized field mode
(vacuum Rabi-frequency) which for simplicity is assumed to be
equal for all atoms.
\newline

In analogy with Sec. \ref{EIT}, in this $\Lambda$--configuration,
if the field is initially in a state with at most one photon, the
simplest dark state is \bgar |D,1\rangle &=& \cos\theta(t)\,
|1,1\rangle -\sin\theta(t)\,
|2,0\rangle\label{dark1}\\
&&\tan\theta(t)=\frac{g\sqrt{N}}{\Omega(t)}\label{theta} \enar
where $|D,1\rangle$ indicates a dark configuration with one
photon\footnote{In this tensorial product of two states, the first
one indicates the atomic state and the second one the number of
photon. For example $|2,0\rangle$ represents a configuration in
which one atom is the level $2$ and there is no photons.}.

By definition the dark states do not contain the excited state and
are thus immune to spontaneous emission. In
\cite{cavity-QED,Mabuchi-PRL-1997,Lukin-PRL-2000,Fl00-OptCom} the
authors show that families of dark states exist and then it is
possible to transfer the quantum state of the single-mode field to
collective atomic excitations. Indeed adiabatically rotating the
mixing angle $\theta$ from $0$ to $\pi/2$ leads to a complete and
reversible transfer of the photonic state to a collective atomic
state if the total number of excitations $n$ is less than the
number of atoms. This can be seen very easily in this way: If
$\theta: 0\to\pi/2$ one has for all $n\le N$ \be |D,n\rangle :\,
|1\rangle|n\rangle \longrightarrow |2^n\rangle |0\rangle \ee where
in the final state there are $n$ atoms in the level $2$ and no
photons.
\par
Thus if the initial quantum state of the single-mode light field
is in any mixed state described by a density matrix
$\hat\rho=\sum_{n,m} \rho_{nm}\, |n\rangle\langle m|$, the
transfer process generates a quantum state of collective
excitations according to \bgar &&\sum_{n,m}\rho_{nm} \,
|n\rangle\langle m|
 \otimes |{1}\rangle\langle{1}|
\longrightarrow |0\rangle\langle 0|\otimes \sum_{n,m} \rho_{nm}\,
|{2}^n\rangle\langle {2}^m| \enar

Let us point out that the quantum-state transfer does not
necessarily constitute a transfer of energy from the quantum field
to the atomic ensemble. Indeed in the Raman process the coherent
``absorption'' of a photon from the quantized mode is followed by
a stimulated emission into the classical control field and then
most of the energy is actually deposited in the latter field.

\newpage
\section{Propagation of a quantum field}

Let us generalize to propagating fields, in three--level media
under conditions of EIT, the adiabatic transfer of the quantum
state from the radiation mode to collective atomic excitations, as
in \cite{Fleischhauer-PRA-2002}.
\par
Consider a quantum field propagating along the radial $x$-axis of
a 10 cm long cell containing rubidium at $35^\circ$C, as in Sec.
\ref{article}. A quantized electromagnetic field with positive
frequency part of the electric component $\hat E^{(+)}$ couples
resonantly the transition between the ground state $|1\rangle$ and
the excited state $|3\rangle$; $\omega_p=\omega_{31}$ is the
carrier frequency of the optical field. The upper level
$|3\rangle$ is furthermore coupled to the stable state $|2\rangle$
via a coherent control field with Rabi-frequency $\Omega_P$.
\par
The interaction Hamiltonian reads \bgar
\hat V&=&-p_{31} \sum_j\Bigl(\hat\rho_{31}^j\, \hat E^{(+)}(x_j)+h.c.\Bigr)\nonumber \\
&& -\hbar \sum_j\Bigl(\hat\rho_{32}^j\, \Omega_P(x_j,t) \, {\rm
e}^{i (k^\parallel x_j -\omega_P t)}
   +h.c.\Bigr)
\enar where $x_j$ denotes the position of the $j$th atom, $p_{31}$
denotes the dipole matrix element between the states $|3\rangle$
and $|1\rangle$, $\hat\rho_{\alpha\beta}^j \equiv |\alpha_j\rangle
\langle \beta_j| $ defines the atomic flip operators and
$k^\parallel = \vec k \cdot \vec {\rm e}_x
=\frac{\omega_P}{c}\cos\vartheta$ is the projection of the
wavevector of the control field to the propagation axis of the
quantum field. Particularly we assume that the carrier frequencies
$\omega_p$ and $\omega_P$ of the quantum and control fields
coincide with the atomic resonances $\omega_{31}$ and
$\omega_{32}$ respectively, i.e. we are on the EIT resonance.
\par
Then we introduce slowly-varying variables according to \bgar \hat
E^{(+)}(x,t) &=& \sqrt{\frac{\hbar \omega_p }{2\varepsilon_0 V}}
\hat{\cal E}(x,t)\, {\rm e}^{i\frac{\omega_p}{c}(x-ct)}\\
\hat\rho_{\mu\nu}^j(t) &=& {\widetilde\rho}_{\mu\nu}^j(t) \, {\rm
e}^{-i\frac{\omega_{\mu\nu}}{c}(x-ct)} \enar where $\varepsilon_0$
is the vacuum electric permittivity and $V$ is some quantization
volume, which for simplicity was chosen to be equal to the
interaction volume.
\par
If the (slowly-varying) quantum amplitude does not change in a
length interval $\Delta x$ which contains $N_x\gg 1$ atoms, we can
introduce continuum atomic variables
 \be
{\widetilde\rho}_{\mu\nu}(x,t) = \frac{1}{N_x}\sum_{x_j\in N_x}\,
{\widetilde\rho}_{\mu\nu}^j(t) \ee
 and make the replacement
$\sum_{j=1}^N \longrightarrow \frac{N}{L}\int {\rm d} x $, where
$N$ is the number of atoms, and $L$ the length of the interaction
volume in propagation direction of the quantized field. This
yields the continuous form of the interaction Hamiltonian \bgar
{\hat V} = - \int\!\! \frac{{\rm d} x}{L} \biggl( \hbar g N
{\widetilde\rho}_{31}(x,t)\, \hat{\cal E}(x,t) +\hbar
\Omega_P(x,t){\rm e}^{i\Delta k x}
N(z){\widetilde\rho}_{32}(x,t)+h.c.\biggr)\nonumber \\ \label{ham}
\enar where $g=p_{31}\sqrt{\frac{\omega_p}{2\hbar\epsilon_0 V}}$
is the atom-field coupling constant and $\Delta
k=k^\parallel-k=\frac{\omega_{32}}{c}
\left(\cos\vartheta-1\right)$.
\par
Therefore, in slowly varying amplitude approximation, the
evolution of the Heisenberg operator corresponding to the quantum
field can be described  by the propagation equation
\begin{eqnarray}
\left(\frac{\partial}{\partial t}+c\frac{\partial}{\partial
x}\right) \hat {\cal E}(x,t)= { i} g N \ws_{13}(x,t)\label{field}
\end{eqnarray}
On the other hand, the atomic evolution is governed by a set of
Heisenberg-Langevin equations
\begin{eqnarray}
\dot{\ws}_{33} &=&-2\gamma_3 {\ws}_{33} -ig\Bigl(\hat{\cal
E}^\dagger{\ws}_{13}-h.c.\Bigr)-i\Bigl(\Omega_P^*{\rm e}^{-i\Delta
k x}{\ws}_{23}-h.c.\Bigr)+F_3
\nonumber \\
\dot{\ws}_{11} &=&\Gamma_1 {\ws}_{33}-\Gamma_{12} \rho_{11}
+ig\Bigl(\hat{\cal E}^\dagger{\ws}_{13}-h.c.\Bigr)+F_1 \nonumber\\
\dot{\ws}_{22} &=&\Gamma_2 {\ws}_{33}+\Gamma_{12} \rho_{11}
+i\Bigl(\Omega_P^*{\rm e}^{-i\Delta k x}{\ws}_{23}-h.c.\Bigr)
+F_2 \\
\dot{\ws}_{13} &=&-(\gamma_{1}+\gamma_{3}) {\ws}_{13}
+ig\hat{\cal E}\Bigl({\ws}_{11}-{\ws}_{33}\Bigr)+i\Omega_P{\rm e}^{i\Delta k x}{\ws}_{12} +F_{13} \nonumber\\
\dot{\ws}_{23} &=&-\gamma_{3} {\ws}_{23} +i\Omega_P{\rm
e}^{i\Delta
k x} \Bigl({\ws}_{22}-{\ws}_{33}\Bigr)+ig\hat{\cal E}{\ws}_{21} +F_{23} \nonumber\\
\dot{\ws}_{12} &=& -\gamma_{1} {\ws}_{12} +i\Omega_P^*{\rm
e}^{-i\Delta k x}{\ws}_{13} -ig\hat {\cal E}{\ws}_{32} \nonumber
\label{rho_bc_eq}
\end{eqnarray}
where, as in Fig. \ref{fig8},
$\gamma_3=\frac{\Gamma_1+\Gamma_2}{2}$ and
$\gamma_1=\frac{\Gamma_{12}}{2}$ are the overall dephasing,
$\Gamma_1$ and $\Gamma_2$ are the respective levels linewidths and
$F_\mu$ and $F_{\mu\nu}$ are $\delta$-correlated Langevin noise
operators.
\par
Using the configuration of fixed population as in Sec.
\ref{article} for weak quantum fields, we can assume
\begin{eqnarray}
\ws_{11}&=&1-\eta \\
\ws_{22}&=&\eta \\
\ws_{33}&\simeq&0
\end{eqnarray}
where $\eta$ is fixed by a incoherent RF field stimulating a
transition between the ground sub-levels. By using the
steady-state solution to the rate equations, we find that $\eta
\simeq 4 \frac{\gamma_1}{\Omega_P}$ for $\eta \ll 1$.
\par
Therefore one has for the coherence terms of the density operator:
\begin{eqnarray}
\dot{\ws}_{13} &=&-(\gamma_{1}+\gamma_{3}) {\ws}_{13} +ig
\hat{\cal
E}(1-\eta)+i\Omega_P{\rm e}^{i\Delta k x}{\ws}_{12} +F_{13} \nonumber\\
\dot{\ws}_{23} &=&-\gamma_{3} {\ws}_{23} +i\Omega_P{\rm
e}^{i\Delta
k x} \eta+ig\hat{\cal E}{\ws}_{13} +F_{23} \label{rho_coh_eq} \\
\dot{\ws}_{12} &=& -\gamma_{1} {\ws}_{12} +i\Omega_P^*{\rm
e}^{-i\Delta k x}{\ws}_{13} -i g \hat{\cal E}{\ws}_{32} \nonumber
\end{eqnarray}

\subsection{Low-intensity approximation}\index{low-intensity approximation}

In order to solve the propagation problem, we now assume that the
Rabi-frequency of the quantum field, $\Omega_p$, is much smaller
than $\Omega_P$ and that the number density of photons in the
input pulse is much less than the number density of atoms. In such
a case the atomic equations can be treated perturbatively in $\hat
{\cal E}$. By using Eqs. (\ref{rho_coh_eq}) and neglecting the
first order terms in $\hat{\cal E}$, for $\gamma_1$ fixed, one
obtains

\begin{eqnarray}
\ws_{13}=-\frac{ i}{\Omega_P^*}  {\rm e}^{i\Delta k x} \big[
\frac{\partial}{\partial t} \ws_{12} + \gamma_1
\ws_{12}\Big]-i\frac{g}{\gamma_3} \eta \hat{\cal E}  \label{12}
\end{eqnarray}

Then the interaction of the probe pulse with the medium can be
described by the amplitude of the probe electric field $\hat {\cal
E}$ and the collective ground-state spin variable $\ws_{12}$:

\begin{eqnarray}
\left(\frac{\partial}{\partial t}+c\frac{\partial}{\partial
x}\right) \hat {\cal E}(x,t)= g N \Big\{\frac{{\rm e}^{i\Delta k
x}}{\Omega_P^*}  \big[ \frac{\partial}{\partial t} \ws_{12} +
\gamma_1 \ws_{12}\Big]+\frac{g}{\gamma_3} \eta \hat{\cal E}(x,t)
\Big\} \ \ \
\end{eqnarray}

and

\bgar \ws_{12}= &-& \frac{g \hat {\cal E}}{B} {\rm e}^{-i\Delta k
x}(1-\eta)-\frac{\gamma_1}{B \Omega_P^{*}}
\frac{\partial}{\partial t} \ws_{12}+ \nonumber \\&-&\frac{
i}{B}\left[ \left(\frac{\partial}{\partial
t}+(\gamma_{1}+\gamma_3)\right)\left(-\frac{i}{\Omega_P^*}
\frac{\partial}{\partial t}\ws_{12} -i \frac{g }{\gamma_3} {\rm
e}^{-i\Delta k x} \eta \hat {\cal E} \right)+{\rm e}^{-i\Delta k
x}
 F_{13}\right] \nonumber
\label{s_cb_2} \enar where \be B=\frac{\Omega^2 +
(\gamma_1+\gamma_3)\gamma_1}{\Omega^*}. \nonumber \ee

\newpage
\subsection{Adiabatic limit}\index{adiabatic limit}

The propagation equations simplify considerably if we assume a
sufficiently slow change of $\Omega_P$, i.e. adiabatic conditions
\cite{Fl-Manka96,Hau99b,lukin}. \par Normalizing the time to a
characteristic scale $T$ via $\tilde t=t/T$ and expanding the
r.h.s. of (\ref{s_cb_2}) in powers of $1/T$ we find in lowest
non-vanishing order

\be \ws_{12}(x,t) \simeq -g\frac{\hat{\cal E}(x,t)}{B}{\rm
e}^{-i\Delta k x} \label{s_cb_ad} \ee where we use also that
$\gamma_3 \simeq\gamma_3+\gamma_1$, because $\gamma_1 \ll
\gamma_3$.

We note that also the noise operator $F_{ba}$ gives no
contribution in the adiabatic limit, since $\langle  F_x(t)
F_y(t')\rangle \sim \delta(t-t') = \delta(\tilde t-\tilde t')/T$.
Thus in the perturbative and adiabatic limit the propagation of
the quantum light pulse is governed by the following equation

\begin{eqnarray}
\left(\frac{\partial}{\partial t}+c\frac{\partial}{\partial
x}\right) \hat {\cal E}(x,t)&=& -\frac{g^2 N}{ \Omega^*}
 \frac{\partial}{\partial t}
\frac{\hat {\cal E}(x,t)}{B} + g^2 N
\Big(\frac{\eta}{\gamma_3}-\frac{\gamma_1}{\Omega^* B}\Big) \hat
{\cal E}(x,t) \label{field_ad} \nonumber \\
\end{eqnarray}

If $\Omega(x,t)=\Omega(x)$ is {\it constant in time} and the
incoherent RF field frequency is constant, the term on the r.h.s.
of the propagation equation (\ref{field_ad}) leads to a
modification of the group velocity of the quantum field according
to

\bgar v_g=v_g(x)=\frac{c}{1+n_g(x)} \ \ \ \ \ \ \ \ \
n_g(x)=\frac{g^2 N}{\Omega_P^* B} =\frac{3}{8\pi^2}
\frac{N}{V}\lambda_{13}^3 \frac{kc\, \gamma}{\Omega_P^* B} \enar

where $\frac{N}{V}$ is the atom density, $n_g$ is the group
velocity index and $\lambda_{13}$ the resonant wavelength of the
$3\to 1$ transition. In the next section we will analyze the
propagation under time-dependent conditions.

\section{Quasi-particle picture}\index{quasi-particle picture}

Up to now we have considered the propagation of a quantum field in
an EIT medium under stationary conditions, i.e. with a constant or
only spatially varying control field. In these conditions, a
coherent process that allows for an uni-directional transfer of
the quantum state of a photon wavepacket to the atomic ensemble
isn't possible because the hamiltonian of the system is {\it
time-independent}.
\par
Now let us show how to realize this transfer by using a {\it
time-dependent} control field. Indeed, for a spatially homogeneous
but time-dependent control field\footnote{A purely temporal change
can be realized by shining the control field perpendicular to the
propagation direction of the probe pulse. This has however a
number of disadvantages. Indeed, because of technical difficulties
to achieve a sufficiently high field strength over an extended
laser-beam cross section, the two-photon resonance would be very
sensitive to Doppler-shifts. Moreover the stored spin-coherence
has a rapidly oscillating spatial phase which would be washed out
by atomic motion very quickly. For these reasons co-propagating
fields are preferable.}, $\Omega_P=\Omega_P(t)$, the propagation
problem can be solved in a very instructive way in a
quasi-particle picture. As a consequence the population of the
level 2 will vary in time also for effect of $\Omega_P(t)$, i.e.
$\eta(t) \simeq 4 \frac{\gamma_1}{\Omega_P(t)}$, when $\gamma_1$
is fixed. \par Therefore we will introduce these quasi-particles,
called dark--state polaritons by the authors in
\cite{Fleischhauer-PRL-2000,Mazets96}, and we will show how it is
possible to transfer the quantum state from the light to the
matter.

\subsection{Definition of dark- and bright-state polaritons}

Let us consider the case of a time-dependent, spatially
homogeneous and real control field
$\Omega_P=\Omega_P(t)=\Omega_P(t)^*$.

\begin{figure} [!ht]
\begin{center}
\includegraphics[width=.7\textwidth ]{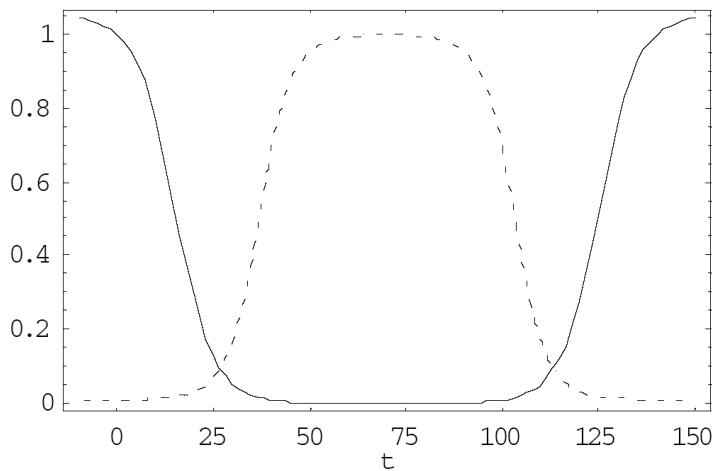}
\caption{Control field (continuous line) $\Omega_P(t)/\Omega_P(0)$
as function of time according to $\Omega_P(t)=0.8(1-0.5
\tanh[0.1(t-15)] + 0.5\tanh[0.1(t-125)])$ (in units of
$\gamma_3$). The mixed angle $\frac{\theta(t)}{\pi/2}$ (dashed
line) is also reported. Parameters are $g^2N=1/100$. }
\label{fig57}
\end{center}
\end{figure}

The physically relevant variables are the electric field $\hat
{\cal E}$ and the atomic spin coherence $\ws_{12}$, so defining a
rotation in the space of these variables, we can introduce two new
quantum fields, $\hat \Psi(x,t)$ and $\hat \Phi(x,t)$
\bgar \hat\Psi(x,t) &=& \cos\theta(t)\, \hat {\cal E}(x,t) -
\sin\theta(t)\, \sqrt{N}\,
\ws_{12}(x,t)\, {\rm e}^{i\Delta k x}\\
\hat\Phi(x,t) &=& \sin\theta(t)\, \hat {\cal E}(x,t) +
\cos\theta(t)\, \sqrt{N}\, \ws_{12}(x,t)\, {\rm e}^{i\Delta k x}
\enar

where the mixing angle $\theta(t)$ is such as

\be \tan^2\theta(t) = \frac{g^2 N}{\Omega_P^2(t)}=n_{\rm g}(t) \ee
\par
Le us to point out that $\hat \Psi$ and $\hat\Phi$ are
superpositions of electromagnetic ($\hat{\cal E}$) and collective
atomic components ($\sqrt{N}\ws_{12}$), whose admixture can be
controlled through $\theta(t)$ by changing the strength of the
external driving field.
\par
Introducing a plain-wave decomposition $\hat\Psi(x,t)=\sum_k
\hat\Psi_k(t)\, {\rm e}^{ikx}$ and
$\hat\Phi(x,t)=\sum_k\hat\Phi_k(t)\, {\rm e}^{ikx}$ respectively,
one finds that the mode operators obey the commutation relations

\begin{eqnarray}
\Bigl[\hat\Psi_k, \hat\Psi_{k'}^{\dagger}\Bigr] &=&
 \delta_{k,k'}\, \Bigl[\cos^2\theta +
\sin^2\theta\frac{1}{N}\sum_j
({\hat \rho}_{11}^j-{\hat \rho}_{22}^j)\Bigr]\\
\Bigl[\hat\Phi_k, \hat\Phi_{k'}^{\dagger}\Bigr] &=&
 \delta_{k,k'}\, \Bigl[\sin^2\theta +
\cos^2\theta\frac{1}{N}\sum_j
({\hat \rho}_{11}^j-{\hat \rho}_{22}^j)\Bigr]\\
\Bigl[\hat\Psi_k, \hat\Phi_{k'}^{\dagger}\Bigr] &=&
 \delta_{k,k'}\sin\theta\cos\theta\, \Bigl[1-\frac{1}{N}\sum_j
({\hat \rho}_{11}^j-{\hat \rho}_{22}^j)\Bigr]
\end{eqnarray}

In our case, ${\hat \rho}_{11}^j = 1-\eta, {\hat \rho}_{22}^j =
\eta$, the new fields possess the following commutation relations

\bgar \Bigl[\hat\Psi_k, \hat\Psi_{k'}^{\dagger}\Bigr]
&=&\delta_{k,k'}(1-2
\eta \sin^2\theta) \nonumber \\
\Bigl[\hat\Phi_k, \hat\Phi_{k'}^{\dagger}\Bigr] &=& \delta_{k,k'}(1-2 \eta \cos^2\theta)\label{commr} \\
\Bigl[\hat\Psi_k, \hat\Phi_{k'}^{\dagger}\Bigr] &=& 2 \eta
\delta_{k,k'}\sin\theta\cos\theta\ \nonumber \enar
\par
Now we analyze the following two case: 1) $\eta=0$ ($\gamma_1=0$),
2) $\eta>0$ ($\gamma_1>0$).
\newline
\newline
\newline
\begin{itemize}

\item[1)]$\eta=0$ ($\gamma_1=0$)

In this case \cite{Fleischhauer-PRA-2002} the quasi-particles
satisfy bosonic commutation relations and therefore one
immediately verifies that all number states created by
$\hat\Psi_k^\dagger$,

\begin{eqnarray}
|n_k\rangle = \frac{1}{\sqrt{n!}} \Bigl(\hat\Psi_k^\dagger\Bigr)^n
|0\rangle |1_1 ...1_N\rangle \label{dark}
\end{eqnarray}

where $|0\rangle$ denotes the field vacuum, are dark--states
\cite{ari,Lukin-PRL-2000}. The states $|n_k\rangle$ do not contain
the excited atomic state and are thus immune to spontaneous
emission. Moreover, they are eigenstates of the interaction
Hamiltonian with eigenvalue zero,

\be {\hat V}\, |n_k\rangle  = 0 \ee

For these reasons the authors in \cite{Fleischhauer-PRA-2002} call
these quasi-particles $\hat \Psi$ ``dark--state polaritons''.
Similarly one finds that the elementary excitations of $\hat \Phi$
correspond to the bright-states in three--level systems.
Consequently these quasi-particles are called ``bright-state
polaritons''.

Besides one can transform the equations of motion for the electric
field and the atomic variables into the new field variables.

First of all we write down $\hat {\cal E}$ and $\ws_{12}$ in terms
of the fields $\hat\Phi$ and $\hat\Psi$:

\bgar
 \hat{\cal E} &=& \cos\theta \hat\Psi + \sin\theta \hat\Phi \nonumber \\
 \ws_{12}&=&\frac{1}{\sqrt{N}}{\rm e}^{-i\Delta k x}(\cos\theta \hat\Phi -\sin\theta \hat\Psi)
\label{ero}\enar

By substituting these expressions in Eq. (\ref{field_ad}) with
$\eta=0$ ($\gamma_1=0$), one finds

\be \biggl[\frac{\partial}{\partial t} +c\cos^2\theta
\frac{\partial}{\partial x}\biggr]\, \hat\Psi =-\dot\theta\,
\hat\Phi -c \sin\theta \cos\theta\, \frac{\partial}{\partial
x}\hat\Phi\label{Psi-full} \ee

 \bgar \hat\Phi &=&\frac{\sin\theta}{g^2 N}
\biggl(\frac{\partial}{\partial t}+\gamma_3\Bigr)\Bigl(\tan\theta
\frac{\partial}{\partial t}\biggr)
\Bigl(\sin\theta\,\hat\Psi-\cos\theta\, \hat\Phi\Bigr)-i
\frac{\sin\theta}{g} F_{13} \nonumber \enar where one has to keep
in mind that the mixing angle $\theta$ is a function of time.

\item[2)]$\eta>0$ ($\gamma_1>0$)

In this case the quasi--particles don't satisfy bosonic
commutation relations and the previous discussion about dark
states is not possible. However, in chapter \ref{para} from the
statistical point of view we will examine this configuration
because these generalized commutation relations play an important
role in parastatistics.
 \par
By using Eqs. (\ref{ero}) and by substituting these expressions in
Eq. (\ref{field_ad}), we have

\end{itemize}

\bgar \biggl[\frac{\partial}{\partial t} &+&c\cos^2\theta
\frac{\partial}{\partial x}-\frac{g^2 N}{\gamma_3}\eta
\cos^2\theta +\gamma_1 \sin^2\theta \biggr]\, \hat\Psi =
\\ &=&-\dot\theta\, \hat\Phi -c \sin\theta \cos\theta\,
\frac{\partial}{\partial x}\hat\Phi +(\gamma_1+\frac{g^2
N}{\gamma_3} \eta) \sin\theta \cos\theta \, \hat\Phi \nonumber
\enar

\bgar \hat\Phi &=& \frac{1}{\cos^2 \theta
+\frac{\Omega_P}{B}\sin^2 \theta }\Big\{ (1-\frac{\Omega_P}{B})
\sin\theta \cos\theta \hat \Psi
+\\
&-&\frac{\gamma_1}{B\Omega_P^*}\cos\theta \frac{\partial}{\partial
t}(\cos\theta \hat \Phi - \sin\theta \hat \Psi)-\sin\theta
\frac{\Omega_P}{B} \frac{\eta}{\gamma_3}\frac{\partial}{\partial
t}(\cos\theta \hat
\Psi + \sin\theta \hat \Phi) +\nonumber\\
&+&\frac{\sin\theta}{g^2 N} \biggl(\frac{\partial}{\partial
t}+(\gamma_1+\gamma_3)\Bigr)\Bigl(\tan\theta
\frac{\partial}{\partial t}\biggr)
\Bigl(\sin\theta\,\hat\Psi-\cos\theta\, \hat\Phi\Bigr)-i
\frac{\sin\theta}{g} F_{13} \nonumber\Big\} \enar

\subsection{Adiabatic limit}\index{adiabatic limit}

Introducing the adiabaticity parameter $\varepsilon\equiv
\Bigl(g\sqrt{N} T\Bigr)^{-1}$ with $T$ being a characteristic
time, one can expand the equations of motion in powers of
$\varepsilon$. In lowest order, i.e. in the adiabatic limit, one
finds

\bgar \hat\Phi &\approx & 0. \enar

Consequently

\bgar
\hat{\cal E}(x,t) &=& \enspace\cos\theta(t)\, \hat\Psi(x,t),\label{E-Psi}\\
\sqrt{N} \ws_{12} &=& -\sin\theta(t)\, \hat\Psi(x,t)\, {\rm
e}^{-i\Delta k x} \label{rho-Psi} \enar

Now let us consider the two cases, $\eta=0$ and $\eta>0$.

\begin{figure} [p]
\begin{center}
\includegraphics[width=1.\textwidth ]{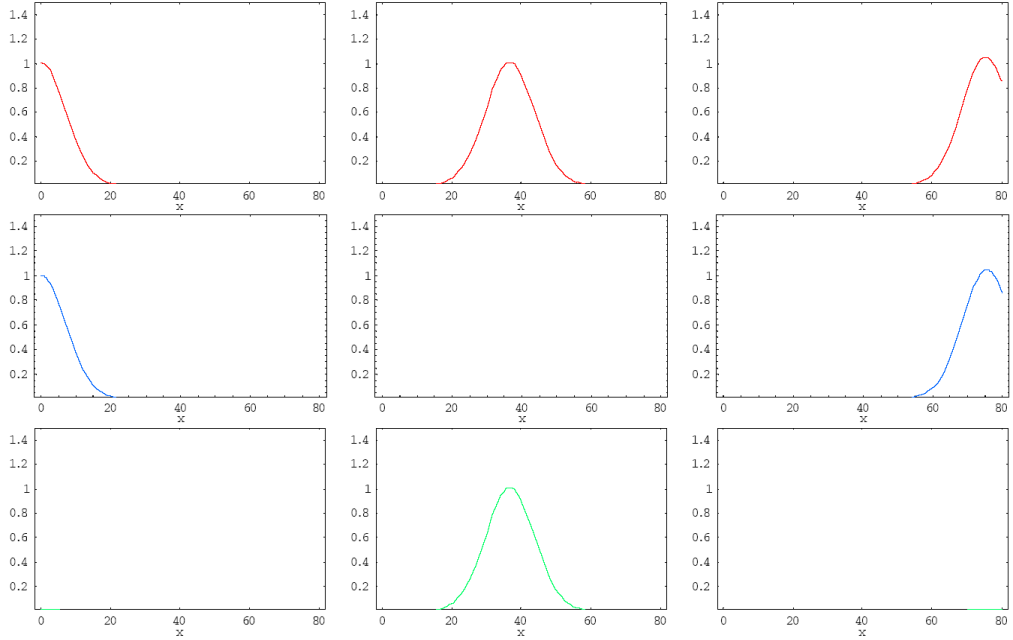}
 \caption{Propagation of a dark--state polariton
with envelope $\exp\{-(x/10)^2\}$ for three time slices
corresponding to $\theta(t)=0,\ \pi/2,\ 0$, in the case $\eta=0$
($\gamma_1=0$), i.e. ideal EIT regime. The mixing angle is rotated
from $0$ to $\pi/2$ and back according to $\cot\theta(t)=0.8(1-0.5
\tanh[0.1(t-15)]+0.5\tanh[0.1(t-125)])$. The coherent amplitude of
the polariton $\Psi=<\hat \Psi>$ (red line), the electric field
$E=<\hat E>$ (blu line) and the matter components
$|\rho_{12}|=|<\hat \rho_{12}>|$ (green line) are plotted. Through
the dynamic rotation of the mixing angle, it is possible to obtain
an adiabatic passage from a pure photon-like to a pure spin-wave
polariton, i.e. decelerating the initial photon wavepacket to a
full stop. Therefore the quantum state of the optical field is
completely transferred to the atoms. Afterwards the photon
wave-packet is regenerated by reversing the rotation. Hence the
extension of EIT to a dynamic group-velocity reduction via
adiabatic following in polaritons can be used as the basis of an
effective quantum memory. Indeed the information stored in the
spin excitations is transferred back to the radiation field and
the original signal pulse shape is perfectly preserved. Parameters
are $g^2N=1/10000$ and axes are in arbitrary units with $c=1$.}
\label{fig58}
\end{center}
\end{figure}

\begin{figure} [p]
\begin{center}
\includegraphics[width=1.\textwidth ]{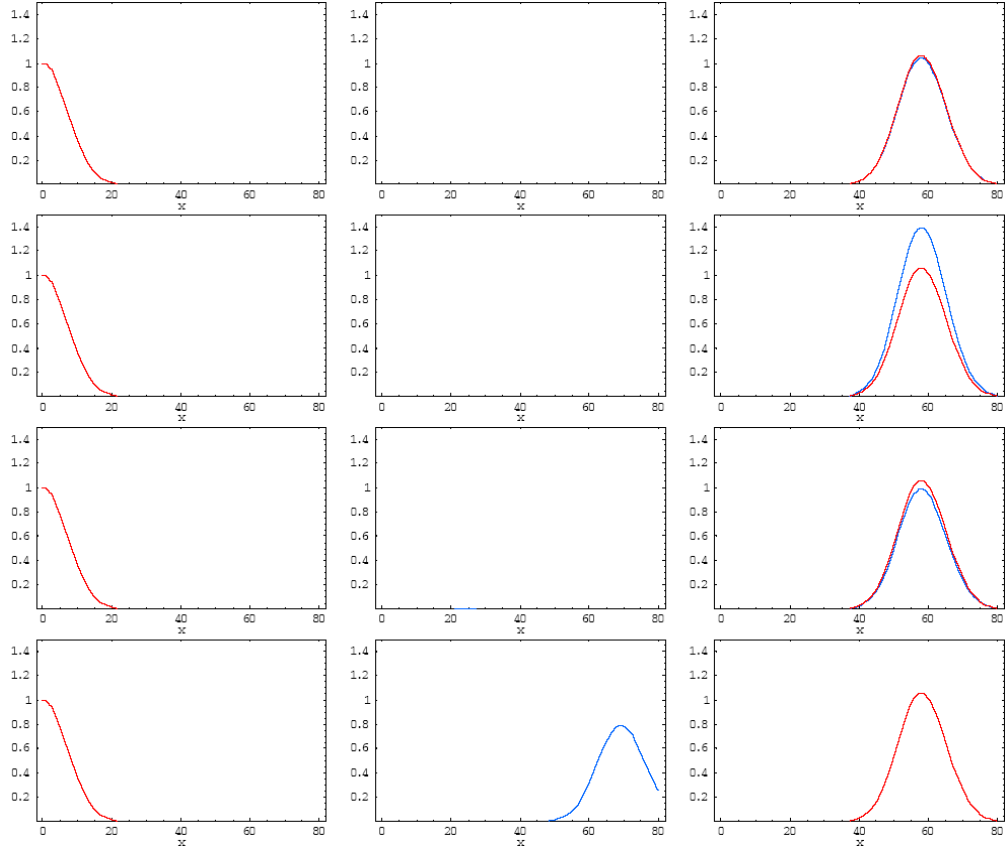}
 \caption{Propagation of the electric field $E=<\hat E>$ (blu line)
with envelope $\exp\{-(x/10)^2\}$ for three time slices
corresponding to $\theta(t)=0,\ \pi/2,\ 0$, in the case $\eta>0$
($\gamma_1>0$), i.e. amplification without inversion regime; it is
also reported the case $\eta=0$ (red line). The mixing angle is
rotated from $0$ to $\pi/2$ and back according to
$\cot\theta(t)=0.8(1-0.5 \tanh[0.1(t-15)]+ 0.5\tanh[0.1(t-125)])$.
By varying the dephasing ($\gamma_1$) between the two
lower-states, there is a balance between two $\gamma_1$-induced
effects: the destruction of the EIT window at higher dephasings
and the amplification without inversion produced by the population
in the level 2. This situation is represented for different values
of $\gamma_1$ in the range $[0,1]$, in units of $\gamma_3$. In
agreement with Fig. \ref{fig47}, for low values of $\gamma_1$
there are unavoidable losses in the process of storing and
releasing; indeed the peak of pulse (blu line) is lower than one
for $\eta=0$ (red line). For higher dephasing there is gain, that
compensates the losses and give also the amplification of
dark--state polariton. When increasing more and more $\gamma_1$,
the dephasing destroy the EIT window, there is absorption and it
is no more possible to stop the pulse. Parameters are $g^2N=1/100$
and axes are in arbitrary units with $c=1$. } \label{fig59}
\end{center}
\end{figure}

\begin{itemize}

\item[-] $\eta=0$ ($\gamma_1=0$)

In this case Eq. (\ref{Psi-full}), in the adiabatic limit, reduces
to the following very simple equation of motion of $\Psi$:

\begin{eqnarray}
\left[\frac{\partial}{\partial t}+c\cos^2\theta(t)
\frac{\partial}{\partial x}\right]\hat\Psi(x,t)=0\label{Psi-eq}
\end{eqnarray}

Eq.(\ref{Psi-eq}) describes a shape- and quantum-state preserving
propagation with instantaneous velocity
$v=v_g(t)=c\cos^2\theta(t)$:

\begin{equation}
\hat\Psi(x,t)=\hat \Psi\biggl(x- c\int^t_0\!\!\!{\rm d}\tau
\cos^2\theta(\tau),0\biggr) \label{sol}
\end{equation}

For $\theta\to 0$, i.e. for a strong external drive field
$\Omega^2\gg g^2 N$, the polariton has purely photonic character
$\hat\Psi =\hat{\cal E}$ and the propagation velocity is that of
the vacuum speed of light. In the opposite limit of a weak drive
field $\Omega^2\ll g^2 N$ such that $\theta\to \pi/2$, the
polariton becomes spin-wave like $\hat\Psi =-\sqrt{N}\ws_{12} {\rm
e}^{i\Delta k x}$ and its propagation velocity approaches zero.
\par This is the essence of the transfer technique of quantum states
from photon wave-packets propagating at the speed of light to
stationary atomic excitations (stationary spin waves).
Adiabatically rotating the mixing angle from $\theta=0$ to
$\theta=\pi/2$ decelerates the polariton to a full stop, changing
its character from purely electromagnetic to purely atomic. Due to
the linearity of Eq. (\ref{Psi-eq}) and the conservation of the
spatial shape, the quantum state of the polariton is not changed
during this process.
\par
Likewise the polariton can be re-accelerated to the vacuum speed
of light; in this process the stored quantum state is transferred
back to the field. This is illustrated in Fig. \ref{fig58}, where
the coherent amplitude of the pulse field is showed during the
storing and the following releasing.

\newpage
\item[-] $\eta>0$ ($\gamma_1>0$)

When there is population ($\eta$) in the level 2, $\hat\Psi$ obeys
the following equation of motion

\bgar \biggl[\frac{\partial}{\partial t} &+&c\cos^2\theta
\frac{\partial}{\partial x}-\frac{g^2 N}{\gamma_3}\eta
\cos^2\theta +\sin^2\theta \gamma_1 \biggr]\, \hat\Psi =0 \enar

By varying the dephasing ($\gamma_1$) between the two
lower-states, there is a balance between two $\gamma_1$-induced
effects: the destruction of the EIT window at higher dephasings
and the amplification without inversion produced by the population
in the level 2. The situation is represented in Fig. \ref{fig59}
for different values of $\gamma_1$. In one case in Fig.
\ref{fig59}, we have an amplification of dark--state polariton,
that at least compensates the unavoidable losses in the process of
storing and releasing. \par One could speculate that in this way
it could be possible to obtain also a quantum cloning
(continuous-variable) of a coherent state into the quantum memory
described above. However this needs to be verified and it will not
be treated in this thesis.

\end{itemize}

\newpage
\section{Decoherence of quantum state transfer}\index{decoherence}

In this section we report some results published in
\cite{Mewes-PRA-2002} regarding the decoherence effects in the
transfer process of a quantum memory. It must be noted that an
extension of these results to the case of gain without inversion
would be necessary to discuss the possible application to quantum
cloning. As already explained this would go beyond the scope of
this thesis.

\subsection{Random spin flips and dephasing}

On the level on individual atoms the storage occurs within the
two-state system consisting of $| 1\rangle$ and $|2\rangle$. If we
assume that all other atomic states including $|3\rangle$ are
energetically much higher, we may safely neglect decoherence
processes involving the excitation of those states. Then
decoherence caused by individual and independent reservoir
interactions can be described by the action of the two-level Pauli
operators
\begin{eqnarray}
 X= \rho_{12}+\rho_{21},\enspace  Z = \left[\rho_{12},
\rho_{21}\right],\enspace  Y= i \rho_{12}-i\rho_{21}
\end{eqnarray}
where $X$ describes a symmetric spin flip of the atom, $Z$ a phase
flip, and $Y$ a combination of both. Any single-atom error can be
expressed in terms of these and we will restrict the discussion
here to the action of $X$ (symmetric spin flip), $X+i Y$
(asymmetric spin flip) and $Z$ (phase flip).

\subsubsection{Spin flip from $|1\rangle \to |2\rangle$}

Consider a quantum memory initially in an ideal storage state,
i.e. $\eta=0$. Suppose that an atom then undergoes a spin flip to
the internal state $|2\rangle$ if it is initially in state
$|1\rangle$. Such a spin flip process could mimic a stored photon.
\par
The authors in \cite{Mewes-PRA-2002} have calculated the fidelity
of the quantum memory after a single spin flip error, which for
the case of an initial pure state $\rho_0=|\psi_0\rangle\langle
\psi_0|$ is defined as \be f\Bigl(|\psi_0\rangle\Bigr)
=\langle\psi_0|\rho_1|\psi_0\rangle = \textrm{Tr}\{\rho_1\rho_0\}
\ee where $\rho_1$ represent the imperfect final state.
\par
They find for a stored Fock-state $|n\rangle$ with $n\ll N$ \be
f_{b\to c}\Bigl(|n\rangle\Bigr) = \frac{1-
\frac{1}{N}}{1-\frac{n}{N}} = 1 - \frac{n+1}{N} +{\cal
O}\left(\frac{1}{N^2}\right) \ee while for a coherent state
$|\alpha\rangle$ they obtain \be f_{b\to
c}\Bigl(|\alpha\rangle\Bigr) = \frac{1-\frac{1}{N} +
\frac{|\alpha|^2}{N}} {1+\frac{|\alpha|^2}{N}} =
 1-\frac{1}{N} +{\cal O}\left(\frac{1}{N^2}\right)
\ee
This reflects the general property of non-classical states to
be more sensitive to decoherence than classical ones.

\subsubsection{Symmetric spin flip}

If instead of the asymmetric spin flip $\rho_{21} = X +i Y$ a
symmetric flip happens, the authors in \cite{Mewes-PRA-2002} find
that the fidelity of the memory reads for a stored Fock and
coherent state
\bgar f_{b\leftrightarrow c}(|n\rangle)  &=& 1 -
\frac{2n+1}{N}
+{\cal O}\left(\frac{1}{N^2}\right)\\
f_{b\leftrightarrow c}(|\alpha\rangle)  &=& 1 - \frac{1}{N} +{\cal
O}\left(\frac{1}{N^2}\right) \enar

\subsubsection{Phase flip}\label{phase-flip}

If after the preparation of an ideal storage state an atom
undergoes a phase flip the fidelity is \cite{Mewes-PRA-2002} \be
f_{\rm deph}(|\psi_0\rangle) = 1-\frac{2 \langle n\rangle}{N}
+{\cal O} \left(\frac{1}{N^2}\right) \ee where $\langle
n\rangle=\langle\Psi^\dagger\Psi\rangle$ is the average number of
dark--state polaritons in the initial state. One recognizes that a
phase flip of a single atom leads to a fidelity reduction which is
of the order of $1/N$. The term $1/N$ again compensates for the
fact that in an $N$-atom ensemble the likelihood that one
arbitrary atom undergoes a phase flip is $N$ times the probability
of a phase flip for a single atom. It is interesting to note that
the fidelity only depends on the average dark--state polariton
number, i.e. dephasing affects in lowest order of $1/N$ classical
and nonclassical states in a similar way.

\newpage
\subsection{One-atom losses}

Another important source of errors in a collective quantum memory
is the loss of an atom from the ensemble. \par The fidelity of the
quantum memory for a Fock state $|n\rangle$ after loss of a single
atom is given by \cite{Mewes-PRA-2002}
 \be f_{\rm
loss}(|n\rangle) = 1-\frac{n}{N} \ee The decrease of the fidelity
again scales only as $1/N$. This result could of course have been
expected as the $n$ excitations are equally distributed over all
atoms. Thus removing one reduces the stored information only by
the amount $n/N$.
\par
For the case of a general state the fidelity after the loss of an
atom reads \cite{Mewes-PRA-2002} \be f_{\rm loss}(|\psi_0\rangle)
= 1-\frac{1}{N} \Bigl(\langle\Psi^\dagger\Psi\rangle
-\langle\Psi^\dagger\rangle \langle\Psi\rangle\Bigr) +{\cal
O}\left(\frac{1}{N^2}\right) \label{loss-2} \ee If the initial
storage state corresponds e.g. to a coherent state, the second and
third term in (\ref{loss-2}) compensate each other and the
fidelity differs from unity only in order $1/N^2$. Here again the
robustness of classical states becomes apparent.

\subsection{Atomic motion}\label{motion}

Until now it has been assumed that the atoms used in the quantum
memory are at a fixed position during the entire storage time.
Since the coupling of the atoms to the quantum as well as control
fields contains however a spatial phase, see Eq. (\ref{ham}),
atomic motion results in an effective dephasing and will lead to a
reduction of the fidelity. Recently Sun et al. have argued that
inhomogeneities of the atom-light interaction strength or in the
control field together with atomic motion lead to an increase of
the characteristic decoherence rate by a factor $\sqrt{N}$
\cite{sunquantph0203072v1}. In this context the authors in
\cite{Mewes-PRA-2002} analyze the effect of atomic motion
\par
To reduce the effect of motion in an atomic vapor one could either
reduce the temperature or use a buffer gas of sufficient density.
In the latter case, which has been used in room temperature
gas-cell experiments
\cite{van-der-Wal-Science-2003,Phillips-PRL-2001},
 the free motion is replaced by a diffusion.
Therefore now we will restrict the discussion to this important
case. Recall that the phase is given by \be \Delta\phi_j(t)\equiv
\Delta \vec k \cdot\vec r_j(t) \ee where $\vec r_j(t)$ denoting
the position of the $j$th atom at time $t$ and $\Delta \vec k =
\vec k_1 -k_0 \vec e_x$ is the wavevector difference between
control field and quantized mode. The authors in
\cite{Mewes-PRA-2002} assume that this phase follows a Wiener
diffusion process \cite{Gardiner}: \bgar
\frac{{\rm d}}{{\rm d}t} \Delta\phi_j(t) &=& \mu_j(t)\\
\overline{\, \mu_j(t)\, } &=& 0,\\
\overline{\, \mu_j(t)\mu_k(t^\prime)\, } &=&
D\delta_{jk}\delta(t-t^\prime) \enar
with $D$ being a
characteristic diffusion rate.
\par
In this approach they find the following fidelity of the quantum
memory for atomic motion \bgar && f_{\rm
motion}\Bigl(|1\rangle\Bigr) = \frac{1}{N}\Bigl(1+(N-1) {\rm
e}^{-D t}\Bigr)\sim {\rm e}^{-D t}\nonumber \enar A generalization
to an arbitrary Fock state $|D,n\rangle$ leads to a fidelity decay
proportional to $\exp\{-n D t\}$. One recognizes that the atomic
motion causes a decay of the fidelity with a rate given only by
the single-atom diffusion rate $D$. In contrast to the results of
Sun, Yi and You \cite{sunquantph0203072v1} they find that there is
no enhancement of the decay with increasing number of atoms.
\par
All these results play a key role to determine the storage time of
these quantum memories and the quantum state transfer fidelity,
necessary to decide if that device is or not good for the desired
task for which the quantum state is kept.

\chapter{Parastatistics in gain medium}\label{para}

\section{Generalized physical statistics}\index{parastatistics}

The notion of statistics is related to the symmetry properties of
the wave function of $N$ identical particles, $\psi(1,2,...,N)$,
the bosons (fermions) corresponding to symmetric (antisymmetric)
wave functions under the exchange of particles; besides in the
second quantization scheme the annihilation and creation operators
for bosons (fermions) obey the canonical (anti-)commutation
relations.
\par
Wigner was the first to remark that the canonical quantization was
not the most general quantization scheme consistent with the
Heisenberg equations of motions \cite{wigner}. Parastatistics was
so introduced by Green \cite{green} as a general quantization
method of quantum field theory different from the canonical Bose
and Fermi quantization, by extending the Bose and Fermi statistics
\cite{ohnuki,leinaas,wilczek1,poly}. Nevertheless for the long
period of time the interest to it was rather academic even if it
was closely related with the discovery of color in the context of
the theory of strong interactions. \par Recently it has found
applications in the physics of fractional Hall
effect\index{fractional Hall effect} and it probably is relevant
to high-Tc superconductivity\index{high-Tc superconductivity}
\cite{wilczek}, so it draws more and more attentions from both
theoretical and experimental physicists. Indeed the experiments on
quantum Hall effect confirm the existence of fractionally charged
excitations.
\par
Since the pioneering work of Gentile\index{Gentile} and
Green\index{Green} \cite{green,gentile}, there have been many
extensions beyond the standard statistics, among the others we may
list the following: parastatistics, fractional
statistics\index{fractional statistics}, quon
statistics\index{quon statistics} \cite{chi}, anyon
statistics\index{anyon statistics} \cite{leinaas,wilczek1,wu} and
quantum groups statistics\index{quantum groups statistics}. Note
that in the literature there are two principal methods of
introducing an intermediate statistical behavior. The first one is
to deform the quantum algebra of the commutation-anticommutation
relations thus deforming the exchange factor between permuted
particles. The second method is based on modifying the number of
ways of assigning particles to a collection of states and thus the
statistical weight of the many-body system. Here we will refer to
the first approach.
\par
In particular in the parastatistics framework the standard bosonic
or fermionic fields, which would create identical particles, are
replaced by composite fields whose components commute with
themselves and anticommute with each other for
parabosons\index{parabosons}, or vice versa for
parafermions\index{parafermions}. The number of components of the
fields, $p$, defines the ``order'' of
parastatistics\index{parastatistics' order}. In general, one can
put at most $p$ parafermions in a totally symmetric wavefunction,
and at most $p$ parabosons in a totally antisymmetric one.
Parastatistics in this approach has been well-studied
\cite{steinmann,stapp,ohnuki1,haag} and the simplest non-trivial
representations arise for $p = 1$ and coincide with the usual Bose
(Fermi) Fock representations.

\section{Parapolariton in EIT}

Let us analyze in detail the effect of the population ($\eta$) in
the level $\ket{2}$, in the three--level scheme studied in the
last chapter, on the commutation relations of the quasi-particles,
i.e. on their statistics.
\par
Recall Eq. (\ref{commr}) in which the dark--state polariton fields
(in the same k-mode) possess the following commutation relations
\bgar \Bigl[\hat\Psi, \hat\Psi^{\dagger}\Bigr] =1-2 \eta
\sin^2\theta \label{commre} \enar where $\eta$ is the population
in level 2 and $\theta$ is the mixing angle of the polariton.
\par
Eq. (\ref{commre}) describe a kind of deformed oscillator or
paraboson, because it presents a deviation from the canonical
bosonic commutation relations; we call them
\textbf{parapolaritons}.\index{parapolaritons}
\par
Many kinds of deformed oscillators\index{generalized deformed
oscillators} (parabosons) have been introduced in the literature
(see \cite{BD100} for a list). All of them can be accommodated
within the common mathematical framework of the {\it generalized
deformed oscillator} \cite{Das789,DY4157}, which is defined as the
algebra generated by the operators $\{\hat 1, \hat a, \hat
a^\dagger, \hat N\}$ and the {\sl structure function} $\Phi(x)$,
satisfying the relations \bgar [\hat a, \hat N]=\hat a \qquad
[\hat a^\dagger, \hat N]=-\hat a^\dagger \enar \bgar \hat
a^\dagger \hat a =\Phi(\hat N) =[\hat N] \qquad \hat a\hat
a^\dagger = \Phi(\hat N+ \hat 1) =[\hat N+ \hat 1] \label{phin}
\enar where $\Phi(x)$ is a positive analytic function with
$\Phi(0)=0$ and $\hat N$ is the number operator. From Eq.
(\ref{phin}) we conclude that \bgar \hat N=\Phi^{-1} (\hat
a^\dagger \hat a) \enar  and that the following commutation and
anticommutation relations are obviously satisfied: \bgar [\hat a,
\hat a^\dagger]=[\hat N+\hat 1]-[\hat N] \qquad \{\hat a,\hat
a^\dagger\}=[\hat N+\hat 1]+[\hat N] \enar The {\sl structure
function} $\Phi(x)$ is characteristic to the deformation
scheme\index{deformation scheme}. In Table \ref{tablepara} the
structure functions\index{structure function} corresponding to
different deformed oscillators are given.
\par
It can be proved that the generalized deformed algebras possess a
Fock space of eigenvectors $\ket{0},\ket{1},\ldots,\ket{n},\ldots$
of the number operator $\hat N$ \bgar \hat N\ket{n}=n\ket{n} \quad
\sca{n}{m}=\delta_{nm}\enar if the {\it vacuum state} $\ket{0}$
satisfies the following relation:\bgar \hat a\ket{0}=0 \enar
 These eigenvectors are generated by the formula:
\bgar \ket{n} {1 \over \sqrt{ [n]!}} {\left( \hat a^\dagger
\right)}^n \ket{0} \enar where
 \bgar [n]!=\prod_{k=1}^n [k]= \prod_{k=1}^n \Phi(k) \enar
The generators, $\hat{a}^\dagger$ and $\hat{a}$, are the creation
and annihilation operators of this deformed oscillator algebra:
\bgar \hat a \ket{n} = \sqrt{[n]}  \ket{n-1} \qquad
 \hat a^\dagger \ket{n} = \sqrt{[n+1]}  \ket{n+1} \enar
These eigenvectors are also eigenvectors of the energy operator
\bgar \hat H={\hbar \omega \over 2} (\hat a\hat a^\dagger +\hat
a^\dagger \hat a) \enar corresponding to the eigenvalues \bgar
E(n)= {\hbar\omega \over 2} (\Phi(n)+\Phi(n+1)) = {\hbar \omega
\over 2} ([n]+[n+1])\enar For $\Phi(n)=n$ one obtains the results
for the ordinary harmonic oscillator. Besides in this generalized
framework we can define two new operators, $\hat b$ and $\hat
b^\dagger$ \bgar \hat b=\hat a \sqrt{\frac{n}{\phi(n)}} \ \ \ \ \
\ \ \ \ \ \ \ \hat b^\dagger=\hat a^\dagger
\sqrt{\frac{n}{\phi(n)}} \enar that obey the canonical bosonic
commutation relations, as one can easily show; indeed $[\hat b,
\hat b^{\dagger}]=1$. According to Eq. (\ref{commre}),
$\Phi(n)=n(1-2 \eta \sin^2{\theta})$ with $0\leq \eta<1/2$; this
restriction in $\eta$ is justified by the requirement that
$\phi(x)$ is a positive analytic function. By this definition of
$\phi$, we have \bgar \Bigl[\hat\Psi, \hat\Psi^{\dagger}\Bigr] =
(1-2 \eta \sin^2\theta) \enar and the eigenvalues are: \bgar E(n)=
{\hbar\omega \over 2} (\Phi(n)+\Phi(n+1)) = \hbar
\omega\Big(n+\frac{1}{2}\Big)(1-2 \eta \sin^2\theta) \enar

\begin{figure} [bh]
\begin{center}
\includegraphics[width=.915\textwidth ]{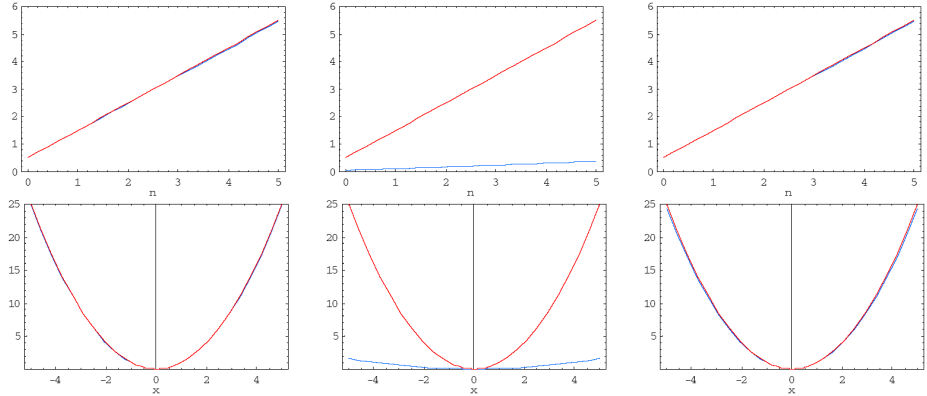}
\caption{Top: the energy of the deformed boson oscillator (blu
line), $E(n)$, as function of $n$ for three time slices
corresponding to $\theta(t) \simeq 0,\ \pi/2,\ 0$, in units of
$\hbar \omega$. During the stopping of light the ideal bosonic
spectrum (red line) is compressed and when releasing the photon
the bosonic energy width is recovered. There is a mapping from
bosons to fermions described by the compression of bosonic energy
spectrum to a fermionic 2-level one. The narrowing of energy
spacing reflects the Pauli principle and an effective repulsive
interaction that describes the mapping
bosons--fermions\index{bosons}\index{fermions}. Bottom: the
relative effective potential, $V(x)$, is reported according to the
results above.} \label{fig60}
\end{center}
\end{figure}

\newpage
When a dark--state polariton propagates through a three--level
atomic system, there is a mapping from bosons to fermions. Indeed,
at the beginning, when the pump field $\Omega_P$ is turned on, the
polariton has only a photonic component and it is a boson.
Afterwards, when turning off the pump frequency, the quantum state
encoded in photon is converted to spin eccitation of atoms. In
other words, during the stopping of light, the dark--state
polariton has only spin-atomic component and so it is a fermion.
In the intermediate time the polariton obey to a particular
parastatistics, as deformed harmonic oscillator.
\par
In terms of energy, during the stopping of light the ideal bosonic
spectrum is compressed and it recovers the original energy width
during the release of photon. There is a mapping from bosons to
fermions described by the compression of bosonic energy spectrum
to a fermionic 2-level one. The narrowing of energy spacing
reflects the Pauli principle\index{Pauli principle} and an
effective repulsive interaction that describes the mapping
bosons--fermions. In other words the commutation relations in Eq.
(\ref{commr}) resemble bosonic commutation relations but in
addition include corrections due to the presence of the Pauli
principle\footnote{This fact has just been used for boson mapping
techniques (see the recent reviews by Klein and Marshalek
\cite{KM375} and Hecht \cite{Hecht87}  and references therein), by
which the description of systems of fermions in terms of bosons is
achieved.}.

\begin{table}[p]
\begin{center}
\caption{Structure functions of special deformation schemes}
\bigskip
\begin{tabular}{|c c p{2.0 in}|}
\hline
\ & $\Phi(x)$ & Reference \\
\hline\hline \romannumeral 1 & $x$ &
harmonic  oscillator, bosonic algebra\\[0.05in]
\romannumeral 2&  ${ {q^x- q^{-x} }  \over {q- q^{-1} } } $ &
$q$-deformed harmonic oscillator \cite{Bie873,Mac4581}
\\[0.05in]
\romannumeral 3& ${ {q^x- 1 } \over {q- 1 } } $ & Arik--Coon,
Kuryshkin, or $Q$-deformed oscillator \cite{AC524,Kur111} \\[0.05in]
\romannumeral 4&  ${ {q^x- p^{-x} }  \over {q- p^{-1} } } $ &
2-parameter deformed oscillator \cite{BJM820,JBM775,CJ711}
\\[0.05in]
\romannumeral 5& $x(p+1-x)$ & parafermionic oscillator
 \cite{ohnuki} \\[0.05in]
\romannumeral 6& $ { \sinh (\tau x) \sinh (\tau (p+1-x) )}\over {
\sinh^2(\tau) } $ & $q$-deformed parafermionic oscillator
\cite{FV1019,OKK591} \\[0.05in]
\romannumeral 7& $x\cos^2(\pi x/2) + (x+p-1)\sin^2(\pi x /2)  $&
parabosonic oscillator \cite{ohnuki} \\[0.05in]
\romannumeral 8& $\begin{array}{c} \frac{\sinh(\tau
x)}{\sinh(\tau)}
\frac{\cosh(\tau (x+2N_0-1))}{\cosh(\tau)} \cos^2 (\pi x/2) +\\
+ \frac{\sinh(\tau (x+2N_0-1))}{\sinh(\tau)} \frac{\cosh(\tau
x)}{\cosh(\tau)} \sin^2 (\pi x/2)
\end{array}$
 & $q$-deformed parabosonic oscillator
 \cite{FV1019,OKK591} \\[0.05in]
\romannumeral 9 &
$\sin^2 {\pi x/2}$ & fermionic algebra \cite{JBSZ123} \\[0.05in]
\romannumeral 10 & $ q^{x-1} \sin^2 {\pi x/2}$ & $q$-deformed
fermionic algebra \cite{Hay129,CK72,FSS179,Gang819,Sci219,PV613}
 \\[0.05in]
\romannumeral 11& $\frac{1-(-q)^x}{1+q}$ & generalized
$q$-deformed fermionic algebra
 \cite{VPJ335} \\[0.05in]
\romannumeral 12& $x^n$ & \cite{Das789} \\[0.05in]
\romannumeral 13& ${ {sn(\tau x)} \over {sn(\tau )} }$ &
\cite{Das789}  \\[0.05in]
\hline
\end{tabular}
 \label{tablepara}
\end{center}
\end{table}

\chapter{Polarization quantum memory}\index{polarization quantum
memory}

Up to now we have analyzed a quantum memory for the number of
photons. Indeed the information that is conserved in the quantum
memory studied above is the number of photons, converted into the
number of collective spin excitations. Actually in quantum
information science many protocols need two--level quantum system
or \textbf{qubit} and, because the most of implementations use the
photons as ideal carriers of information, the quantum information
is encoded in some physical freedom degrees of a photon, i.e its
polarization. For example we can use the two polarizations,
$\sigma_{+}$ and $\sigma_{-}$, to encode two orthogonal quantum
states, i.e. to treat a photon as a qubit.

\begin{figure}[th]
\begin{center}
\includegraphics[scale=.4]{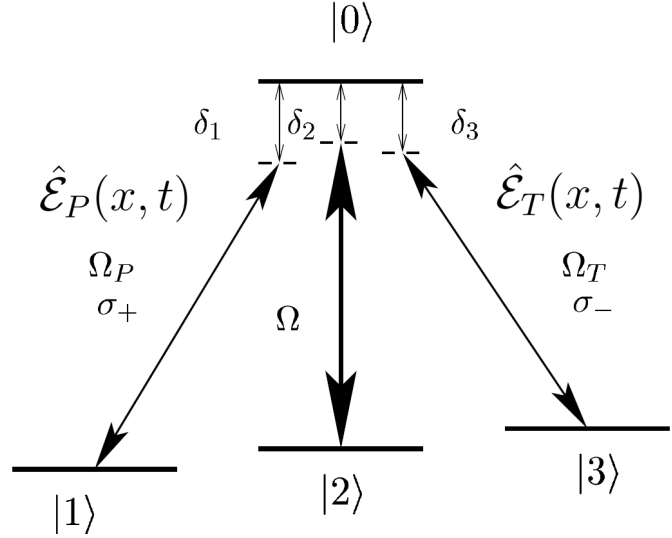}
\caption{Energy level scheme for a tripod. Detunings $\delta_j=
\omega_0-\omega_j-\omega_j^{(L)}$  denote the laser frequency
($\omega_j^{(L)}$) detunings from the respective transitions
$|j\rangle \leftrightarrow |0\rangle$. States $|1\rangle$,
$|2\rangle$ and $|3\rangle$ correspond to the states $|5S_{1/2}, F
= 1, m = \{ -1, 0, 1 \} \rangle$ of $^{87}$Rb, while state
$|0\rangle = |5P_{3/2}, F = 0 \rangle$. \label{fig61}
\cite{Rebic03}}
\end{center}
\end{figure}

For this reason now we examine a four-level atomic
\textit{tripod}\index{tripod configuration} configuration in Fig.
\ref{fig61}, already studied in \cite{Rebic03} for quantum phase
gate\index{quantum phase gate} (off--resonance). This has often
been used as an extension of a $\Lambda$ scheme, e.g. by
Paspalakis who suggested its potential for nonlinear optical
processes~\cite{Paspalakis02}. It should be mentioned that while
Paspalakis considered the case of a single weak probe, the authors
in \cite{Rebic03} adopt a setup with two weak fields namely a
$probe$ and a $trigger$ in the presence of a strong pump that
creates EIT.
\par
The advantage of this scheme is that the two probe, $\hat {\cal
E}_P(x,t)$ and $\hat {\cal E}_T(x,t)$, have different
polarization, i.e. $\sigma_{+}$ and $\sigma_{-}$. In such a way,
we so introduced a quantum memory for polarized photons. This kind
of quantum memory has been investigated only in cavity QED and
proposed recently by Lukin and Fleischhauer in a different atomic
scheme \cite{Lukin-PRL-2000}. Then the proposal of this thesis
would represent a new interesting polarization quantum memory for
light, useful in different quantum information protocols.

\section{EIT effect in a Tripod System \label{sec:dress}}

The energy level scheme of a tripod system is given in Fig.
\ref{fig61}. Probe and trigger fields have Rabi frequencies
$\Omega_P$ and $\Omega_T$ and polarizations $\sigma_+$ and
$\sigma_-$. The pump Rabi frequency is $\Omega$.

Among the four eigenstates in the system two contain no
contribution from the excited state $|0\rangle$ and hence they are
\textit{dark states} \cite{Unanyan98}. For the special case of
atomic detunings $\delta_j = \delta = 0$, these are

\begin{eqnarray}
    |e_1\rangle &=& \frac{\Omega_T |1\rangle -
\Omega_P|3\rangle}{\sqrt{\Omega_P^2+\Omega_T^2}}\\
    |e_2\rangle &=& \frac{\Omega \Omega_P |1\rangle + \Omega \Omega_T
|3\rangle - \left( \Omega_P^2 + \Omega_T^2 \right)
|2\rangle}{\sqrt{\left( \Omega_P^2+\Omega_T^2\right)\left(
\Omega_P^2+\Omega^2+\Omega_T^2 \right)}}
\end{eqnarray}

The two other states (\textit{bright states})
\begin{equation}
    |e_\pm\rangle = \frac{\Omega_P |1\rangle \pm |0\rangle + \Omega_T
|3\rangle + \Omega
|2\rangle}{\sqrt{\Omega_P^2+\Omega^2+\Omega_T^2}}
\label{eq:brightstates}
\end{equation}
contain $\ket{0}$ and have energies
$\pm\sqrt{\Omega_P^2+\Omega^2+\Omega_T^2}$. \par The atomic
evolution is governed by a set of Heisenberg-Langevin equations
(including atomic spontaneous emission and dephasing)
\begin{eqnarray}
     i\dot{\rho}_{00} &=& -i(\gamma_{11}+\gamma_{22}+\gamma_{33})
\rho_{00} + \Omega_P^*\rho_{10} - \Omega_P\rho_{01} \nonumber \\
&\ &+ \Omega^*\rho_{20} - \Omega\rho_{02} + \Omega_T^*\rho_{30} -
\Omega_T\rho_{03} \nonumber \\
     i\dot{\rho}_{11} &=& i\gamma_{11} \rho_{00} + i\gamma_{12}
\rho_{22} + i\gamma_{13} \rho_{33} + \Omega_P\rho_{01} -
\Omega_P^*\rho_{10} \nonumber \\
     i\dot{\rho}_{22} &=& i\gamma_{22} \rho_{00} - i\gamma_{12}
\rho_{22} + i\gamma_{23} \rho_{33} + \Omega\rho_{02} -
\Omega^*\rho_{20} \nonumber \\
     i\dot{\rho}_{33} &=& i\gamma_{33} \rho_{00} -
i(\gamma_{13}+\gamma_{23}) \rho_{33} + \Omega_T\rho_{03} -
\Omega_T^*\rho_{30} \nonumber \\
     i\dot{\rho}_{10} &=& -\Delta_{10} \rho_{10} + \Omega_P \rho_{00}
- \Omega_P \rho_{11} - \Omega\rho_{12} - \Omega_T\rho_{13} \\
     i\dot{\rho}_{20} &=& -\Delta_{20} \rho_{20} + \Omega \rho_{00} -
\Omega \rho_{22} - \Omega_P\rho_{21} - \Omega_T\rho_{23}
\nonumber\\
     i\dot{\rho}_{30} &=& -\Delta_{30} \rho_{30} + \Omega_T \rho_{00}
- \Omega_T \rho_{33} - \Omega_P\rho_{31} - \Omega\rho_{32}
\nonumber \\
     i\dot{\rho}_{12} &=& -\Delta_{12} \rho_{12} + \Omega_P\rho_{02} -
\Omega^*\rho_{10} \nonumber \\
     i\dot{\rho}_{13} &=& -\Delta_{13} \rho_{13} + \Omega_P\rho_{03} -
\Omega_T^*\rho_{10} \nonumber \\
     i\dot{\rho}_{23} &=& -\Delta_{23} \rho_{23} + \Omega\rho_{03} -
\Omega_T^*\rho_{20} \nonumber
   \end{eqnarray}
where the decay rates $\gamma_{ij}$ describe decay of populations
and coherences, $\Delta_{j0} = \delta_j +i\gamma_{j0}$ and
$\Delta_{ij} = \delta_j - \delta_i - i\gamma_{ij}$, with $i,j =
1,2,3$.
\par
When all detunings $\delta_j$ are vanishing (on resonance), probe
and trigger share a strong control field with Rabi frequency
$\Omega$ and exhibit all effects associated with ideal EIT
including strong group velocity reductions. In this regime the
system behaves as a perfect polarization quantum memory for two
independent pulses. Indeed if a probe (trigger) field propagates
into this medium then the polarization $\sigma^{+}$ ($\sigma^{+}$)
is mapped to atomic spin-excitation. Besides when a photon in any
polarization state enters upon this memory, also this
superposition of $\sigma^{+}$ and $\sigma^{-}$ is saved in this
quantum register because, for perfectly equal detunings
degeneracy, there is a common transparency window for both fields
and our tripod system is linear \cite{Rebic03}. Therefore, apart
from the quantum memories analyzed in the last chapters, here with
this four-level configuration any quantum state of a polarized
photon is coherently memorized.
\par
\begin{figure} [!ht]
\begin{center}
\includegraphics[width=.98\textwidth ]{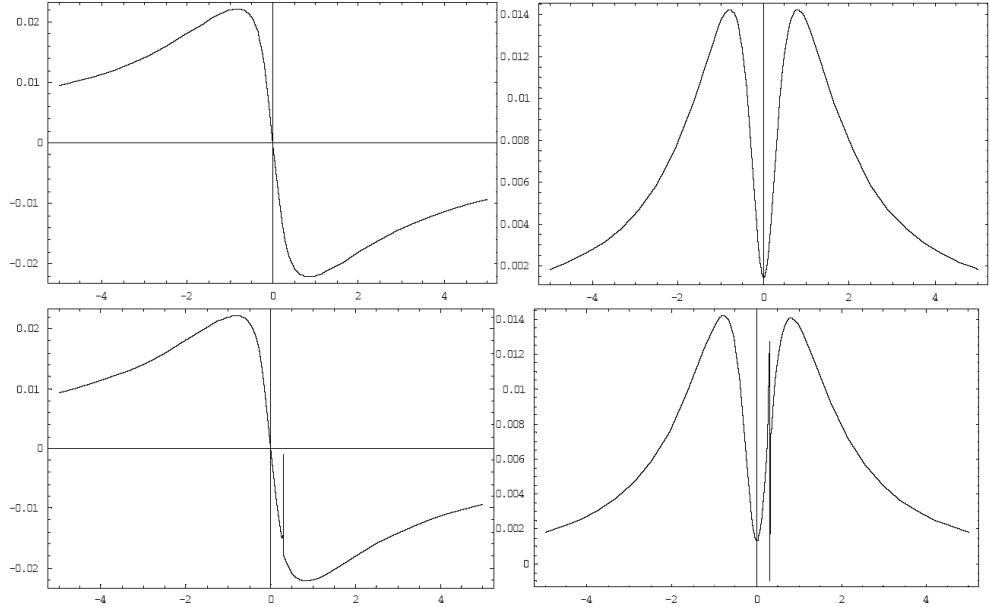}
\caption{Real (left) and imaginary (right) part of $\rho_{10}$ are
reported when the pump field is on the resonance. In the top the
trigger field is also on the resonance and then the system is
linear, there is no dephasing between the two field and an ideal
EIT regime for both of pulses. In the bottom we analyze the
situation in which there is a mismatch between the detunings and a
cross-phase modulation appears \cite{Rebic03}. Indeed the probe
detuning dependence of $\rho_{10}$ is perturbed by this
off--resonance Kerr non-linearity.}\label{fig62}
\end{center}
\end{figure}

In order to verify this ideal EIT effect in presence of two
fields, now we consider the steady state solutions to the Bloch
equations. When $|\Omega|^2 \gg |\Omega_{P,T}|^2$ and $\Omega_P
\approx \Omega_T$, the steady-state population distribution will
be symmetric with respect to the $1 \leftrightarrow 3$ exchange,
i.e. $\rho_{11} \approx \rho_{33} \approx \frac{1}{2}$, with
vanishing population in the other two levels. This population
assumption allows to decouple the equations for the populations
from those of the coherences and to obtain the steady state
solution. In the following we show the steady-state expression for
$\rho_{10}$:
\begin{subequations}
\label{eq:bigsuscgen}
\begin{eqnarray}
\frac{\left(\rho_{10}\right)_{ss}}{\Omega_P} &=& \left(
1+\frac{1}{4}
\frac{\left(\Delta_{12}\Delta_{23}/\Delta_{13}^2\right)|\Omega_P|^2|\Omega_T|^2}{\left(\Delta_{10}\Delta_{12}-
|\Omega|^2\right)\left(\Delta_{30}^*\Delta_{23}-|\Omega|^2\right)}\right)^{-1}
\label{eq:rho10}
\nonumber \\
&\times&
\left\{-\frac{1}{2}\frac{\Delta_{12}\Delta_{13}}{\Delta_{10}\Delta_{12}\Delta_{13}-\Delta_{13}|\Omega|^2-\Delta_{12}|\Omega_T|^2}\right.
\nonumber \\
&\ &\left.
-\frac{1}{2}\frac{\Delta_{12}\Delta_{13}\Delta_{23}|\Omega_T|^2}{\Delta_{30}^*\Delta_{13}\Delta_{23}-\Delta_{13}|\Omega|^2-\Delta_{23}|\Omega_P|^2}
\right\}
\end{eqnarray}
\end{subequations}
In Fig. \ref{fig62} we report the real and imaginary part of
$\rho_{10}$ when the pump field is on the resonance. When the
trigger field is also on the resonance, the system is linear,
there is no dephasing between the two field and an ideal EIT
regime for both of pulses: in these conditions we propose the
polarization quantum memory. Instead, when there is a mismatch
between the detunings, an effective Kerr non-linearity appears
\cite{Rebic03}.

\section{Scattering of dark--state polaritons}\index{DSP scattering}

Let us consider the case in which probe and trigger are quantum
fields. The pump is still considered much stronger than both of
them so as to neglect its quantum fluctuations as for a classical
field. We specifically adopt a recently developed formalism
\cite{lukin,Petrosyan,Fleischhauer-PRA-2002} and apply it to our
tripod atomic configuration. The relevant interaction Hamiltonian
is
\begin{eqnarray}
H_{int} &=& -\int \frac{dz}{L} N \left[ \hbar \delta_1 \rho_{00} +
 \hbar (\delta_1-\delta_2) \sigma_{22} + \hbar (\delta_1-\delta_3) \rho_{33} \right. \nonumber \\
&\ &+ \hbar g_P \left(\hat{{\cal E}}_P\rho_{01} + \hat{{\cal
E}}_P^\dagger \rho_{10}\right) + \hbar g_T \left(\hat{{\cal
E}}_T\rho_{03} + \hat{{\cal E}}_T^\dagger \rho_{30}\right)\left .
+ \hbar \Omega (\rho_{02} + \rho_{20}) \right] \nonumber
\end{eqnarray}
where $L$ is the interaction length along the propagation axis
$x$, $g_{P,T}$ denotes coupling strengths of probe and trigger
fields $\hat{{\cal E}}_{P,T}$ to the respective atomic
transitions, and $\rho_{ij} = \frac{1}{N}\sum_k \rho_{ij}^{(k)}$
are the collective operators for the atomic populations and
transitions for $N$ atoms in a medium. The equations for probe and
trigger pulses propagating along $x-$axis through the tripod-media
are given by
\begin{eqnarray}
\left( \frac{\partial}{\partial t} + c \frac{\partial}{\partial x}
\right)\hat{{\cal E}}_P (x,t) &=& i g_P N \rho_{10} \\
\left( \frac{\partial}{\partial t} + c \frac{\partial}{\partial x}
\right)\hat{{\cal E}}_T (x,t) &=& i g_T N \rho_{30}
\end{eqnarray}
while the equations for atomic transition operators (with all
detunings vanishing) are
\begin{eqnarray}
\dot{\rho}_{10} &=& -\gamma_{10}\rho_{10} + i g_P \hat{{\cal E}}_P
(\rho_{11}-\rho_{00})+ i g_T \hat{{\cal E}}_T \rho_{13} + i \Omega  \rho_{12} \nonumber \\
\dot{\rho}_{20} &=& -\gamma_{20}\rho_{20} + i g_P \hat{{\cal E}}_P
\rho_{21} +
i g_T \hat{{\cal E}}_T \rho_{23} + i \Omega (\rho_{22} - \rho_{00}) \nonumber \\
\dot{\rho}_{30} &=& -\gamma_{30}\rho_{30} + i g_P \hat{{\cal E}}_P
\rho_{31}+ i g_T \hat{{\cal E}}_T (\rho_{33} - \rho_{00}) + i \Omega \rho_{32}  \\
\dot{\rho}_{12} &=& -\gamma_{12}\rho_{12} - i g_P \hat{{\cal E}}_P \rho_{02} + i \Omega \rho_{10} \nonumber \\
\dot{\rho}_{13} &=& -\gamma_{13}\rho_{13} - i g_P \hat{{\cal E}}_P \rho_{03} + i g_T \hat{{\cal E}}_T^\dagger \rho_{10} \nonumber \\
\dot{\rho}_{23} &=& -\gamma_{23}\rho_{23} + i g_T \hat{{\cal
E}}_T^\dagger \rho_{20} - i \Omega \rho_{03} \nonumber
\end{eqnarray}
where $\gamma_{\mu \nu}$ are the relative dephasing between the
levels $\ket{\mu}$ and $\ket{\nu}$.
\par
We now proceed by assuming the low intensity probe and trigger,
$g_j \langle \hat{{\cal E}}_{P,T} \rangle \ll \Omega$ and strong
pump $|\Omega|^2/\gamma_{0j}\gamma_{ij} \gg 1$. The latter
condition also implies that the EIT resonances for both, probe and
trigger fields, are strongly saturated. Furthermore, if $g_P \sim
g_T \sim g$, for a probe and trigger fields of equal mean
amplitudes $\langle \hat{{\cal E}} \rangle$, we can assume
$\langle \rho_{00} \rangle \approx \langle \rho_{22} \rangle
\approx 0$ and $\langle \rho_{11} \rangle \approx \langle
\rho_{33} \rangle \approx \frac{1}{2}$. \par Besides, following
the procedure in chapter \ref{qm}, in the low-intensity
approximation we have \bgar \rho_{10}= -\frac{i}{\Omega}
\frac{\partial}{\partial t}
\rho_{12} \\
\rho_{30}= -\frac{i}{\Omega} \frac{\partial}{\partial t} \rho_{32}
\enar
Hence in the adiabatic limit one finds
\bgar \rho_{12}= -\frac{g}{\Omega} \Big[\frac{\hat{{\cal E}}_P}{2}+\hat{{\cal E}}_T \rho_{13}\Big] \\
\rho_{32}= -\frac{g}{\Omega} \Big[\frac{\hat{{\cal
E}}_T}{2}+\hat{{\cal E}}_P \rho_{31}\Big] \enar Therefore one
arrives at the equations for the pulse propagation
\begin{eqnarray}
\label{ep} \left( \frac{\partial}{\partial t} + c
\frac{\partial}{\partial x} \right) \hat{{\cal E}}_P (x,t) &\cong&
-\frac{{g}^2 N}{\Omega} \frac{\partial}{\partial
t}\Big[\frac{\hat{{\cal E}}_P}{2 \Omega}+\frac{\hat{{\cal
E}}_T}{\Omega} \rho_{13}\Big]
\\
\left( \frac{\partial}{\partial t} + c \frac{\partial}{\partial x}
\right) \hat{{\cal E}}_T (x,t) &\cong& -\frac{{g}^2 N}{\Omega}
\frac{\partial}{\partial t}\Big[\frac{\hat{{\cal E}}_T}{2
\Omega}+\frac{\hat{{\cal E}}_P}{\Omega} \rho_{31}\Big] \label{et}
\end{eqnarray}
where \bgar \rho_{13}=-\gamma_{13} \rho_{13}-\frac{\hat{{\cal
E}}_P}{N} \left( \frac{\partial}{\partial t} + c
\frac{\partial}{\partial x} \right) \hat{{\cal E}}_T (x,t)
-\frac{\hat{{\cal E}}_T}{N}\left( \frac{\partial}{\partial t} + c
\frac{\partial}{\partial x} \right) \hat{{\cal E}}_P (x,t)
\nonumber \enar Note that the Eqs. (\ref{ep}-\ref{et}) emphasize
the full symmetry between the probe and the trigger dynamics, a
symmetry which is intimately linked with the atomic population
being equally distributed between levels $\ket{1}$ and $\ket{3}$.
Besides in the regime considered here the coupling between two
fields is substantially zero. Therefore two field propagates
through atomic system independently.
\par
Let us introduce again a rotation in the space of physically
relevant variables --  the electric field $\hat {\cal E}$ and the
atomic spin coherences -- defining four new quantum fields $\hat
\Psi_1(x,t)$, $\Phi_1(x,t)$ for the probe and $\Psi_2(x,t)$ and
$\Phi_2(x,t)$ for the trigger field, as follows \bgar
\hat\Psi_1(x,t) &=& \cos\theta(t)\, \hat {\cal E}_P(x,t) -
\sin\theta(t)\, \sqrt{N}\,
\rho_{10}(x,t)\, {\rm e}^{i\Delta k x}\\
\hat\Phi_1(x,t) &=& \sin\theta(t)\, \hat {\cal E}_P(x,t) +
\cos\theta(t)\, \sqrt{N}\, \rho_{10}(x,t)\, {\rm e}^{i\Delta k x}
\enar \bgar \hat\Psi_2(x,t) &=& \cos\theta(t)\, \hat {\cal
E}_T(x,t) - \sin\theta(t)\, \sqrt{N}\,
\rho_{30}(x,t)\, {\rm e}^{i\Delta k x}\\
\hat\Phi_2(x,t) &=& \sin\theta(t)\, \hat {\cal E}_T(x,t) +
\cos\theta(t)\, \sqrt{N}\, \rho_{30}(x,t)\, {\rm e}^{i\Delta k x}
\enar where the mixing angle $\theta(t)$ is defined as \be
\tan^2\theta(t) = \frac{g^2 N}{\Omega^2(t)} \ee In the
low-intensity approximation and in the adiabatic limit, we obtain
(as for one three--level system) these two PDE's
\begin{eqnarray}
\left[\frac{\partial}{\partial t}+c\cos^2\theta(t)
\frac{\partial}{\partial x}\right]\hat\Psi_1(x,t)=0 \qquad
\left[\frac{\partial}{\partial t}+c\cos^2\theta(t)
\frac{\partial}{\partial x}\right]\hat\Psi_2(x,t)=0\nonumber
\end{eqnarray}
that describe the propagations of the two polaritons associated to
two polarized photons, i.e. with $\sigma_{-}$ and $\sigma_{+}$
polarization.
\par
Therefore it confirms that this system represents a good tool to
realize a polarization quantum memory for photons. Moreover, as
seen in chapter \ref{qm}, by turning on two incoherent RF field
($\gamma_1^{(1)}$ and $\gamma_1^{(2)}$) we can put population in
the level 2 and have amplification without inversion to compensate
the unavoidable losses. So perhaps it would be possible to obtain
also a quantum (continuous-variable) cloning of a quantum state of
polarized photons into a quantum memory. However, the relative
fidelity of cloning process will be less than one according to
No-Cloning Theorem (see \ref{noclon}).

\begin{figure} [!ht]
\begin{center}
\includegraphics[width=.94\textwidth ]{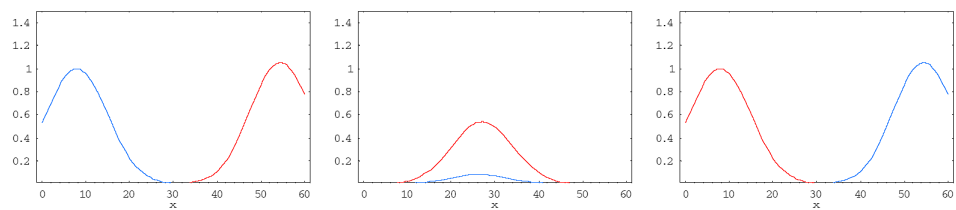}
\caption{Scattering between two dark--state polaritons.
Propagation of the electric fields, ${\cal E}_P=<\hat {\cal E}_P>$
(blu line) and ${\cal E}_T=<\hat {\cal E}_T>$ (red line), with
envelope $\exp\{-(x/10)^2\}$ for three time slices corresponding
to $\theta(t)=0,\ 3\pi/7,\ 0$, in ideal EIT regime. The mixing
angle is rotated from $0$ to $\pi/2$ and back according to
$\cot\theta(t)=0.8(1-0.5 \tanh[0.1(t-15)]+ 0.5\tanh[0.1(t-125)])$.
Parameters are $g^2N=1/100$ and axes are in arbitrary units with
$c=1$.}\label{fig63}
\end{center}
\end{figure}
In Fig. \ref{fig63} we show the scattering between two DSP in this
tripod configuration and we show that they present a solitonic
behavior, i.e. they propagate unperturbed in polarization quantum
memory. Let us point out that off-resonance and in presence of a
detuning mismatch a relative phase between two dark state
polaritons appears, as shown for polarization quantum phase gate
in \cite{Rebic03}. Then in different regimes this nice system
behaves as or quantum phase gate or polarization quantum memory.

\chapter*{Conclusions and Outlook} \pagestyle{fancy}\chaptermark{Conclusions and Outlook}
\addcontentsline{toc}{chapter}{Conclusions and Outlook}

In this thesis we have investigated some proposals for quantum
memories by analyzing the propagation of a coherent light pulse
through a three-level atomic system. Under particular conditions
we have showed that it is possible to obtain anomalous and
retarded light propagation and that a very slow propagation
involves a very large amplification of a probe field. This
achievement is based on the phenomenon of electromagnetically
induced transparency (EIT), which leads to steep normal dispersion
of the medium accompanied by vanishing absorption.
\newline
EIT happens for a weak probe field resonant or quasi-resonant with
an atomic transition when a strong drive field is applied to an
adjacent atomic transition under condition of two--photon
resonance. Therefore ultraslow light propagation is possible when
the two--photon atomic coherence decays much slower than
one-photon coherence. This behavior is similar to that of a pulse
propagating in any ``two--level'' dispersive medium and it can be
explained in terms of polaritonic modes. However, unlike usual
polaritons, almost no energy is stored in the coherent media where
ultraslow light is observed. Instead, the coherent media work as a
transducer between a slow light pulse and coupling electromagnetic
wave. \par Ultraslow light and atomic coherence have already found
many important applications in low-intensity nonlinear optics and
metrology. Novel applications in quantum nonlinear optics and
quantum information processing seem to be feasible. Indeed, for
example, the coherence allows one to store information about the
probe light and transport it in space. It allows time reversing of
the light and allows one to increase coupling between light fields
such that it becomes possible to study the interaction between
single photons. There are a lot of new applications of this
coherence and the list is growing fast.
\par
Here, in particular, we have discussed how to couple the light to
an atomic system, i.e. a dense gas of three--level atoms, and to
imprint the information carried by the photons onto the atoms,
specifically as a coherent pattern of atomic spins. This procedure
is reversible and the information stored in the atomic spins can
later be transferred back to the light field, reconstituting the
original pulse. In this context we have analyzed the propagation
of a quantum field in an EIT medium sustaining ``dark state
polaritons'' in the quasi-particle picture; then we have studied
the decoherence effects in this quantum memory for photons, by
recalling some results about the fidelity of the quantum state
transfer. We have shown how it is possible to extend the DSP
concept in presence of inversionless gain. In this novel regime we
have also discussed the possibility of simulating parastatistics
showing how the bosonic spectrum is compressed while being mapped
onto a fermionic spin $1/2$ system.
\par
Besides we have introduced a polarization quantum memory for
photons in EIT regime, by using a four--level atomic
configuration, and investigated the DSP solitonic behavior in the
scattering of two dark--state polaritons in a tripod atomic
configuration.
\par
As regards the experimental implementations of these ideas, an
intriguing question is whether EIT effects can be induced in solid
state media. In general the high dephasing rates in solid state
systems will prevent the development of strong coherence unless a
large coupling, sufficiently large that it may risk causing
optical damage to the medium, is employed. Given this limitation
it remains to be seen how solid state EIT effects can be further
developed. To an extent the beauty of EIT is that it leads to
exceptionally high efficiency non-linear optical processes and
potentially high gains (non-inverted) in a gas phase medium. The
high nonlinear conversion efficiencies are of a magnitude normally
associated with non-linear frequency mixing in optical crystals.
Thus a renewed interest in gas phase non-linear optical devices,
possessing unique capabilities (e.g. high conversion efficiencies
into the XUV and far-IR), seems likely. \par Important potential
applications such as quantum communications over long distances
might be implemented by combining an EIT-based memory with linear
optical elements. Moreover quantum information processing with
continuous variables represents an interesting alternative to the
traditional qubit-based approach. Indeed, regarding quantum
communication applications, as for example quantum teleportation
\cite{Furusawa98} or quantum key distribution (QKD)
\cite{Grosshans03}, continuous variables (CV) are particularly
promising. Another important feature of CV is the feasibility of
the light-atoms quantum interface
\cite{Kuzmich03,Julsgaard-Nature-2001}, which unlike its qubit
analogue does not require strongly coupled cavity QED regime for
deterministic operations. For these reasons, the idea of a quantum
memory for light with macroscopic atomic ensembles has been
explored in the last years
\cite{Hald-PRL-1999,Kuzmich-PRL-2000,Schori-PRL-2002,Kozhekin}.
Such a quantum memory is crucial for applications such as quantum
repeaters or quantum secret sharing.
\par
In this field macroscopic atomic ensemble could be also a good
tool to realize an optimal cloning machine. We have mentioned that
it could be possible realize a quantum cloning into a quantum
memory, but in the same device, by using the amplification in an
EIT medium.
\par
Finally let us cite an other important application of a quantum
memory: quantum cryptography. In the field of quantum information,
quantum key distribution (QKD) is the application which is more
developed, to the point that already commercial prototypes exist.
This fact is a good indicator of just how much attention the
subject has received in the last years. In this field it would be
interesting implement experimentally an optimal eavesdropping
strategy in a quantum channel between two communication agents,
Alice and Bob \cite{caruso}. If the cloning is used as an
eavesdropping attack, then the eavesdropper, Eve, would like to
store her clone in the memory while the second clone should be
sent as a light pulse down the communication line. Afterwards Eve
needs to transfer one of the clones from the atomic memory back to
light in order to extract the maximal information by a von Neumann
measurement of the quantum state, encoded in photon. In this
context the quantum memory for polarized photons play a key role
for implementing some QKD schemes and for testing the theoretical
security of the quantum key distribution protocols, necessary in
our global communication era.

\begin{figure} [!ht]
\begin{center}
\includegraphics[width=.65\textwidth ]{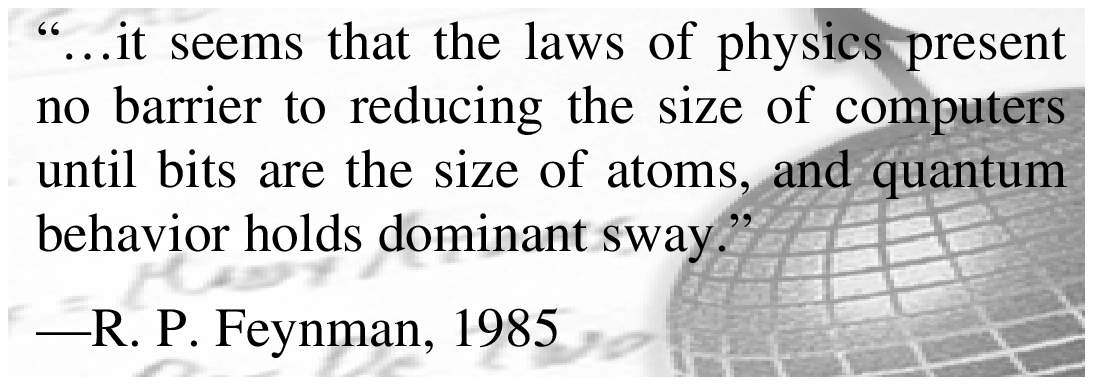}
\label{fig64}
\end{center}
\end{figure}

\appendix
\chapter{Appendix}

\section{Lasing without inversion}\label{lwi}

A \textbf{laser}\index{laser} is a light source that produces a
beam of highly coherent and very nearly monochromatic light as
result of cooperative emission from many atoms. The name laser is
an acronym for \textbf{light amplification by stimulated emission
of radiation} \cite{laser}. The main ingredients for a laser are
an optical cavity and ``gain medium'', i.e. a system capable of
amplifying electromagnetical radiation. When we think of gain
media, we automatically associate \textit{stimulated emission} and
\textit{population inversion} \index{population inversion} with
the lasing process. Both of these processes are considered
necessary in order to accomplish ``light amplification''; indeed
all traditional laser systems require a population inversion
because the probability of an initially unexcited atom absorbing a
photon is equal to the probability of stimulated emission from an
excited atom [\ref{einstein}]. As a result we need to create a
non-equilibrium situation in which the number of atoms in a
higher-energy state is greater than the number in the lower-energy
state, such a situation is called \textbf{population
inversion}\index{population inversion}. Then the rate of energy
radiation by stimulated emission can exceed the rate of absorption
and the system acts as a net source of radiation. The photons
being the result of stimulated emission have all the \textit{same
frequency}, \textit{phase}, \textit{polarization} and
\textit{direction of propagation}. Hence the resulting radiation
is much more coherent than the light form ordinary sources. From
this argument it would be acceptable to suggest that, without a
population inversion, lasing would be impossible. Is it really
true?
\par
More to the point we need an atom that is more likely to emit
radiation than absorb it. This is where the effect of EIT
\index{EIT} and the huge benefits of it come into play. Therefore
it is possible to lase without population inversion (LWI) in EIT
regime. Nevertheless it is noted that although lasing is achieved
without requiring an inversion of the population, another laser
has to be used in order to achieve this result. With relevance to
existing experiments not a lot has been gained but, for the
future, a number of exciting possibilities and opportunities have
been opened up.
\par
The advantages of lasing without inversion are
\begin{itemize}

\item[1)] Much lower power pump beams required.

\item[2)] Transitions which normally shouldn't lase can be made to
lase.

\item[3)] X-Ray \index{X-Ray} and XUV resonant transitions can be
made to lase.

\end{itemize}

Finally, one could make to laser transitions at short wavelengths,
which is impossible with existing systems. This is due to the fact
that it is very difficult to create population inversions at X-ray
wavelengths (below 200nm) as result of the rapid decay of the
excited states via spontaneous emission. However using EIT and
quantum interference ideas an alternative may be discovered with
the eventual goal of a conventional laser working at X-ray
wavelengths.

\subsection{Einstein's coefficients} \label{einstein}
\index{Einstein's coefficients}

At the beginning of the century, the combination of Bohr's atomic
model \index{Bohr's atomic model} with a stochastic conception of
the light-matter interaction allowed Einstein to establish the
existence of three basic processes in the interaction of light
with matter: \textbf{absorption}, \textbf{stimulated emission} and
\textbf{spontaneous emission}. In the context of black-body
radiation and from thermodynamical arguments, Einstein determined
(in terms referred to as Einstein \textit{\textbf{A}} and
\textit{\textbf{B}} coefficients) \index{Einstein's coefficients}
the relationship between the rates of absorption and stimulated
emission of light by a gas molecule possessing discrete energy
levels and the spontaneous emission rate (given by the Einstein
\textit{\textbf{A}} coefficient). For a two-level system with
ground level $|g \rangle$ and excited level $|e \rangle$ with
unperturbed transition frequency $\omega$ probed by an
electromagnetic field, the relationships between the Einstein
coefficients read

\be B_{abs}=B_{est} \label{eq1} \ee \be \label{eq2}
\frac{A}{B}=\frac{\hbar \omega^{3}}{\pi^{2} c^{3}} \ee
 with
\be A=\frac{\omega^{3}}{3 \pi \epsilon_{0} \hbar
c^{3}}|\mu|^{2}\ee where $\mu$ is the electric dipole moment of
the two-level transition, $\hbar$ Planck's constant,
$\epsilon_{0}$ the vacuum electric permittivity, and $c$ the speed
of light in vacuum. Thus, the rates of absorption and stimulated
emission of light by an atomic/molecular medium will be
proportional to the corresponding Einstein B coefficient times the
population $\rho_{ii}~(i=g,e)$ of the initial state of the
process. Then,

\bgar \frac{absorption \cdot rate}{stimulated \cdot emission \cdot
rate}=\frac{B_{abs}}{B_{est}} \frac{\rho_{gg}}{\rho_{ee}}
\label{eq3} \enar From the above equations, it can be deduced
straightforwardly that since $B_{abs}=B_{est}$, then population
inversion, i.e. $\rho_{ee}>\rho_{gg}$, is a necessary condition
for light amplification and it is not possible to invert a
two-level system with, for instance, resonant light, since atomic
excitation is always accompanied with de-excitation through the
stimulated emission process. As a consequence, in order to create
the required population inversion, laser systems operate on three-
or more level configurations (i.e. the three--level configuration
of the \textbf{Ruby laser} \index{Ruby laser} or the four-level
configuration of the \textbf{He-Ne laser}\index{He-Ne laser}),
where the inversion is created by pumping to an excited level
which rapidly decays to the upper level of the lasing transition.
\par
In these schemes, in order to reach laser action, the incoherent
pump power $\textit{P}$ should be strong enough to create a
population inversion such that the associated gain overcomes
cavity losses. A simple estimation of the lower bound $P_{th}$ for
the pump power required to reach laser oscillation gives

\bgar P_{th}=\frac{\hbar \omega \Delta N_{th}}{\tau} \enar

where $\Delta N_{th}>0$ is the population difference at the first
laser threshold and $\tau$ the lifetime of the upper level of the
lasing transition. This threshold population can be written as
$\Delta N_{th}=\frac{2\kappa}{\hbar \omega B(\omega)}$, $\kappa$
being the cavity losses and $B(\omega)$ the Einstein $B$
coefficient at frequency $\omega$.
\par
For a broadened atomic system with normalized spectral line given
by $g(\omega)$ such that $B(\omega)=Bg(\omega)$, the above
expression becomes: \bgar \label{eq5}
P_{th}=\frac{2\kappa}{g(\omega)}\frac{A}{B}=\frac{2\kappa\hbar
\omega^{3}}{\pi^{2} c^{3} g(\omega)} \enar where use of relation
(\ref{eq2}) has been made. In general, and due to the
normalization of $g(\omega)$, one has $g(\omega) \cdot \Delta
\omega \sim 1$. On the other hand, for natural broadening and
using the Heisenberg principle one has $\Delta \omega \sim 1/
\tau$ with $1/\tau=A$, while for Doppler broadening it is well
known that $\Delta \omega$ scales with $\omega$ and does not
depend on the electric dipole moment. Therefore, according to
equation (\ref{eq5}) the threshold pump power scales with
$\omega^{6}$ and $|\mu|^{2}$ for natural broadening, while for
Doppler broadening it scales with $\omega^{4}$ and does not depend
on the dipole moment. All these arguments clearly show that as we
increase the frequency of the laser transition it becomes harder
to attain the required population inversion. Therefore, in
continuous-wave x-ray lasing, the main obstacle for the
achievement of coherent oscillation is the required pump power.
\par
LWI (\textit{\textbf{lasing without inversion}}) \index{LWI}
differs from conventional lasing in that the reciprocity between
absorption and stimulated emission is broken. In LWI the
absorption of light is reduced or even cancelled and lasing is
possible even with a small fraction of the atoms in the upper
level (less than in the lower level) of the lasing transition.
Therefore, LWI is not subject to the limitations of conventional
lasing, i.e. the required incoherent pump power can be drastically
reduced and, in this sense, LWI \index{LWI} opens a new way
towards the generation of continuous-wave short-wavelength lasers.

\subsection{Quantum-jump approach to LWI}
\index{quantum-jump approach}

The quantum-trajectory \index{quantum-trajectory formalism} or
quantum-jump formalism could be used in order to calculate the
individual contribution of the different physical processes
(one-photon and two-photon gain/loss processes) responsible for
inversion-less amplification. This formalism gives the same
results as the standard density-matrix formalism
\index{density-matrix formalism} but provides new insights into
the underlying physical mechanisms.
\par
In this formalism, the time evolution of the atom plus lasers
system is pictured as consisting of a series of coherent evolution
periods separated by quantum-jumps occurring at random times, i.e.
a so-called \textit{\textbf{quantum trajectory}}
\index{quantum-trajectory formalism}. The quantum-jumps are
determined by dissipative processes, such as spontaneous emission
or incoherent pumping, while the continuous evolution is governed
by the coherent laser fields. Thus, a one-photon gain (loss)
process is a coherent evolution period between two consecutive
quantum-jumps such that the probe field photon number increases
(decreases) by one with no change in the driving field photon
number. A two-photon gain process corresponds to a coherent
evolution period for which the probe field photon number increases
by one and the driving field photon number decreases by one in $V$
and $\Lambda$ schemes or increases by one in the cascade
schemes.\newline
\par
Using this technique, Arimondo \index{Arimondo} \cite{arimondo}
and Cohen-Tannoudji et al.\index{Cohen-Tannoudji} \cite{cohen}
revealed that AWI (\textbf{\textit{amplification without
inversion}}) \index{AWI} in folded schemes results from the fact
that, for appropriate parameter values, two-photon gain processes
overcome one-photon and two-photon loss processes even without
one-photon or two-photon inversion. By contrast the quantum-jump
formalism shows that the one-photon gain is the physical process
responsible for inversionless gain in cascade schemes.
\par
One of the benefits of the quantum-jump formalism relies on the
fact that the knowledge of the particular physical processes
responsible for inversionless gain allows one to select
appropriate probe and driving field detunings to stimulate these
processes while at the same time preventing the unfavorable ones.
\par
As discussed above, the population inversion requirement for
conventional lasers to operate is a direct consequence of the
symmetry between the Einstein B coefficients for one photon
processes, in two-level lasers, and for two-photon processes, in
Raman lasers. The quantum-jump formalism has been used very
recently to define generalized Einstein B coefficients for one-
and two-photon gain and loss processes \cite{mompart}. Thus, it
has been shown that in three--level systems coherently driven
close to resonance there is a symmetry breaking between the
Einstein $B$ coefficients for one-photon and two-photon processes.

\section{Causality and Einstein's
relativity}\label{einstein1}\index{Einstein}\index{causality}\index{relativity}

Einstein's theory of special relativity and the principle of
causality imply that the speed of any moving object cannot exceed
that of light in vacuum, $c$. Nevertheless, there exist various
proposals for observing faster-than-c propagation of light pulses,
using anomalous dispersion near an absorption line, nonlinear and
linear gain lines, or tunnelling barriers. In particular, the
group velocity of a laser pulse in EIT region exceeds c and can
even become negative; in other words, the pulse appears at the
exit side so much earlier than if it had propagated the same
distance in vacuum that the peak of the pulse appears to leave the
cell before entering it (while the shape of the pulse is
preserved). However the observed superluminal light pulse
propagation is not at odds with causality, being a direct
consequence of the interference between its different frequency
components in an anomalous dispersion region.
\par
When a light pulse of frequency $\nu$ and bandwidth $\Delta \nu$
enters a dispersive linear medium of an optical refractive index
$n(\nu)$, the light pulse propagates at the group velocity
\index{group velocity} $v_{g}=\frac{c}{n_{g}}$, where
$n_{g}=n(\nu)+\nu \frac{dn(\nu)}{\nu}$ is the group velocity
index. If the group velocity index remains constant over the pulse
bandwidth $\Delta \nu$, the light pulse maintains its shape during
propagation. In recent experiments involving electromagnetically
induced transparency (EIT), the group velocity index was greatly
enhanced using the lossless normal dispersion region between two
closely spaced absorption lines; thus the group velocity of light
was dramatically reduced to as slow as $8 \ m/s$ \cite{hau}.
Conversely, between two closely spaced gain lines, an anomalous
dispersion region appears where $\nu \frac{dn(\nu)}{\nu}$ is
negative and its magnitude can become large. In this situation,
the group velocity of a light pulse can exceed c and can even
become negative \cite{chiao}.
\par
A negative group velocity of light is counterintuitive but can be
understood as follows. For a medium of a length $L$, it takes a
propagation time $\frac{L}{v_{g}}=\frac{n_{g}L}{c}$ for a light
pulse to traverse it. Compared with the propagation time for light
to traverse the same distance in vacuum, that is the vacuum
transit time $\frac{L}{c}$, the light pulse that enters the medium
will exit at a moment that is delayed by a time difference $\Delta
T=L/v_{g}-L/c=(n_{g}-1)L/c$. When $n_{g}<1$, the delay time
$\Delta T$ is negative, resulting in an advancement. In other
words, when incident on a medium with group velocity index
$n_{g}<1$, a light pulse can appear on the other side sooner than
if it had traversed the same distance in vacuum. Furthermore, in
contradiction to traditional views that a negative group velocity
of light has no physical meaning, when the group velocity index
becomes negative, the pulse advancement $-\Delta T=(1-n_{g})L/c$
becomes larger than the vacuum transit time $L/c$. In other words,
it appears as if the pulse is leaving the cell even before it
enters. This counterintuitive phenomenon is a consequence of the
wave nature of light.
\par
In this thesis we have showed that in the classical theory of wave
propagation in an anomalous dispersion region the interference
between different frequency components produces this rather
counterintuitive effect. Let us point out that the observed
superluminal light pulse propagation is not at odds with causality
or special relativity. Indeed the very existence of the lossless
anomalous dispersion region is a result of the
\textit{\textbf{Kramers-Kronig relation}}\index{Kramers-Kronig
relation} which itself is based on the causality requirements of
electromagnetic responses. Remarkably, the signal velocity of a
light pulse, defined as the velocity at which the half point of
the pulse front travels, also exceeds the speed of light in
vacuum, $c$, as it is showed in the experiment in \cite{wang}. It
has also been suggested that the true speed at which information
is carried by a light pulse should be defined as the ``front''
velocity \index{front velocity} of a step-function-shaped signal
which has been shown not to exceed c \cite{somm}.

\section{The \textit{fidelity}}\index{fidelity}\label{fidelity}

The \textit{fidelity} measures the \textit{distance} between two
quantum states, $\rho$ and $\sigma$, in a Hilbert space. Uhlmann
\cite{uhlmann} defines it as follows:
 \be
 F(\rho,\sigma) \equiv \Big(\tr{\sqrt{\rho^{1/2} \sigma
 \rho^{1/2}}}\Big)^2
 \ee
It assumes values in the range $[0,1]$ and, for example, when the
\textit{fidelity} is one then two quantum states are equal.
\par
Let us restrict to two special cases in which it is possible to
give the fidelity a more explicit form. The first one is when
$\rho$ e $\sigma$ commute, i.e. diagonal in the same basis,
 \bgar
 \rho=\sum_i{r_i \proj{i}} \ \ \ \ \  \ \ \ \ \ \sigma = \sum_i{s_i
 \proj{i}}
 \enar
where $\{\ket{i}\}$ is an orthonormal basis in the Hilbert space
associated to a particular quantum system.
 \par
In this case the \textit{fidelity} is
  \bgar
  F(\rho,\sigma) = \Bigg(\tr{\sqrt{\sum_i{r_i s_i \proj{i}}}}\Bigg)^2= \nonumber \\
  = \Bigg(\tr{\sum_i{\sqrt{r_i s_i} \proj{i}}}\Bigg)^2= \nonumber \\
  = \Big(\sum_i{\sqrt{r_i s_i}}\Big)^2=F(r_i, s_i)
  \enar
It is easy to show that this quantity is the \textit{classical
fidelity}, $F(r_i,s_i)$, between the distributions of eigenvalues,
$r_i$ and $s_i$, respectively, of $\rho$ and $\sigma$. Indeed, if
we consider two any classical probability distributions, $\{p_x\}$
and $\{q_x\}$, the \textbf{classical fidelity}\index{classical
fidelity} is defined as:
  \be
 F(p_x,q_x) \equiv \Big(\sum_x{\sqrt{p_x q_x}}\Big)^2
  \ee \par
The second example, in which a more explicit form for the
\textit{fidelity} does exists, is represented by \textit{fidelity}
between a pure state, $\ket{\psi}$, and a generic quantum state,
$\rho$. In this circumstance we have \bgar
F(\ket{\psi},\rho)=\Big(\tr{\sqrt{\bra{\psi} \rho \ket{\psi}
\proj{\psi}}} \Big)^2=\bra{\psi}\rho \ket{\psi} \enar that is the
mean value of $\rho$ in the state $\ket{\psi}$.

\section{No-Cloning Theorem}\label{noclon}\index{no-cloning theorem}

\newtheorem{teo1}{Theorem (no-cloning)}[chapter]
\begin{teo1} \label{teo1}
\
\newline
Quantum Mechanics forbids any device that clones perfectly any
unknown quantum state; nevertheless cloning only orthogonal states
is possible.
 \end{teo1}
 \textit{Proof}
 \par
Consider a cloning machine with two slots labelled with $A$ and
$B$, where $A$ is used for the unknown state, $\ket{\psi}$, and in
$B$ one will obtain the clone. Initially a certain pure state,
$\ket{s}$ (normalized), is in $B$ and therefore the initial state
of the cloning machine is
 \be
 \ket{\psi} \otimes \ket{s} \otimes \ket{a}
 \ee
where $\ket{a}$ is a normalized state of a possible auxiliary
system, i.e. \textit{ancilla}.
\par
In order to have a cloning, without loss of generality, we
consider a generic unitary evolution, as follows
 \be
\ket{\psi} \otimes \ket{s} \otimes \ket{a} \longrightarrow U(
\ket{\psi} \otimes \ket{s} \otimes \ket{a})= \ket{\psi} \otimes
\ket{\psi} \otimes \ket{a_\psi}
 \ee
where $\ket{a_\psi}$ is the normalized final state of
\textit{ancilla}, eventually depending on $\ket{\psi}$.
\par
If this procedure does work for two particular normalized quantum
states, $\ket{\psi}$ and $\ket{\phi}$, we have
 \bgar
U( \ket{\psi} \otimes \ket{s} \otimes \ket{a}) &=& \ket{\psi}
\otimes \ket{\psi} \otimes \ket{a_\psi}
\nonumber \\
U( \ket{\phi} \otimes \ket{s} \otimes \ket{a}) &=& \ket{\phi}
\otimes \ket{\phi} \otimes \ket{a_\phi} \enar

If now we take the scalar products between these two equations,
terms by terms, one obtains
 \be
 \sca{\psi}{\phi}=(\sca{\psi}{\phi})^2 \ \sca{a_\psi}{a_\phi}
 \ee
Therefore, if the two states, $\ket{\psi}$ and $\ket{\phi}$, are
not orthogonal, the following equation is satisfied
 \be
1/\sca{\psi}{\phi}=\sca{a_\psi}{a_\phi}
 \ee
that is wrong because, for the normalization of the states,
$|\sca{a_\psi}{a_\phi}|<1$ while $|1/\sca{\psi}{\phi}|>1$.
\par
So this theorem is proved; this result can be extended also to
no-unitary transformations and to mixed states \cite{barnum}.

\chapter*{Acknowledgements} \pagestyle{fancy}\chaptermark{Acknowledgements}
\addcontentsline{toc}{chapter}{Acknowledgements}

Sintetizzare quattro anni alla Scuola Superiore di Catania \`e
certamente impresa ardua, ma \`e tuttavia possibile e doveroso
ringraziare quanti hanno creduto ed hanno permesso la sua nascita
ed il suo buon funzionamento.
\par
Entrare nella Scuola al suo secondo anno di vita mi ha permesso di
seguire da vicino la sua crescita con le inevitabili difficolt\`a
ma anche apprezzare i grandi passi che ha fatto grazie agli sforzi
di quanti ci hanno lavorato ma anche per merito degli studenti che
con le loro assemblee hanno avanzato tante proposte per
migliorarla continuamente.
\par
Certamente un grazie particolare spetta alla persona che ha
``fatto nascere'' la Scuola Superiore di Catania e ne ha guidato
con la sua \hyphenation{e-spe-rien-za}esperienza lo sviluppo in
questi anni, il Presidente prof. Emanuele Rimini; soprattutto
grazie a lui ho potuto sfruttare al meglio le varie opportunit\`a
offerte e riuscire ad entrare nel mondo della ricerca, seguendo
varie conferenze internazionali e producendo diverse
pubblicazioni, gi\`a durante gli anni universitari: questo \`e il
vero e pi\`u importante bagaglio che mi porto dietro dopo questi
quattro anni. A tal \hyphenation{pro-po-si-to}proposito devo
ringraziare quanti mi hanno insegnato ed inizializzato al mondo
della ricerca \hyphenation{scien-ti-fi-ca}scientifica: i
professori dei corsi interni ed in particolare il prof. Andrea
\hyphenation{Rapisarda}Rapisarda ed il dott. Vito Latora. Con loro
ho condiviso tante belle esperienze, ho imparato diverse tecniche
di metodi numerici, programmare in Fortran e tanti aspetti
interessanti della Meccanica Statistica e della Fisica dei Sistemi
Complessi, oltre a scrivere alcune pubblicazioni su
riviste\hyphenation{ri-vi-ste}
\hyphenation{scien-ti-fi-che}scientifiche internazionali. Vorrei
ringraziare anche il tutor dell'area scientifica, il dott.
Giovanni Piccitto, che mi ha seguito durante gli anni
universitari, allontanando ogni timore, proponendo validi
approfondimenti di studio e dando sempre importanti consigli.\par
Inoltre ringrazio il relatore di questa tesi, il prof. Francesco
\hyphenation{Saverio}Saverio Cataliotti, che con la sua bravura mi
ha insegnato molti argomenti di Ottica Quantistica e soprattutto
mi ha dato la grande opportunit\`a di sviluppare un'attivit\`a di
ricerca teorica a fianco dell'interessante attivit\`a
\hyphenation{spe-ri-men-ta-le}sperimentale svolta nel Laboratorio
di Informazione Quantistica della Scuola Superiore di Catania. Ne
approfitto anche per ringraziare il prof. Giuseppe La Rocca, il
prof. Maurizio Artoni, la dott.ssa Chiara Macchiavello e la
dott.ssa Helle Bechmann-Pasquinucci per le utili discussioni e per
i suggerimenti riguardo allo sviluppo di questa tesi.
\par
Infine \`e doveroso ringraziare quanti lavorano in questa
struttura e sono stati al mio fianco in questi anni e lo far\`o
percorrendo i vari piani della
Residenza\hyphenation{re-si-den-za}. Innanzitutto devo ringraziare
di cuore la dott.ssa Lorella Alfieri e la dott.ssa Gabriella Lo
Re, che sono state con me sempre molto gentili ed
\hyphenation{af-fet-tuo-se}affettuose e con i loro sforzi mi hanno
aiutato a vivere al meglio questi quattro anni e ad usufruire
appieno dei vari servizi offerti. Un'altra persona per me
indimenticabile \`e il simpatico ragioniere Alberto Teodoru, che,
oltre a ``sostenerci'' economicamente con i vari pagamenti e
rimborsi, incontrandolo in Residenza ha sempre allietato con la
sua grande simpatia e le sue battute scherzose le varie giornate
intense di studio e qualche cena trascorsa assieme.
\par Comunque un grazie spetta a tutto il personale che lavora
intensamente nei vari uffici, importante per la crescita ed il
buon funzionamento della Scuola, ed anche alle simpatiche donne
delle pulizie, fondamentali per risparmiarci pure la ``faticosa''
pulizia della stanza. Per concludere con il piano terra, meritano
sentitamente un grosso grazie i tre custodi tutto-fare,
indispensabili per risolvere tanti problemi pratici, inevitabili
in una struttura residenziale. In particolare sono particolarmente
legato affettivamente ad Arturo e Fabrizio, che, oltre ad essere
simpaticissimi, mi sono stati vicino ed hanno fatto di tutto per
accontentarmi e rendere serena e confortevole la mia permanenza
nella Residenza.
\par
Andando al secondo piano, ringrazio particolarmente la direzione
scientifica, il prof. Agatino Russo, e il direttore
amministrativo, l'avv. Fabio Lo Presti, che, con il loro lavoro,
hanno permesso di organizzare al meglio le attivit\`a della Scuola
e che hanno supportato la mia attiva partecipazione a vari
congressi scientifici internazionali, permettendomi anche di
svolgere il lavoro di tesi a Pavia. Poi vorrei ringraziare la
dott.ssa Laura Vagnoni, che mi ha anche aiutato affettuosamente a
districarmi nelle varie faccende burocratiche e mi ha messo nelle
condizioni per svolgere al meglio i miei studi universitari.
\par
Salendo fino al terzo piano, ringrazio il simpatico e bravo
\hyphenation{re-spon-sa-bi-le}responsabile dei servizi
informatici, Ennio Li Volsi, il cui lavoro \`e fondamentale nella
gestione del laboratorio informatico, indispensabile per lo studio
di ogni allievo. Infine, ma non perch\`e meno importante,
ringrazio \hyphenation{af-fet-tuo-sa-men-te}affettuosamente il
responsabile della Residenza, il dott. Federico Manitta, che,
vivendo con noi, in questi anni ci \`e stato vicino pi\`u come un
fratello maggiore ed ha lavorato per noi anche a
\hyphenation{di-sca-pi-to}discapito della sua vita privata.
\par Quindi, giunti al terzo e quarto piano, ringrazio per la
simpatia e per la compagnia tutti gli studenti, destinati ad
essere i veri protagonisti del successo della Scuola. In
particolare ringrazio i miei due simpaticissimi ``compagnetti di
stanza'', Michele ed Ignazio, con i quali ho condiviso tanti bei
momenti ed anche la stanza per un anno; ci siamo divertiti tanto
assieme e mi hanno dato l'allegria e la serenit\`a anche per
affrontare al meglio il terzo anno universitario in Fisica, forse
il pi\`u difficile del mio corso di studi.
\par
Lasciando la Scuola, un ringraziamento particolare spetta ad
Elisa, che mi \`e stata tanto vicino ed ha sopportato con molta
comprensione tutti i miei impegni, ed un grazie di tutto cuore ai
miei genitori per tutto ci\`o che hanno fatto per me in questi
anni; un pensiero va pure a mio fratello Gianpiero che sta vivendo
una simile esperienza a Pisa, augurandogli un corso di studi
eccellente. Il loro affetto mi ha messo nelle condizioni ottimali
per svolgere al meglio i miei studi, la mia vita in Residenza e
quindi mi ha permesso serenamente di arrivare fin qui a poter
concludere questa tesi e scrivere questa pagina di ringraziamenti.

\pagestyle{fancy}\chaptermark{Analytical Index}
\addcontentsline{toc}{chapter}{Analytical Index}
\printindex

\pagestyle{fancy}\chaptermark{Bibliography}

\end{document}